%% file: main-arxiv.tex
\documentclass[letterpaper,11pt]{article}

\input{Arxiv-macro}

\input{math}

\newcommand{\revcolor}[1]{{\color{black}#1}}

\title{Dynamic Pricing and Advertising with Demand Learning}
\author{
Shipra Agrawal \thanks{Columbia University, 
\texttt{sa3305@columbia.edu}} \and
Yiding Feng\thanks{Hong Kong University of Science and Technology, 
\texttt{ydfeng@ust.hk}} \and
Wei Tang\thanks{The Chinese University of Hong Kong, \texttt{weitang@cuhk.edu.hk}}  
}
\date{\today}

\begin{document}

\maketitle

\begin{abstract}

\input{Paper/0-abstract}

\end{abstract}
\newpage

\section{Introduction}
\label{sec:intro}

\input{Paper/1-intro-revision}

\subsection{Related Work}
\label{subsec:related-work}
\input{Paper/1-2-related.tex}

\section{Problem Formulation}
\label{sec:setting}

\input{Paper/2-setting-new}

\section{Advertising as the Revenue Lever}
\label{sec:voa meta}

\input{Paper/3-limits}

\section{Dynamic Pricing and Advertising with Demand Learning}
\label{sec:alg design}
\input{Paper/4-algo}

\section{Regret Bound of \texorpdfstring{\Cref{algo:dynamic pricing and advertising}}{Algorithm 1} and  Proof Overview}
\label{sec:analysis}

\input{Paper/5-analysis}

\section{Conclusions and Future Directions}
\label{sec:conclusion}
\input{Paper/conclusions}

\bibliographystyle{plainnat}
\bibliography{mybib}

\newpage

\makeatletter

\renewcommand{\theHsection}{appendix.section.\thesection}
\providecommand*{\theHsubsection}{\thesubsection}
\renewcommand{\theHsubsection}{appendix.subsection.\thesubsection}

\providecommand*{\theHequation}{\theequation}
\renewcommand{\theHequation}{appendix.equation.\theequation}

\providecommand*{\theHfigure}{\thefigure}
\renewcommand{\theHfigure}{appendix.figure.\thefigure}

\providecommand*{\theHtable}{\thetable}
\renewcommand{\theHtable}{appendix.table.\thetable}

\providecommand*{\theHtheorem}{\thetheorem}
\renewcommand{\theHtheorem}{appendix.theorem.\thetheorem}

\providecommand*{\theHassumption}{\theassumption}
\renewcommand{\theHassumption}{appendix.assumption.\theassumption}

\providecommand*{\theHlemma}{\thelemma}
\renewcommand{\theHlemma}{appendix.lemma.\thelemma}

\providecommand*{\theHproposition}{\theproposition}
\renewcommand{\theHproposition}{appendix.proposition.\theproposition}

\providecommand*{\theHcorollary}{\thecorollary}
\renewcommand{\theHcorollary}{appendix.corollary.\thecorollary}

\providecommand*{\theHexample}{\theexample}
\renewcommand{\theHexample}{appendix.example.\theexample}

\makeatother

\appendix
\revcolor{\section{Illustrations of Real-World Applications}
\label{apx:example}
\input{Paper/apx-example}}

\section{Numerical Studies}
\label{sec:numerical}
\input{Paper/7-numerical}

\revcolor{\section{The Revenue Gap due to Misspecified Type Distribution}
\label{apx:rev-gap}

\input{Paper/apx-rev-gap-miss}}

\revcolor{\section{Comparisons with Prior Works with Known Type Distribution}
\label{apx:comparisons}
\input{Paper/apx-comparisons}}

\revcolor{\subsection{Sufficient Conditions for No-Information Advertising to be Optimal}
\label{apx:no-info-optimality}

\input{Paper/apx-sufficient-conditions-noinfor}}

\revcolor{\section{Missing Analysis and Proof of \texorpdfstring{\Cref{thm:voa bound}}{Theorem 1}}
\label{apx:voa proof}

\input{Paper/apx-proof-voa-new}}

\revcolor{
\section{Demand Learning against Alternative Benchmarks}
\label{apx:alternative-benchmark-learning}

\input{Paper/apx-alternative-benchmark-learning}
}

\section{Missing Proof of Section~\ref{sec:alg design}}
\label{apx:proof of complexity}

\input{Paper/apx-proof-complexity}

\section{Detailed Discussions and Proofs of Section~\ref{sec:analysis}}
\label{apx:main proof}
\input{Paper/apx-proof-analysis}

\section{Improved Regret Bounds for Additive Valuations} 
\label{subsec:improvement}
\input{Paper/6-improvement}

\label{apx:proof of improvement}
\input{Paper/apx-proof-improvement}

\end{document}

%% file: Arxiv-macro.tex
\usepackage[margin=1in]{geometry}
\usepackage{amsfonts,amsmath,amssymb,amsthm,graphicx,bm}
\usepackage{mathtools}
\usepackage{mathrsfs}
\usepackage{ifthen}
\usepackage{changepage}
\usepackage{enumerate}
\usepackage{scalerel}
\usepackage{accents}
\usepackage{enumitem}
\usepackage{float}      
\usepackage[utf8]{inputenc} 
\usepackage[T1]{fontenc} 
\usepackage{csquotes}
\usepackage{comment}
\usepackage{bbm}
\usepackage{array}
\usepackage{makecell}
\usepackage{booktabs}
\usepackage{xfrac}
\usepackage{thm-restate}
\usepackage{tikz}
\usepackage{pgfplots}
\usepackage{subcaption}
\usepackage{graphicx}
\usepackage{multirow}
\usetikzlibrary{patterns}
\usepgfplotslibrary{fillbetween}
\usetikzlibrary{intersections}
\usepackage{wrapfig}
\usetikzlibrary{positioning,fit,calc,arrows.meta}
\usepackage{fix-cm}
\usepackage[ruled,vlined,linesnumbered,algo2e]{algorithm2e}
\newcommand{\algcomment}[1]{\textcolor{blue}{\footnotesize{/* \texttt{{#1}}} */}}

\usepackage{xcolor}
\newenvironment{myprocedure}[1][htb]
{  
\begin{algorithm2e}[#1]%
}{\end{algorithm2e}}

\usepackage{natbib}
 \bibpunct[, ]{(}{)}{,}{a}{}{,}%

\usepackage[suppress]{color-edits}  
\addauthor{yf}{purple}    
\addauthor[Wei]{wt}{red}
\addauthor[Wei]{wtres}{red}

\allowdisplaybreaks
\providecommand{\halmos}{\qedsymbol}
\theoremstyle{plain}
\newtheorem{theorem}{Theorem}[section]
\newtheorem{lemma}[theorem]{Lemma}
\newtheorem{claim}[theorem]{Claim}

\newtheorem{proposition}[theorem]{Proposition}

\theoremstyle{plain}
\newtheorem{definition}{Definition}[section] 
\newtheorem{example}[definition]{Example}
\newtheorem{remark}[definition]{Remark}

\theoremstyle{plain}
\newtheorem{assumption}{Assumption}[section]
\makeatletter
\providecommand*{\theHtheorem}{}
\renewcommand*{\theHtheorem}{\theHsection.\arabic{theorem}}
\providecommand*{\theHlemma}{}
\renewcommand*{\theHlemma}{\theHtheorem}

\providecommand*{\theHproposition}{}
\renewcommand*{\theHproposition}{\theHtheorem}
\providecommand*{\theHcorollary}{}
\renewcommand*{\theHcorollary}{\theHtheorem}

\providecommand*{\theHexample}{}
\renewcommand*{\theHexample}{\theHdefinition}

\providecommand*{\theHassumption}{}
\renewcommand*{\theHassumption}{\theHsection.\arabic{assumption}}

\makeatother
\usepackage[colorlinks=true, linkcolor=blue!70!black, citecolor=blue!70!black, urlcolor=blue, breaklinks=true]{hyperref}
\usepackage{cleveref}
\Crefname{algocf}{Algorithm}{Algorithms}
\crefname{claim}{claim}{claims}
\crefname{assumption}{assumption}{assumptions}
\Crefname{assumption}{Assumption}{Assumptions}

\newcommand{\squishlist}{
\begin{list}{{{\small{$\bullet$}}}}
{\setlength{\itemsep}{3pt}      \setlength{\parsep}{1pt}
\setlength{\topsep}{1pt}       \setlength{\partopsep}{0pt}
\setlength{\leftmargin}{1em} \setlength{\labelwidth}{1em}
\setlength{\labelsep}{0.5em} } }
\newcommand{\squishend}{  \end{list}}

\newcommand{\indicator}[2][]{\one\ifthenelse{\not\equal{}{#1}}{_{#1}}{}\!\left[{\def\givenn{\middle|}#2}\right]}
\newcommand{\Reg}[2][]{\text{Regret}\ifthenelse{\not\equal{}{#1}}{_{#1}}{}\!\left[{\def\givenn{\middle|}#2}\right]}

\newcommand{\timeHorizon}{T}
\newcommand{\qualitySpace}{\Omega}
\newcommand{\quality}{\omega}
\newcommand{\qualityVal}{\bar{\quality}}

\newcommand{\type}{\theta}

\newcommand{\typeSpace}{\Theta}
\newcommand{\typeCDF}{F}

\newcommand{\demand}{D}
\newcommand{\decision}{a}

\newcommand{\buyerUtility}{v}

\newcommand{\posteriorMean}{q}
\newcommand{\posteriorDist}{\mu}

\newcommand{\signal}{\sigma}
\newcommand{\signalSpace}{\Sigma}
\newcommand{\scheme}{\phi}
\newcommand{\prior}{\lambda}
\newcommand{\stateNum}{m}

\newcommand{\price}{p}
\newcommand{\Rev}{\text{Rev}}
\newcommand{\optPricing}{\price^*}
\newcommand{\optPrice}{\optPricing}

\newcommand{\distOfPost}{\rho}
\newcommand{\optAdver}{\distOfPost^*}
\newcommand{\distOfMean}{\rho}
\newcommand{\adverSpace}{\mathcal{A}}

\newcommand{\highestVal}{\bar{v}}
\newcommand{\lowestVal}{\underline{v}}
\newcommand{\plower}{0}
\newcommand{\pupper}{U}

\newcommand{\inverseVal}{\kappa}

\newcommand{\criticalValue}{\text{critical type}}
\newcommand{\realizedQuality}{{\quality}}
\newcommand{\lipschitzconst}{L}

\newcommand{\SC}{\textsc{SC}}
\newcommand{\SCRev}{\Rev^{\SC}}
\newcommand{\pricePolicy}{\mathbf{\price}}
\newcommand{\optSCPrice}{\mathbf{\price}^*}

\newcommand{\sellerCost}{c}
\newcommand{\degType}{\bar{\type}}

\newcommand{\optNI}{\mathrm{OPT}^{\textsc{PO}}}
\newcommand{\optU}{\mathrm{OPT}}
\newcommand{\optSC}{\mathrm{OPT}^{\SC}}
\newcommand{\signalPrice}{\mathbf{p}}

\newcommand{\posteriorAuxOne}{A}
\newcommand{\posteriorAuxTwo}{B}
\newcommand{\priorAuxOne}{\bar{A}}
\newcommand{\priorAuxTwo}{\bar{B}}
\newcommand{\typeRevenue}{R_{\typeCDF}}
\newcommand{\condRev}{r}

\newcommand{\boundedDemand}{\demand_{\eps}}
\newcommand{\tightAlpha}{\alpha_{\eps}}
\newcommand{\tightDelta}{\delta_{\eps}}
\newcommand{\tightTau}{\tau_{\eps}}
\newcommand{\tightVal}{\buyerUtility_{\eps}}
\newcommand{\tightSignalMean}{q_{\signal}}
\newcommand{\tightM}{m_{\signal}}
\newcommand{\tightMBar}{\bar{m}}
\newcommand{\tightThreshold}{t}

\newcommand{\discPrice}{\mathcal{P}}
\newcommand{\discPostMeanSpace}{\mathcal{Q}}
\newcommand{\discTyepSpace}{\mathcal{S}}
\newcommand{\discTypeSpace}{\discTyepSpace}

\newcommand{\discretizedOptPrice}{\widetilde{p}^*}
\newcommand{\discretizedOptAdver}{\widetilde{\distOfMean}^*}

\newcommand{\averageDemand}{\bar{\demand}}

\newcommand{\UCB}{\textsc{UCB}}
\newcommand{\UCBDemand}{\demand^{\UCB}}
\newcommand{\trueUCBDemand}{\demand^{\textsc{UCB}}}

\newcommand{\UCBRev}{\Rev^{\textsc{UCB}}}

\newcommand{\chosenSet}{\mathcal{N}}
\newcommand{\chosenCounter}{N}

\newcommand{\badEvent}{\textsc{E}}
\newcommand{\badEventComple}{\textsc{E}^{\text{c}}}

\newcommand{\distOfThreshold}{\beta}

\newcommand{\offeredPrice}{p}
\newcommand{\offeredAdver}{\distOfMean}

\newcommand{\numDiscType}{|\discTyepSpace|}

\newcommand{\Round}{\textsc{Rounding}}

\newcommand{\newPrice}{p^\dagger}
\newcommand{\newAdver}{\distOfMean^\dagger}

\newcommand{\leftPosteriorMean}{\posteriorMean_L}
\newcommand{\rightPosteriorMean}{\posteriorMean_R}

\newcommand{\capLeftPosteriorMean}{\posteriorMean_L^\dagger}
\newcommand{\capRightPosteriorMean}{\posteriorMean_R^\dagger}

\newcommand{\discrePrecision}{\varepsilon}

\newcommand{\instance}{\mathcal{I}}
\newcommand{\newInstance}{\instance^\dagger}
\newcommand{\newQualitySpace}{\qualitySpace^\dagger}
\newcommand{\newPrior}{\prior^\dagger}
\newcommand{\newPosteriorMean}{\posteriorMean^\dagger}
\newcommand{\pooledQuality}{\qualityVal^\dagger}

\newcommand{\targetPrecision}{(\sfrac{\log T}{T})^{\sfrac{1}{3}}}
\newcommand{\threeForthsRegret}{T^{\sfrac{3}{4}}(\log T)^{\sfrac{1}{4}}}

\newcommand{\poolThreshold}{(\sfrac{\log T}{T})^{\sfrac{1}{4}}}

\newcommand{\poolPrecision}{\widehat{\varepsilon}}

\newcommand{\RevProb}[2][]{\Rev\ifthenelse{\not\equal{}{#1}}{_{#1}}{}\!\left({\def\givenn{\middle|}#2}\right)}

\newcommand{\prog}{\textsc{P}}

\newcommand{\ratio}{\Gamma}

\newcommand{\noinfor}{\scheme^{\textsc{\NI}}}

\newcommand{\priorMean}{\quality_0}

\newcommand{\fullinfor}{\scheme^{\FI}}

\newcommand{\typePDF}{f}

\newcommand{\reals}{\mathbb{R}}

\newcommand{\eps}{\varepsilon}
\newcommand{\auxfunc}{g}

\newcommand{\primed}{^\dagger}

\newcommand{\simplePriceGrid}{\discPrice^{\NI}}
\newcommand{\roundedPrice}{\price^{\discrePrecision}}

\newcommand{\postMeanSignal}{q_{\signal}}
\newcommand{\postedMeanPrice}{p_0}

\newcommand{\ExpModel}{\mathrm{E}}
\newcommand{\GaussModel}{\mathrm{G}}
\newcommand{\MisModel}{\mathrm{Mis}}

\newcommand{\highQualityProb}{\alpha}
\newcommand{\expRate}{\eta}
\newcommand{\gaussStd}{s}
\newcommand{\gaussVar}{\gaussStd^2}
\newcommand{\normalCDF}{\Phi}

\newcommand{\trueTypeCDF}{\typeCDF^{\ExpModel}}
\newcommand{\gaussTypeCDF}{\typeCDF^{\GaussModel}_{\gaussStd}}
\newcommand{\trueDemand}{\demand^{\ExpModel}}
\newcommand{\gaussDemand}{\demand^{\GaussModel}_{\gaussStd}}

\newcommand{\truePriceRevenue}{V^{\ExpModel}}

\newcommand{\trueOptRev}{\optU^{\star,\ExpModel}}

\newcommand{\gaussOptPrice}{\price^{\GaussModel}_{\gaussStd}}
\newcommand{\gaussLeftPostMean}{\posteriorMean^{\GaussModel}_{L,\gaussStd}}
\newcommand{\gaussRightPostMean}{\posteriorMean^{\GaussModel}_{H,\gaussStd}}
\newcommand{\gaussPostWeight}{t^{\GaussModel}_{\gaussStd}}

\newcommand{\misspecifiedRev}{\Rev^{\ExpModel\leftarrow\GaussModel}_{\gaussStd}}
\newcommand{\misspecifiedGap}{\Delta^{\MisModel}_{\gaussStd}}
\newcommand{\relativeMisspecifiedGap}{\mathcal{L}^{\MisModel}_{\gaussStd}}

\newcommand{\filtration}{\mathcal{F}}
\newcommand{\smallIndex}{i^\dagger}
\newcommand{\CR}{\cc{CR}}

\newcommand{\newSupp}{\discPostMeanSpace}

\DeclareMathOperator*{\argmin}{arg\,min}
\DeclareMathOperator*{\argmax}{arg\,max}

%% file: math.tex
\newcommand{\xhdr}[1]{\vspace{6pt}\noindent{\bf {#1.}}}

\newcommand{\omt}[1]{}

\newcommand{\cc}[1]{\ensuremath{\mathsf{#1}}} 

\newcommand{\prob}[2][]{\mathbb{P}\ifthenelse{\not\equal{}{#1}}{_{#1}}{}\!\left[{\def\givenn{\middle|}#2}\right]}
\newcommand{\expect}[2][]{\mathbb{E}\ifthenelse{\not\equal{}{#1}}{_{#1}}{}\!\left[{\def\givenn{\middle|}#2}\right]}

\newcommand{\dd}{{\mathrm d}}

\newcommand{\R}{\mathbb{R}}

\newcommand{\N}{\mathbb{N}}

\newcommand{\integers}{\N}

\newcommand{\supp}{\cc{supp}}

\newcommand{\one}{\boldsymbol{1}}

\newcommand{\NI}{\textsc{NI}}
\newcommand{\FI}{\textsc{FI}}

\newcommand{\poly}{\cc{poly}}

%% file: Paper/0-abstract.tex
We consider a novel pricing and advertising framework in which a seller not only sets the product price but also designs flexible advertising schemes to influence customers' valuations of the product.
We impose no structural restriction on the seller's feasible advertising strategies and allow her to advertise the product by disclosing or concealing any information.
Following the information design literature,
we model this fully flexible advertising as the seller choosing an arbitrary information policy that signals the product quality to customers.
Customers observe the advertising signal and form a Bayesian posterior belief over the product quality.
We investigate two questions in this work:
(1) What is the value of advertising? To what extent can advertising enhance a seller's revenue?
(2) Without any a priori knowledge of the customers' demand function, how can a seller adaptively learn and optimize both pricing and advertising strategies using past purchase responses?

\revcolor{To study the first question, we quantify the value of advertising by comparing the optimal revenue from jointly designing advertising and a single posted price with the optimal revenue from pricing alone. We show that advertising can increase revenue by at most a factor of two, and this bound is tight.}
For the second question, we study the seller's dynamic pricing and advertising problem under demand uncertainty.
Our main result for this question is a computationally efficient online algorithm
that achieves the optimal $O(\timeHorizon^{\sfrac{2}{3}}
(\stateNum\log\timeHorizon)^{\sfrac{1}{3}})$ regret rate when the valuation function is linear in the product quality. Here, $\stateNum$ is the cardinality of the discrete product quality domain and $\timeHorizon$ is the time horizon.
This result requires some mild regularity assumptions
on the valuation function, but no Lipschitz or smoothness assumption on the customers' demand function.
We also obtain several improved results for the widely studied special case of additive valuations.
\footnote{A preliminary conference version of this work appeared in the proceeding of the 37th Conference on Neural Information Processing Systems \citep{AFT-23}.}

%% file: Paper/1-intro-revision.tex
One of the central questions in modern economics is how a seller should optimally price its products. When customer valuations and demand responses for a product are a priori unknown, price variation can also be used to observe and learn the demand function, enabling the seller to adaptively optimize prices and revenue over time. This learning and optimization problem has been a focus of the recent literature that applies exploration--exploitation and multi-armed bandit techniques to dynamic pricing (e.g., see \citealp{KL-03,BZ-09,KZ-14,BDKS-15}).

In practice, sellers also have access to another important tool---{\em advertising}---to inform and shape customers' valuations of a product.
It has been shown theoretically \citep{N-70,N-74} and empirically \citep{SN-20,MDGC-06} that advertisements can serve as a credible signal of the quality or characteristics of the advertised product.
Sellers can use advertising to provide partial information about a product, better position the product in the market, and potentially increase customers' likelihood of purchasing it.
\revcolor{As reported by \citet{stat-digital-ads}, worldwide spending in the digital advertising market is projected to reach US\$854.9 billion in 2026, up from US\$740.3 billion in 2024.}
To give a few examples,
\revcolor{sellers of recurring opaque products, such as rescued-produce boxes, may disclose selected information (e.g., featured items, broad product categories, an expected-value range, or an aggregate quality score) about an upcoming box;
on-demand service platforms may disclose selected information about a matched provider, such as ratings, expected waiting time, or service attributes, to shape a customer's assessment of the service.
\footnote{\revcolor{In \Cref{apx:example}, we provide more discussions about the real-world applications.}}}

However, advertising must be carefully designed to achieve the desired gains. At first glance, upselling or inflating the perceived product quality by selectively disclosing only favorable information might appear to be a profitable advertising strategy. However, such strategies may be ineffective at shifting customer beliefs, since customers may not trust that the disclosed information accurately reflects the product's true quality. Moreover, the advertising strategy must be designed jointly with the pricing strategy and must account for the demand function. For example, when selling high-priced products, or under heavy competition or low demand, customers may need to be convinced of a good match through richer information and deeper insights into the product's characteristics.
On the other hand, in markets with high demand or for very low-priced goods, revealing very little information may suffice. An extreme example of this phenomenon is the mystery box sold by some retailers such as Amazon and Woot, in which customers are not even told the exact contents of the low-cost box they are purchasing.\footnote{Similarly, for opaque products, the precise product features or characteristics are hidden from customers \citep{EH-21}.}

Despite the widespread use of advertising in practice, there is limited understanding of how it directly contributes to improving a seller's revenue.
Furthermore, how should a seller design its pricing and advertising strategies when facing market uncertainty (e.g., the seller may not fully know the customers' preferences or their willingness to pay)?
This work seeks to answer two key questions:
\begin{displayquote}
   (Question 1) \emph{What is the value of advertising? To what extent can advertising enhance a seller's revenue?}
   \\
   (Question 2) \emph{How can a seller learn and optimize both pricing and advertising strategies when initially uncertain about market demand?}
\end{displayquote}
\noindent In this paper, we impose no structural restriction on the set of the seller's feasible advertising strategies and allow the seller to advertise the product by disclosing or concealing any information.
Following the information design literature \citep{KG-11,RS-17,HKB-19,BM-19}, we model this fully flexible advertising as the seller choosing an arbitrary information policy that prescribes, upon observing the realized product quality, a distribution over signals to send to customers. Upon observing the seller's advertising signal, customers form a posterior belief over the product quality, and each customer's purchase decision then maximizes their expected utility under this posterior.

The goal of this work, under the proposed pricing and advertising framework, is to (1) understand and characterize how advertising benefits the seller's revenue;
and (2) design, without any a priori knowledge of the demand function, an efficient online algorithm that adaptively learns a joint pricing and advertising strategy to maximize the seller's revenue.
Below, we detail the contributions of this paper and position it within the literature.

\subsection{The High-level Problem Formulation and Our Contributions} 
\label{subsec:contribution}

\input{Paper/1-1-contributions}

%% file: Paper/1-1-contributions.tex
In this section, we explain, at a high level, both the {\em conceptual} and {\em technical} contributions of this paper.
Our contributions are threefold.
\revcolor{First, we introduce a pricing-and-advertising framework that combines posted pricing with advertising: the seller observes the product quality, customers do not, and advertising shapes customers' posterior beliefs about the product quality before they make their purchase decisions.
Second, we characterize the static interaction between advertising and pricing.
We quantify the value of advertising relative to pricing alone.
En route, we also quantify the revenue impact of restricting the seller to a simple quality-independent price rather than allowing more general pricing strategies. Third, we study the learning setting in which the seller does not know the customer type distribution and must learn the demand from purchase responses while jointly optimizing pricing and advertising.}
We detail these contributions below.

\xhdr{The high-level problem formulation}
A main {\em modeling contribution} of our work is to integrate a canonical information design framework---Bayesian persuasion---with pricing, in order to model the effect of advertising on customers' beliefs about product quality and, consequently, on their purchase decisions.
This framework allows us to quantify how the joint design of pricing and advertising strategies affects revenue outcomes.
To the best of our knowledge, we are the first to introduce an advertising component into the well-studied dynamic pricing problem and to study the seller's joint pricing and advertising problem.

At a high level, we consider a setting with natural information asymmetry, where a monopolist seller sells a product of varying quality to a market of heterogeneous customers.
We consider a Bayesian model in which the product is associated with a \emph{quality $\quality$} drawn from a publicly known prior distribution over a given quality space.
Customers have heterogeneous preferences for the product, and their heterogeneity is captured by a \emph{private type} $\type$ (not observed by the seller), drawn independently and identically from a type distribution with CDF~$\typeCDF$.
The seller observes the realized product quality, but customers do not.
With this information advantage, the seller can advertise the product by strategically disclosing information about its quality to customers.
We formulate the seller's advertising strategy as an information policy that prescribes a conditional distribution over a set of signals for each realized product quality.

Each customer forms a Bayesian posterior over the product quality based on the observed signal and decides whether to purchase the product. This decision depends on the customer's posterior quality distribution and private type, which jointly determine the customer's valuation of the product (i.e., the willingness to pay).
In particular, if a customer infers a posterior $\mu$ over the product quality $\quality$, has private type~$\type$, and faces price $\price$, then the customer purchases the product if the expected net utility is nonnegative (i.e., if the expected valuation $\expect[\quality\sim \mu]{\buyerUtility(\type, \quality)} - \price \ge 0$, where $\buyerUtility(\cdot, \cdot)$ is the customers' valuation function), assuming that not purchasing yields zero utility.
In our static model, the seller knows the underlying demand (i.e., the type CDF $\typeCDF$), while
in our dynamic learning model, customers arrive sequentially over time and the seller faces demand uncertainty, i.e., does not know the customers' type CDF $\typeCDF$.

With the above formulation, our main {\em technical contributions} are summarized as follows.

\revcolor{\xhdr{Advertising as the revenue lever}
We first analyze the static model to understand how advertising affects revenue and how this effect depends on pricing.
In \Cref{sec:voa meta},
we first present examples that demonstrate how different advertising strategies affect the seller's revenue and why accurately learning the customer type distribution matters for designing these strategies.}
In particular, we consider a simple example in which the product quality is binary (i.e., a high-quality product with $\quality = 1$ or a low-quality product with $\quality = 0$) and customers' valuations are additive, given by $\buyerUtility(\type, \quality) = \type + \quality$.
When customers have binary types (i.e., a high type with $\type = 1$ and a low type with $\type = 0$) and high-type customers form a minority of the market, the seller achieves higher revenue by fully disclosing the product information than by not advertising at all.
When customers' types follow the continuous distribution with CDF $\typeCDF(\type) = \sqrt{\type}$, a carefully designed advertising strategy that selectively discloses product information increases revenue by 31.6\% and 25.1\% relative to not advertising and to fully disclosing the product quality, respectively.
\revcolor{We further use these two examples with different type distributions to illustrate the \emph{cost of misspecifying demand}: if the true type distribution has continuous CDF $\typeCDF(\type) = \sqrt{\type}$, but the seller incorrectly treats it as a mean-matched binary distribution, then the seller chooses full disclosure, which is the best advertising strategy under the binary-type environment.
Evaluated under the true continuous-type environment, this misspecified policy incurs a relative revenue loss of about $24.0\%$.
Overall, these examples highlight how selective advertising can significantly improve the seller's revenue. They also show that the benefit of advertising, and the form of the optimal advertising strategy, depend strongly on the shape of the customer type distribution.

Motivated by these examples, we then provide a tight quantitative characterization of the \emph{value of advertising}.
Let $\optNI$ denote the best revenue achievable with an optimized quality-independent price and no advertising (equivalently, with no information disclosed about the product quality),
$\optU$ denote the best revenue from jointly optimizing advertising and a single quality-independent price, and $\optSC$ denote the stronger benchmark that allows the price to depend on the realized quality and advertising signal.
We show in \Cref{thm:voa bound} that, under common assumptions on the valuation function (\Cref{assump:valuation}), all three benchmarks satisfy
\begin{align*}
    \optNI
    \le
    \optU
    \le
    \optSC
    \le
    2\optNI,
\end{align*}
and each of the three factor-two comparisons is tight: there exist instances with $\optSC = 2\optNI$, $\optSC = 2\optU$, and $\optU = 2\optNI$, respectively.
The ratio $\optU/\optNI$ isolates the value of advertising because both sides optimize over the price, while only $\optU$ allows the seller to additionally design an advertising policy.
The theorem therefore shows that advertising can nearly double revenue relative to no-information advertising, but it cannot improve revenue by more than a factor of two once the price is optimized.
At the same time, the comparison with $\optSC$ shows that jointly optimized advertising and a quality-independent price are always within a factor of two of the more flexible benchmark that allows quality- and signal-contingent pricing.
This provides a revenue justification for focusing on simple quality-independent pricing, which is also more common in practice due to its transparency and operational simplicity.

Finally, we illustrate the importance of joint optimization over pricing and advertising. If the advertising policy is fixed (i.e., not jointly optimized), signal-contingent pricing (allowed in $\optSC$) can be unboundedly better than a single quality-independent price (in $\optNI$ and $\optU$).
Conversely, if a quality-independent price is fixed (i.e., not jointly optimized) when comparing advertising with no advertising, then informative advertising can be unboundedly better than no-information advertising.
Thus, the $2$-factor result is meaningful precisely because the comparison allows the seller to conduct the joint optimization.
Together, these results justify our problem formulation and $\optU$ as the central benchmark of the paper: it captures the revenue value of optimally designed advertising while preserving the operational simplicity of a single price that does not depend on the realized quality or advertising signal.}

\xhdr{Designing efficient demand learning algorithm}
We next study the dynamic pricing and advertising problem in the presence of demand model uncertainty. This problem is motivated by the practical consideration that the customers' type distribution $\typeCDF$ may not be known to the seller a priori.
Instead, the seller may have to learn this type distribution (implicitly or explicitly) from observed customer responses in sales data, in order to design and improve the pricing and advertising policy.

In line with the online learning and multi-armed bandits literature, we evaluate the algorithm's performance via cumulative regret, defined as the total expected revenue loss relative to a clairvoyant benchmark.
Our main contribution in this setting is a computationally efficient learning algorithm that achieves a regret upper bound of $O(\timeHorizon^{\sfrac{2}{3}}
(\stateNum\log\timeHorizon)^{\sfrac{1}{3}})$ over a time horizon $\timeHorizon$, where $\stateNum$ is the cardinality of the product quality space ({\Cref{thm:regret upper bound}}).
Importantly, we achieve this result without any Lipschitz or smoothness assumption on the demand function $\demand(\cdot)=1-\typeCDF(\cdot)$.
Our results do, however, require certain assumptions on the valuation function.
Following the literature, we assume that the valuation function $\buyerUtility(\type, \quality)$ is linear in the product quality $\quality$, that $\buyerUtility(\type, \cdot)$ is monotone over the quality space for every private type $\type$, and that $\buyerUtility(\cdot, \quality)$ is Lipschitz and monotone over the customer type space. Since our model generalizes the classic dynamic pricing problem (without advertising), the regret lower bound of $\Omega(\timeHorizon^{2/3})$ established by \citet{KL-03} certifies that our algorithm's dependence on the time horizon $\timeHorizon$ is optimal up to logarithmic factors. Applying the same algorithmic and analytical framework, we obtain several improved results for the widely studied special case of additive valuations, i.e., $\buyerUtility(\type, \quality) = \type + \quality$ (\Cref{subsec:improvement}).

\xhdr{High-level descriptions of the proposed algorithm and challenges}
The seller's demand learning problem can be viewed as a very high-dimensional combinatorial multi-armed bandit problem in which each arm represents a combination of a price and an advertising strategy.  The set of feasible advertising strategies consists of all possible conditional distributions over a pre-defined signal space.
As a first step toward a more tractable setting, we present an equivalent reformulation of the problem that exploits the observation that advertising affects the customer's decision only through the posterior distribution over quality.
By the linearity of the valuation function $\buyerUtility(\cdot, \cdot)$ in the
product quality $\quality$, the seller's choice of advertising strategy in every round can be further simplified to selecting a distribution over posterior means, subject to a Bayes-consistency feasibility constraint.

This reformulation structures the seller's decision space into two core components:
setting a price and choosing a distribution over posterior means.
Viewing the seller's expected revenue as an unknown (nonlinear) function over this decision space,
one might consider applying techniques from Lipschitz bandits or contextual bandits to obtain sublinear regret.
However, significant challenges prevent a direct application of such approaches.
First, the decision space remains infinite-dimensional due to the continuous nature of posterior mean distributions.
Second, it is unclear whether a Lipschitz property can be established at all: we impose no Lipschitz or smoothness assumption on the demand function, and the feasibility constraints on the advertising space are complex.
Instead, our algorithm takes a \emph{model-based approach}: we use customers' purchase responses to learn the demand function over a discretized type space and to jointly optimize pricing and advertising.
To efficiently explore the demand function over the one-dimensional discretized type space, we propose a novel \emph{instance-dependent non-uniform} discretization scheme that supports a near-optimal price and advertising strategy without relying on Lipschitz or smoothness assumptions, while accommodating the complex feasibility constraints of the advertising space.
Through these design choices, our algorithm achieves the optimal (up to logarithmic factors)  $O(\timeHorizon^{\sfrac{2}{3}}
(\stateNum\log\timeHorizon)^{\sfrac{1}{3}})$ regret bound.

%% file: Paper/1-2-related.tex
Our work is related to several streams of research.
Below, we briefly review some of these connections.

\xhdr{Signaling as advertising}
In our setting, the seller can exploit her information advantage to design advertising that signals the product quality to customers.
A long line of research, from both empirical and theoretical perspectives, studies how advertising can serve as a signal that steers
customers' evaluations of advertised goods
\citep{N-70,N-74,KR-84,MR-86,JR-94,SN-20,KOU-22}.
In our problem, we follow the information design (a.k.a.\ Bayesian persuasion) literature \citep{KG-11}
\citep[see also the recent surveys by][]{D-17,Kamenica-19,BM-19}, in which the seller commits to an information policy that selectively discloses product information in order to influence customers' beliefs about the product quality.
Similar formulations of advertising have also appeared in \cite{BS-12,EFGPT-14,AB-19,HKB-19}.
Our work differs from these works in several ways.
First, the seller's offline problem in our setting
is a joint pricing and advertising problem.
Second, we also study an online setting in which the seller has no a priori knowledge of the demand function and must use past customers' purchase responses to adaptively learn the optimal joint pricing and advertising strategy.

\revcolor{Several closely related studies with known customer type or valuation distributions examine information disclosure jointly with monopoly pricing or richer mechanisms. \citet{RS-17} study information design and posted pricing; \citet{ES-07} and \citet{BHM-26} study disclosure together with auction or screening mechanisms; and \citet{BCW-22} compare third-degree price discrimination with uniform pricing. 
Our model differs from their settings significantly and we provide more detailed comparisons in \Cref{apx:comparisons}.}

When the seller faces demand uncertainty, the seller's dynamic pricing and advertising problem also relates to the growing line of information design research that uses online learning approaches to study regret minimization when the sender is uncertain about the receivers' utility functions \citep{CCMG-20,CMCG-21,FTX-22,ZIX-21}.
These works typically study how to learn an optimal information policy, whereas in our setting the optimal policy is a joint pricing and advertising strategy.

\xhdr{Dynamic pricing and learning}
When there is no uncertainty in the product quality, the
seller's problem in our setting reduces to a standard dynamic pricing and learning problem with an unknown demand function \citep{KL-03,BZ-09,AC-09,HKZ-12,BR-12,KZ-14,BZ-15,BSX-20,CSWZ-22,CSW-22,CWZ-24}.
\wtedit{More closely related, \citet{GKP-26} study demand learning when a firm jointly chooses prices, advertising expenditures, and inventory levels; unlike our information-design formulation, their advertising decision affects customer awareness and demand rather than selectively disclosing product-quality information.}
For any non-trivial product quality space and prior distribution, the seller in our problem must jointly determine a price and a non-trivial advertising strategy in order to maximize total revenue.
This makes the seller's decision space high-dimensional and (as we elaborate shortly) introduces significant challenges for the seller's learning task, since the standard techniques (such as uniform or adaptive discretization) used in the dynamic pricing and continuous/combinatorial multi-armed bandit literatures cannot be applied directly.

\xhdr{Selling information}
Lastly, our problem shares similarities with the problem of selling information in the economics and computer science literatures \citep{BKP-12,BB-15,BBS-18,CXZ-20,CV-21,LSX-21,BCVZ-22,ZC-21,L-22,CZ-20}.
In particular, as in the sale-of-information problem,
the seller in our setting commits to an information structure that reveals information about the realized state to the decision maker (i.e., the customer); and the customer's payment is based on the declared information structure rather than on specific realizations of the seller's informative signals \citep{BCVZ-22,CV-21,LSX-21}.
However, in contrast to these works, the seller in our setting sells a product with inherent value, not just information, and the product's valuation can be shaped by the information provided. This gives rise to new tradeoffs between information and revenue in our problem that are absent in settings where only information is sold.
Moreover, the literature on selling information typically assumes that the customers' type distribution is known. We consider a more practical, data-driven setting in which the underlying customers' type distribution (demand function) is a priori unknown to the seller and must be learned from observations.

%% file: Paper/2-setting-new.tex
We begin by describing the static model of our pricing and advertising problem, and then introduce its online counterpart.

\xhdr{The static pricing and advertising model}
In the static model, a monopolist seller sells a product of varying quality to a market of heterogeneous customers.
We consider a Bayesian model in which the product is associated with a quality $\quality \in \qualitySpace \subseteq [0,1]$, drawn from a finite or infinite set $\qualitySpace$ according to a prior distribution $\prior\in\Delta(\qualitySpace)$ that is common knowledge to both the seller and the customers.\footnote{\revcolor{For the learning results in \Cref{sec:alg design,sec:analysis}, we specialize to a finite quality space; \Cref{subsec:improvement} discusses extensions to large or continuous quality spaces.}}
In this model, the quality of the product represents all payoff-relevant characteristics, including its features and attributes.
Customers have heterogeneous preferences (i.e., different willingness to pay) for the product.
In particular, each customer's preference is captured by a \emph{private} type $\type$, drawn independently and identically (i.i.d.) from a type distribution with CDF $\typeCDF(\cdot)$ and support $\typeSpace = [0, 1]$.\footnote{\revcolor{Following the convention in the literature, we define the CDF $\typeCDF$ of a distribution as $\typeCDF(x)\triangleq \prob[z\sim \typeCDF]{z < x}$.}}
In the static model, the seller does not observe the customers' individual private types but knows the type distribution $\typeCDF$.

\revcolor{A customer with type $\type \in \typeSpace$ makes a binary purchase decision: he either buys the product or does not. If the customer buys the product, his payoff is quasilinear: $\buyerUtility(\type, \quality) - \price$, where $\buyerUtility(\type, \quality) \in\reals_+$ represents the customer's valuation for a product of quality $\quality$, and $\price$ is the price of the product, which is set by the seller.
If the customer does not buy the product, his payoff is normalized to $0$ without loss of generality.
Following the literature,
we make the following assumptions on the valuation function.
\begin{assumption}
\label{assump:valuation}
Customers' valuation function $\buyerUtility(\cdot, \cdot)$ satisfies:
\begin{enumerate}[leftmargin=*, label=\textbf{1\alph*}, topsep=0pt, itemsep = -2pt]
    \item
    \label{assump:1b}
    For each customer type $\type$, the function
    $\buyerUtility(\type, \quality)$
    is non-decreasing and linear in the quality $\quality$.
    \item
    \label{assump:1a}
    For each quality $\quality$,
    the function $\buyerUtility(\type, \quality)$
    is increasing, $1$-Lipschitz,\footnote{\label{footnote:Lip-learning-only}The Lipschitz constant $1$ is chosen for simplicity of exposition; any arbitrary constant $\lipschitzconst \in \reals_+$ can be handled similarly. \revcolor{Indeed, this Lipschitz property is used only in our dynamic pricing and learning results.}}
    and affine in the type $\type$.
\end{enumerate}
\end{assumption}}
We note that such assumptions are in fact common in the literature and natural in many economic situations in which the customers' valuation of a product increases with both the product quality and the customer's type (e.g., ability or willingness to pay).
The existing literature
often makes even stronger assumptions about the customers' valuation function.
For example, both the additive function $\buyerUtility(\type, \quality) =\type + \quality$ \citep[cf.][]{IMSZ-19,KMZL-17,CIMS-17,SVZ-22} and the multiplicative function $\buyerUtility(\type, \quality) =\type\quality$ \citep[cf.][]{CS-21,LSX-21,BHM-26} satisfy these assumptions.

We consider a natural information asymmetry
between the seller and the customers: the seller can observe the product's realized quality $\quality$, but the customers cannot. 
Before the product quality is realized, the seller can determine a price $\price \in \reals_+$, 
as well as an advertising strategy.
We impose no structural restriction on feasible advertising strategies,
so that the seller can disclose or conceal any product information.
In particular, the seller can design an advertising strategy $\scheme = \{\scheme(\cdot|\quality)\}_{\quality\in\qualitySpace}$ (also known as an information strategy or signaling scheme) along with a \revcolor{measurable} signal space $\signalSpace$, where $\scheme(\signal|\quality)\in[0, 1]$ prescribes the conditional probability of sending signal $\signal\in\signalSpace$ to customers when the product quality $\quality \in \qualitySpace$ is realized.\footnote{\label{footnote:signal}\revcolor{We note that the advertising strategy is allowed to have a continuous signal space, and in this case, $\scheme(\cdot|\quality)$ shall be interpreted as a probability density function over $\signalSpace$.}}
In \Cref{sec:voa meta}, we give a few examples of the advertising strategies, including no-information advertising strategy, full-information advertising strategy, and partial-information advertising strategy.

Customers cannot observe the realized product quality $\quality$, but they observe a signal $\signal \sim \scheme(\cdot|\quality)$ sent by the seller.
Customers use this signal, together with the prior $\prior$, to form a Bayesian posterior distribution over the product quality, given by $\posteriorDist(\quality|\signal) \propto \scheme(\signal|\quality)\cdot \prior(\quality)$.
Because customers update their beliefs in a Bayesian manner, the posterior beliefs must satisfy the following Bayes-consistency condition \citep{KG-11}:
\begin{align}
    \label{bayes consistency vanilla}
    \tag{{\textsc{BC}}}
    \sum\nolimits_{\signal\in\signalSpace} \posteriorDist(\quality|\signal) \cdot \sum\nolimits_{\quality'\in\qualitySpace} \scheme(\signal|\quality') \prior(\quality')
    = \prior(\quality), \quad
    \forall \quality\in\qualitySpace~.
\end{align}
For notational simplicity, we sometimes also write $\posteriorDist(\signal)\in\Delta(\qualitySpace)$ for the customer's posterior induced by observing the signal $\signal$.
A customer with type $\type$ then purchases the product if and only if the expected valuation under the updated posterior, $\expect[\quality \sim \posteriorDist(\cdot|\signal)]{\buyerUtility(\type, \quality)}$, is at least the price $\price$.
\revcolor{Let $\instance = (\buyerUtility, \prior, \typeCDF)$ denote a static pricing-and-advertising instance with valuation function $\buyerUtility$, prior distribution $\prior$ over the product quality, and type distribution $\typeCDF$.
Given a price $\price$ and an advertising strategy $\scheme$, we define the seller's expected revenue $\Rev_\instance(\price, \scheme)$ as
\begin{align*}
    \Rev_\instance(\price, \scheme) \triangleq\expect[\quality\sim \prior, \signal\sim\scheme(\cdot\mid \quality), \type\sim\typeCDF]{\price \cdot \decision_{\buyerUtility,\prior}(\type,\signal)}~,
\end{align*}
where $\decision_{\buyerUtility,\prior}(\type,\signal) \in \{0, 1\}$ denotes the purchase decision of a customer with type $\type\in\typeSpace$ upon observing the signal $\signal$. When the context is clear, we also write $\Rev_\instance(\price, \scheme)$ as $\Rev(\price, \scheme)$. Given an instance $\instance$ of the static pricing-and-advertising model, we denote by
\begin{align*}
    \optU(\instance)
\triangleq
\sup\nolimits_{\price\geq 0,\scheme}
\Rev_{\instance}(\price,\scheme)
\end{align*} 
the optimal revenue obtained by jointly optimizing an advertising strategy and a single posted price.}

\xhdr{The seller's dynamic pricing and advertising problem}
In the dynamic learning model, the seller interacts sequentially with customers whose private types are i.i.d.\ draws from an {\em unknown} type distribution $\typeCDF$.
In particular, the interaction unfolds over $\timeHorizon$ rounds, with each round $t \in [\timeHorizon]$ proceeding as follows:
\revcolor{\begin{enumerate}[itemsep=0pt, topsep=4pt]
    \item
    The seller sets a price \revcolor{$\price_t$} and commits to
    an advertising strategy $\scheme_t$ with signal space $\signalSpace_t$.
    \item
    A customer with private type $\type_t\sim_{\text{iid}} \typeCDF$ arrives.
    \item
    A product quality $\quality_t\sim_{\text{iid}}\prior$
    is realized, and
    the seller sends a signal $\signal_t \sim \scheme_t(\cdot|\quality_t)$ to the customer.
    \item
    The customer forms the Bayesian posterior $\posteriorDist_t(\cdot\mid \signal_t)\in\Delta(\qualitySpace)$, and purchases the product, generating revenue \revcolor{$\price_t$} for the seller, if and only if the expected valuation under this posterior is at least the price, i.e.,
    $\expect[\quality \sim \posteriorDist_t(\cdot|\signal_t)]{\buyerUtility(\type_t, \quality)} \ge \revcolor{\price_t}$.
\end{enumerate}}
We investigate how to design an online learning algorithm that, in each round $t$, chooses the price and advertising strategy $(\price_t, \scheme_t)$ based on customers' responses in previous rounds, so as to maximize the total expected revenue over the time horizon $\timeHorizon$ without a priori knowledge of the type distribution $\typeCDF$.
Note that since product qualities and customer types are i.i.d.\ across rounds, the expected per-round revenue $\Rev(\price, \scheme)$ under any fixed price $\price_t=\price$ and advertising strategy $\scheme_t=\scheme$ does not depend on $t$.
Thus, a static price and advertising strategy maximize the total expected revenue over the time horizon $\timeHorizon$.
We therefore measure the performance of an algorithm by its \emph{expected regret}, henceforth regret, which compares the total expected revenue of the algorithm with that of the best static price and advertising strategy:
\begin{align}
    \Reg[\instance]{\timeHorizon}
    \triangleq
    \revcolor{\timeHorizon\cdot \optU(\instance)}
    - \sum\nolimits_{t\in[\timeHorizon]} \expect{\Rev_\instance(\price_t, \scheme_t)}~.
    \label{eq:vanilla regret}
\end{align}
We aim to design an online algorithm that minimizes the worst-case regret $\sup_{\instance} \Reg[\instance]{\timeHorizon}$ over all instances under consideration.
When the context is clear, we omit $\instance$ and write the regret as $\Reg{\timeHorizon}$.

\revcolor{
\subsection{Modeling Discussions}
\label{subsec:model-discussions}

We next explain several key modeling choices made in our problem formulation.

\xhdr{Knowledge of the valuation function}
In our dynamic learning model, we assume that the seller knows the customers' valuation function $\buyerUtility(\cdot,\cdot)$, but does not know their private type distribution $\typeCDF$.
This distinction follows the standard dynamic pricing and demand learning literature, in which the seller typically knows the purchase/choice structure (or a parametric/semi-parametric demand model) but must learn the unknown demand curve or its parameters from sales data \citep{BZ-09,AC-09,HKZ-12,BR-12,KZ-14,BZ-15}.
Our formulation is analogous: the valuation function $\buyerUtility$ serves as the structural mapping from a latent customer type and product quality to willingness to pay, while the unknown type distribution $\typeCDF$ is the demand primitive that must be learned.\footnote{\revcolor{Learning both the valuation function $\buyerUtility$ and type distribution $\typeCDF$ is an important but distinct problem; our focus is on the additional challenge that arises when a seller who does not know the type distribution must jointly learn demand and design information policies.}}

This knowledge separation is natural in our applications. 
For a rescued-produce box, the seller may have learned from historical transactions, customer surveys, or prior experiments how observable box attributes, such as composition, freshness, variety, or aggregate retail value, translate into willingness to pay for a customer with a given baseline preference. 
These relationships can therefore be treated as a known valuation function $\buyerUtility(\type,\quality)$, even though the seller may not know how preferences $\type$ are distributed among future customers. 
Similarly, an on-demand service platform may have extensive historical data on how provider ratings, reliability, waiting time, and other service attributes affect willingness to pay for customers with different baseline preferences, while remaining uncertain about the distribution of such preferences among arriving users. 
Thus, knowing $\buyerUtility$ does not mean that the seller knows an individual customer's utility or type; rather, the seller knows how type and product quality jointly map into valuation, while the unknown $\typeCDF$ captures the market composition that must be learned.
}


\revcolor{\xhdr{Commitment assumption}
Our model follows the Bayesian-persuasion convention that the seller can commit to the information policy used to generate the signal observed by the customer. We acknowledge that commitment is a substantive assumption in information-design models.
In our dynamic learning setting, however, customers are short-lived. Each arriving customer observes the current price and the current advertising rule, receives the signal generated by that rule, makes a purchase decision, and then exits.
Thus, the customer's Bayesian updating requires credibility of the current-round advertising rule, rather than commitment to the full future path of the seller's learning algorithm. The seller's policy may change across rounds as demand is learned, but such changes affect future customers rather than the current customer's inference problem.

Inspired by the literature on calibration and decision-making (see, e.g., \citealp{FV-98,KLST-23,JP-24,FT-25a}), a less demanding implementation has the seller report a calibrated product-quality score, rather than announce the full details of the advertising policy in every round.
The key observation is that, under \Cref{assump:valuation}, the customer's purchase decision depends on the posterior belief about quality only through the posterior mean $\posteriorMean_t\triangleq\expect{\quality \mid \signal_t}$. (See more discussion about this posterior mean transformation in \Cref{subsec:equiv}.)
Thus, the seller does not need to communicate the full conditional signaling rule $\scheme(\cdot\mid\quality)$ to the customer in every round.
Instead, the seller can report a product-quality score $\posteriorMean_t\in[0,1]$, interpreted as a calibrated prediction of the realized product quality.
When the report is calibrated, products assigned score $\posteriorMean$ have an average realized quality equal to $\posteriorMean$, so the announced quality $\posteriorMean_t$ can serve directly as the posterior mean quality used by the customer.
In this sense, the customer can evaluate the product using $\posteriorMean_t$ without needing to know the entire signaling rule that generated it.
This interpretation replaces commitment to a full advertising policy with commitment to a calibration discipline.\footnote{\revcolor{We note that this calibration interpretation does not eliminate all credibility requirements. Rather, it shifts the relevant requirement from ex ante commitment to a complete advertising policy to a calibration constraint.\label{footnote:commitment}}}
Rather than asking customers to verify the entire map from realized qualities to signals, the seller announces a quality score that matches the posterior mean quality customers would form if they knew the full advertising policy and updated their beliefs in a Bayesian manner.
Given such a calibrated score, a type-$\type$ customer evaluates the product as $\buyerUtility(\type, \posteriorMean_t)$ and purchases at price $\price_t$ if and only if $\buyerUtility(\type, \posteriorMean_t) \ge \price_t$. This is exactly the decision statistic used by our learning algorithm.

Thus, the information-policy representation in the model can be viewed as a theoretical device for characterizing which distributions of posterior mean qualities are feasible under Bayes consistency.
In applications, one may implement such an advertising policy by reporting calibrated product qualities.
This calibrated-report interpretation is particularly natural in digital and platform markets (see, e.g., \citealp{AmazonCTR,YQWBN-22}), where prediction scores, ratings, quality indices, relevance scores, or recommendation scores are routinely shown to users and can be evaluated against realized outcomes. }

\revcolor{\xhdr{Quality-independent pricing}
Our baseline model restricts the seller to a single posted price that does not vary with the realized product quality or advertising signal. Consequently, the price is not used as an additional channel for communicating the realized product quality; instead, advertising is the instrument through which the seller shapes customers' beliefs. This restriction is not without loss: if the seller could set quality- or signal-contingent prices, she could tailor prices to different posterior market segments and generally obtain higher revenue.

We impose this restriction for both practical and conceptual reasons.
From a practical perspective, it provides a reasonable first-order approximation to markets in which prices are posted at the product, subscription-tier, box-category, or service-transaction level and are adjusted less frequently than the information shown to customers.
For example, a recurring produce-box service may post a common price for a given box size or subscription tier while varying the disclosed assortment, featured items, or quality information across delivery periods. Similarly, an on-demand service platform may quote an upfront price before a provider is assigned and subsequently disclose provider-specific information about the realized match. We provide more detailed illustrations of these applications in \Cref{apx:example}. More generally, a single posted price is often more transparent and operationally simpler to implement in settings where advertising messages, previews, recommendations, or quality scores may vary across contexts.

From a conceptual perspective, separating pricing from the realized quality or advertising signal allows us to isolate the role of advertising as an information-design instrument. The seller can shape customers' posterior beliefs about product quality, but does not additionally price-discriminate across the posterior segments induced by the advertising policy. Finally, this restriction also admits a theoretical justification. As we show in \Cref{subsec:voa} (see \Cref{thm:voa bound}), once the seller jointly optimizes the advertising policy and the single posted price, the resulting revenue is always within a factor of two of the stronger benchmark that allows the pricing strategy to depend on the realized quality or signal. Thus, the single-price model is both practically motivated and theoretically justified as a robust approximation to the more flexible benchmark.}

%% file: Paper/3-limits.tex
\revcolor{In this section, we study our static pricing-and-advertising model, where the seller knows the customer type distribution $\typeCDF$. Our goal is twofold. First, we use simple examples to illustrate how advertising interacts with pricing and why the optimal advertising strategy depends on the underlying type distribution (see \Cref{subsec:example}). These examples also show that misspecifying the type distribution can lead the seller to choose the wrong advertising policy and incur a nontrivial revenue loss. Second, we provide a tight quantitative characterization of both the {\em value of advertising} and the {\em impact of quality-independent pricing} (see \Cref{thm:voa bound} in \Cref{subsec:voa}). In particular, we compare the revenue from jointly optimized advertising and a single posted price with two natural benchmarks: the weaker benchmark that only offers a single quality-independent price without advertising, and the stronger benchmark that allows prices to depend on the realized quality and advertising signal. We show that these benchmarks are always within a factor of two of one another, and that all factor-two comparisons are tight. These results not only quantify how much advertising can improve revenue, but also provide a theoretical justification for focusing on single-price policies in the paper.}

\subsection{Two Illustrative Examples}
\label{subsec:example}

\input{Paper/3-1-example}

\revcolor{
\subsection{The Value of Advertising and Impact of Quality-Independent Pricing}
\label{subsec:voa}

\input{Paper/3-2-voa-new}

}

%% file: Paper/3-1-example.tex
We consider two simple examples for the static pricing and advertising model. In both examples, the product has binary quality $\quality\in\{0, 1\}$ realized from the Bernoulli distribution with $\prior(0) = \frac{3}{4}$ and $\prior(1) = \frac{1}{4}$. We refer to $\quality = 0$ and $\quality = 1$ as the \emph{low-quality} and \emph{high-quality} products, respectively. The customers have an additive valuation function $\buyerUtility(\type, \quality) = \type + \quality$. The customers' type distributions $\typeCDF$ differ in the two examples, which we detail below.

\xhdr{Binary customer type}
In this example, the customers also have binary types $\type\in\{0, 1\}$ realized from the Bernoulli distribution $\typeCDF$. We let $\typePDF$ denote the probability of a \emph{high-type} customer (i.e., $\type = 1$), so that $1 - \typePDF$ is the probability of a \emph{low-type} customer.



\begin{figure}
    \centering
    \subfloat[Binary customer type example. The x-axis is the high(-customer)-type probability $\typePDF$. The blue (black) curve is the expected revenue from the no-information (full-information) advertising strategy $\noinfor$ ($\fullinfor$) and its corresponding optimal price $\price^\NI$ ($\price^\FI$).]{
\input{Paper/plots/example-binary-binary-total}
\label{fig:example-binary-binary-total}
}~~~~
    \subfloat[Continuous customer type example. The x-axis is the posted price $\price$. The blue (black, red) curve is the expected revenue from the no-information (full-information, partial-information) advertising strategy $\noinfor$ ($\fullinfor$, $\scheme^*$) and given price $\price$.
    \revcolor{The orange dot represents the revenue $\frac{1}{4}$ obtained using the misspecified policy in the true continuous-type environment.}]{
\input{Paper/plots/example-binary-sqrt-total}
\label{fig:example-binary-sqrt-total}
}
\caption{Graphical illustration of binary product quality examples in \Cref{subsec:example}.}
    \label{fig:example-binary}
\end{figure}


First, suppose the seller does not use any advertising strategy, or equivalently, employs a \emph{no-information advertising strategy}, denoted by $\noinfor$.
Under this strategy, the signal conveys no information about the underlying product quality. In other words, no matter which signal the customer observes, his posterior belief about the underlying product quality remains equal to the prior. Consequently, for any customer of type $\type$, his willingness to pay is $\type + \expect[\quality\sim\prior]{\quality} = \type + \frac{1}{4}$. Fixing the no-information advertising strategy $\noinfor$, the optimal expected revenue is $\frac{1}{4}$ with optimal price $\price^\NI = \frac{1}{4}$ if the high-type probability $\typePDF \leq \frac{1}{5}$, and is $\frac{5}{4}\cdot\typePDF$ with optimal price $\price^\NI = \frac{5}{4}$ if the high-type probability $\typePDF \geq \frac{1}{5}$. See the blue curve in \Cref{fig:example-binary-binary-total}.

Second, suppose the seller uses a \emph{full-information advertising strategy}, denoted by $\fullinfor$, that always fully discloses the product quality to the customers. Under this strategy, the signal $\signal$ fully reveals the quality $\quality$ of the product, i.e., $\signal = \quality$. Consequently, given any realized type~$\type$ and signal $\signal$, the customer's willingness to pay is $\type + \signal$. Fixing the full-information advertising strategy $\fullinfor$, the optimal expected revenue is $\frac{1}{4} + \frac{3}{4}\cdot \typePDF$ with optimal price $\price^\FI = 1$. See the black curve in \Cref{fig:example-binary-binary-total}.

\Cref{fig:example-binary-binary-total} illustrates that \emph{it is important for the seller to decide whether to reveal the product quality, as this decision has a significant impact on her revenue, and the optimal choice depends on the problem instance}. Specifically, when the high-type probability $\typePDF$ is small (i.e., $\typePDF<1/2$), fully revealing the product quality $\fullinfor$ achieves strictly higher revenue than hiding the product quality $\noinfor$. The intuition is as follows: When the probability $\typePDF$ is small, most of the customers' willingness to pay comes from the product quality. In this example, by fully revealing the product quality (and optimizing price $\price^\FI$), the seller can extract the full customer surplus unless the customer and product \emph{simultaneously} have high type and high quality. In contrast, by hiding the product quality (and optimizing price $\price^\NI$), the seller suffers a revenue loss when the customer has high (resp.\ low) type for $\typePDF \leq 1/5$ (resp.\ $\typePDF > 1/5$), \emph{regardless of the product quality}, which thus occurs with relatively higher probability. In contrast, when the high-type probability $\typePDF$ is large (i.e., $\typePDF>1/2$), both the product quality and customer type contribute significantly to the customers' willingness to pay. By the same reasoning, hiding the product quality now leads to a revenue loss with smaller probability in this parameter regime. Thus, the seller prefers the no-information advertising strategy.

So far, we have observed the potential benefit of an advertising strategy that naively fully reveals the product quality. One natural follow-up question is \emph{whether the seller can further improve her revenue with a more carefully designed advertising strategy}. To answer this question, we next consider our second example with a continuous type distribution $\typeCDF$.\footnote{For the first example with binary customer types, it can be shown that it is optimal for the seller to use either the no-information advertising strategy $\noinfor$ or the full-information advertising strategy $\fullinfor$ for every high-type probability $\typePDF\in[0, 1]$.}

\xhdr{Continuous customer type} 
In this example, the customers have a continuous type distribution $\typeCDF$ with CDF $\typeCDF(\type) = \sqrt{\type}$ for every $\type \in [0, 1]$. As we will show below, a carefully designed advertising strategy can indeed improve the seller's revenue compared to the revenue when the seller chooses not to advertise her product or the revenue when she simply fully reveals her product quality.

In this example, fixing the no-information advertising strategy $\noinfor$ or the full-information strategy $\fullinfor$, the expected revenues from all prices are illustrated in \Cref{fig:example-binary-sqrt-total}. In particular, the optimal revenue is equal to $\frac{1}{4}$ with price~$\price^\NI =\frac{1}{4}$ under strategy $\noinfor$, and is equal to $\frac{64}{243}\approx 0.263$ with price $\price^\FI=\frac{64}{81}\approx 0.790$ under strategy $\fullinfor$.

Now suppose the seller uses a more carefully designed advertising strategy $\scheme^*$ that partially reveals the product quality using binary signals. Specifically, when quality $\quality = 1$, the seller sends signal $\signal = 1$. When quality $\quality = 0$, the seller sends signal $\signal = 1$ with probability $\frac{2}{9}$ and signal $\signal = 0$ with probability $\frac{7}{9}$. Under this advertising strategy, the customer's posterior belief $\posteriorDist(\signal)$ given signal~$\signal$ satisfies
\begin{align*}
    \expect[\quality\sim \posteriorDist(\signal)]{\quality} = 
    \left\{
    \begin{array}{ll}
        0 &  \text{ if }\signal = 0\\
        3/5 & \text{ if }\signal = 1
    \end{array}
    \right.
\end{align*}
Consequently, given any realized type $\type$ and signal $\signal$, the customer's willingness to pay is $\type + \frac{3}{5}\signal$. Fixing this strategy $\scheme^*$, the expected revenue from all prices is also illustrated in \Cref{fig:example-binary-sqrt-total}. The optimal revenue given $\scheme^*$ is equal to $\frac{1}{100}(60-7\sqrt{15})\approx 0.329$ with price $\price^* = \frac{3}{5}$, which is 31.6\% and 25.1\% higher than the revenue from the no-information advertising strategy $\noinfor$ and the full-information advertising strategy $\fullinfor$, respectively. 

\revcolor{The high-level intuition behind this revenue improvement can be summarized as follows. In this example, hiding quality information enables the seller to achieve high revenue from low-quality products but results in low revenue from high-quality products. Conversely, fully disclosing quality information allows the seller to obtain high revenue from high-quality products but reduces revenue from low-quality products. By implementing the partial-information advertising strategy $\scheme^*$, the seller maximizes revenue by striking an optimal balance between the revenues from low- and high-quality products.

Importantly, the optimal advertising policy also depends on the posted price. To illustrate this interaction, suppose that the price $\price$ is fixed and the seller optimizes only the advertising policy. In the above example with binary quality and continuous type CDF $\typeCDF(\type)=\sqrt{\type}$, an advertising policy is equivalently represented by a distribution over posterior means $\posteriorMean\in[0,1]$ satisfying $\expect{\posteriorMean}=\frac{1}{4}$ (see the formal argument for this equivalence in \Cref{subsec:equiv}). For any fixed price $\price\in[\frac{1}{4},1]$, the optimal advertising policy generates posterior means $\posteriorMean=0$ and $\posteriorMean=\price$, with posterior mean $\posteriorMean=\price$ occurring with probability $\frac{1}{4\price}$. Equivalently, the seller always sends the favorable signal when $\quality=1$ and sends it with probability $\frac{1-\price}{3\price}$ when $\quality=0$. Consequently, when $\price=\frac{1}{4}$, this policy reduces to the no-information advertising strategy $\noinfor$; when $\price=\frac{3}{5}$, it coincides with the partial-information advertising strategy $\scheme^*$ described above; and when $\price=1$, it reduces to the full-information advertising strategy $\fullinfor$. Thus, as the posted price increases, the optimal advertising policy becomes progressively more informative. This illustrates that advertising and pricing are jointly determined: the information policy shapes demand at a given price, while the price determines how much quality information must be disclosed to induce purchases.}


\revcolor{\xhdr{Revenue loss under misspecification}
The two type distributions above also illustrate why knowing the correct customer type distribution can be important: consider the continuous-type distribution with CDF
$\typeCDF(\type)=\sqrt{\type}$ for every $\type\in[0, 1]$ as the ground truth. 

Now suppose the seller incorrectly believes that customer types follow the binary type distribution in the first example. To make the comparison fair, we choose the high-type probability in the binary distribution to match the mean of the true distribution; that is, $\typePDF = \int_0^1 \type\, \dd (\sqrt{\type}) = \frac{1}{3}$.
Under this misspecified binary belief, the seller solves the static pricing-and-advertising problem as in the binary-type example above. As discussed above, when $\typePDF = \frac{1}{3}$, the full-information advertising strategy $\fullinfor$ with price $\price^\FI= 1$ is optimal (see the orange dot in \Cref{fig:example-binary-binary-total}).
However, when the ground truth is the continuous distribution with CDF $\typeCDF(\type)=\sqrt{\type}$, using the full information advertising strategy $\fullinfor$ with price $\price^\FI= 1$ only guarantees a revenue of $1/4$ (see the orange dot in \Cref{fig:example-binary-sqrt-total}).

In contrast, the true optimal policy uses the partial-information advertising strategy $\scheme^*$ and price $\price^*=\frac{3}{5}$, which yields the optimal revenue approximately equal to $0.329$. Therefore, the absolute revenue loss induced by the misspecified type distribution is $0.329 - \frac{1}{4} = 0.079$, and the relative revenue loss is $\frac{0.079}{0.329}\approx 24\%$.

Motivated by the revenue loss under type-distribution misspecification, \Cref{sec:alg design} develops an online learning algorithm that optimizes the revenue by repeatedly interacting with the uncertain environment to learn the underlying type distribution. (In \Cref{apx:rev-gap}, we further illustrate the revenue loss that arises when the underlying true type distribution is exponential, but the seller misspecifies it as a Gaussian distribution with the same mean.)}


%% file: Paper/plots/example-binary-binary-total.tex
\begin{tikzpicture}[scale=0.75, transform shape]
\begin{axis}[
axis line style=gray,
axis lines=middle,
xlabel = $f$,
ylabel = $\Rev$,
xtick={0, 0.3333, 0.5, 1},
ytick={0},
xticklabels={0, $\frac{1}{3}$, $\frac{1}{2}$, 1},
yticklabels={0},
xmin=0,xmax=1.05,ymin=-0.0,ymax=1.3,
width=0.65\textwidth,
height=0.5\textwidth,
samples=500]

\addplot[domain=0:0.2, blue, line width=0.5mm] (x, {0.25});
\addplot[domain=0.2:1, blue, line width=0.5mm] (x, {5 / 4 * x});

\addplot[domain=0:1, black, line width=0.5mm] (x, {(0.25 + 0.75 * x});

\addplot[dotted, gray, line width=0.3mm] (0.5, 0) -- (0.5, 0.625);

\addplot[dotted, orange!80!black, line width=0.3mm] coordinates {
    (0.3333333333, 0)
    (0.3333333333, 0.5)
    (0, 0.5)
};
\addplot[only marks, mark=*, mark size=3pt, orange!80!black] coordinates {
    (0.3333333333, 0.5)
};

\end{axis}

\end{tikzpicture}

%% file: Paper/plots/example-binary-sqrt-total.tex
\begin{tikzpicture}[scale=0.75, transform shape]
\begin{axis}[
axis line style=gray,
axis lines=middle,
xlabel = $\price$,
ylabel = $\Rev$,
xtick={0,  0.25, 0.6, 0.79012, 1},
ytick={0, 0.25},
xticklabels={0, $\price^\NI$, $\price^*$, $\price^\FI$, $1$},
yticklabels={0, $\frac{1}{4}$},
xmin=0,xmax=2.1,ymin=-0.0,ymax=0.39,
width=0.7\textwidth,
height=0.5\textwidth,
samples=500]

\addplot[domain=0:0.25, blue, line width=0.5mm] (x, {x});
\addplot[domain=0.25:1.25, blue, line width=0.5mm] (x, {x * (1 - (x - 0.25)^0.5)});

\addplot[domain=0:1, black, line width=0.5mm] (x, {x * (0.25 + 0.75 * (1 - (x)^0.5))});
\addplot[domain=1:2, black, line width=0.5mm] (x, {x * (0.25 * (1 - (x - 1)^0.5))});

\addplot[domain=0:0.6, red, line width=0.5mm] (x, {x * (5 / 12 + 7 / 12 * (1 - (x^0.5)))});
\addplot[domain=0.6:1, red, line width=0.5mm] (x, {x * (5 / 12 * (1 - (x - 3 / 5)^0.5) + 7 / 12 * (1 - (x^0.5)))});
\addplot[domain=1:1.6, red, line width=0.5mm] (x, {x * (5 / 12 * (1 - (x - 3 / 5)^0.5))});

\addplot[dotted, gray, line width=0.3mm] (0.25, 0) -- (0.25, 0.25) -- (0, 0.25);

\addplot[dotted, gray, line width=0.3mm] (0.6, 0) -- (0.6, 0.328891) -- (0, 0.328891);

\addplot[dotted, gray, line width=0.3mm] (0.79012, 0) -- (0.79012, 0.26337) -- (0, 0.26337);







\addplot[only marks, mark=*, mark size=3pt, orange!80!black] coordinates {
    (1, 0.25)
};

\addplot[dotted, orange!80!black, line width=0.3mm] (1, 0) -- (1, 0.25);
\addplot[dotted, orange!80!black, line width=0.3mm] (1, 0.25) -- (0, 0.25);

\end{axis}

\end{tikzpicture}

%% file: Paper/3-2-voa-new.tex
The examples in \Cref{subsec:example} demonstrate that informative advertising can substantially enhance the seller's revenue (by over 30\% in some instances) and that the optimal advertising strategy depends sensitively on the underlying customer type distribution. These observations motivate a systematic investigation of two fundamental questions relative to our base model, which studies the optimal revenue under jointly optimized advertising and pricing. First, \emph{to what extent can advertising improve revenue over the simpler strategy of posting an optimized quality-independent price with no advertising at all?} Second, \emph{how much revenue does the seller sacrifice by committing to a single quality-independent price, rather than allowing the price to depend on the realized quality or advertising signal?}

To answer these questions, we compare our original benchmark  $\optU(\instance)$ (i.e., jointly optimized advertising and pricing) against two natural alternatives:\footnote{\label{footnote:related-work-benchmark}\revcolor{In \Cref{apx:comparisons}, we discuss several related prior works that study these two benchmarks, and we compare their models and results with ours. In \Cref{apx:alternative-benchmark-learning}, we also extend our algorithmic and analytical framework to design learning algorithms for two variants of the model corresponding to these two benchmarks.}}
\begin{enumerate}
    \item[(i)] \textbf{(Pricing-only benchmark)} $\optNI(\instance) \triangleq \sup\nolimits_{\price\geq 0} \Rev_{\instance}(\price,\noinfor)$, which corresponds to the best revenue achievable under the optimal quality-independent price without advertising (or equivalently, under the no-information advertising strategy $\noinfor$).
    \item[(ii)] \textbf{(Signal-contingent-pricing benchmark)} $\optSC(\instance)$, which allows the seller to jointly optimize advertising and a price menu that can vary with the realized signal.\footnote{\label{footnote:signal-equal-quality}\revcolor{We remark that this is equivalent to allowing the seller to set the price as a function of both the realized quality and the advertising signal. To see this equivalence, note that for any pricing strategy in which the posted price $\price$ is correlated with both the realized} \revcolor{quality~$\quality$ and the signal $\signal$, we can construct a new advertising strategy in which each signal is encoded as a pair of the original signal and the original price, i.e., $\signal\primed = (\signal,\price)$, with the probability of each signal $\signal\primed$ set according to the joint distribution of the original triple $(\quality,\signal,\price)$. One can verify that, upon observing signal $\signal\primed$ and price $\price$ in the constructed advertising scheme, the customer's posterior belief is the same as upon observing signal $\signal$ and price $\price$ in the original policy. Thus, the purchase decisions and revenues coincide.}}
    Formally, for an advertising strategy $\scheme$ with signal space $\signalSpace$, let $\signalPrice:\signalSpace\to\reals_+$ denote a signal-contingent pricing strategy, where $\signalPrice(\signal)$ is the price posted after the signal $\signal\in\signalSpace$ is realized.
    Given $\signalPrice$ and $\scheme$, the corresponding signal-contingent revenue is
    $\SCRev_{\instance}(\signalPrice,\scheme)
    \triangleq
    \expect[\quality\sim\prior,\signal\sim\scheme(\cdot\mid\quality),\type\sim\typeCDF]
    {\signalPrice(\signal)\cdot a(\type,\signal)}$~,
    where the customer purchases, i.e., $a(\type,\signal)=1$, if and only if
    $\expect[\quality\sim\posteriorDist(\cdot\mid\signal)]
    {\buyerUtility(\type,\quality)}
    \geq
    \signalPrice(\signal)$.
    The signal-contingent-pricing benchmark is
    $\optSC(\instance)
    \triangleq
    \sup\nolimits_{\signalPrice,\scheme}
    \SCRev_{\instance}(\signalPrice,\scheme)$.
\end{enumerate}
By construction, the three benchmarks satisfy
$\optNI(\instance)
\leq
\optU(\instance)
\leq
\optSC(\instance)$.
The first inequality holds because no-information advertising is a feasible advertising strategy in the pricing-and-advertising benchmark. The second holds because a quality-independent price is a special case of a signal-contingent pricing strategy.
As the main result of this section, we establish that all three benchmarks are within a factor of two of one another.

\begin{restatable}[Revenue comparisons]{theorem}{voabound}
\label{thm:voa bound}
For every instance $\instance$
where the customers' valuation function satisfies \Cref{assump:valuation},\footnote{\label{footnote:lipschitz}\revcolor{We note that the upper-bound part of \Cref{thm:voa bound} holds under a valuation class broader than the one defined in \Cref{assump:valuation}. 
    Specifically, we prove the upper bound assuming that the valuation function admits an affine-in-type representation
$\buyerUtility(\type,\quality) = \type\cdot \auxfunc_1(\quality)+\auxfunc_2(\quality)$,
where $\auxfunc_1,\auxfunc_2:\qualitySpace\to\reals_+$ are measurable functions. Thus, the theorem requires affinity in the customer type $\type$ but allows arbitrary nonlinear dependence on the product quality $\quality$. 
(The linear-in-quality assumption in \Cref{assump:valuation} is used later in the dynamic pricing and learning analysis.) Moreover, 
the analysis does not rely on the $1$-Lipschitz normalization of $\buyerUtility(\cdot,\quality)$. Consequently, the same upper bound holds for any finite Lipschitz constant, and the factor $2$ is independent of it.}}
the following sandwich inequality holds
\begin{align*}
    \optNI(\instance)
    \leq
    \optU(\instance)
    \leq
    \optSC(\instance)
    \leq
    2\optNI(\instance)~.
\end{align*}
Moreover, all three factors are tight:
\begin{align*}
    \sup\nolimits_{\instance}
    \frac{\optU(\instance)}{\optNI(\instance)}
    =
    2~,\quad
    \sup\nolimits_{\instance}
    \frac{\optSC(\instance)}{\optNI(\instance)}
    =
    2~,\quad
    \sup\nolimits_{\instance}
    \frac{\optSC(\instance)}{\optU(\instance)}
    =
    2~,
\end{align*}
where the supremum is over all instances $\instance$ where customers' valuation function satisfies \Cref{assump:valuation}.
\end{restatable}

\xhdr{Implications of the theorem} 
\Cref{thm:voa bound} delivers a crisp economic message regarding the \emph{value of advertising} and the \emph{impact of quality-independent pricing}.

The value of advertising, defined as the ratio $\optU(\instance)/\optNI(\instance)$, measures the revenue gain from optimally designing an advertising strategy while keeping quality-independent pricing on both sides of the comparison. \Cref{thm:voa bound} shows that this value is always at most $2$, and that the factor $2$ is tight. Thus, advertising can substantially increase revenue, and can nearly double the revenue achieved by the no-information strategy for some instances. At the same time, optimizing only the price without advertising is a robust fallback: it guarantees at least half of the revenue achieved by the best jointly optimized advertising and pricing policy, for every instance (satisfying \Cref{assump:valuation}). This yields a managerial implication: when informative advertising is costly to implement, the $2$-factor provides a benchmark for evaluating the tradeoff between the potential revenue gain from advertising and the associated advertising expenditures.

The impact of quality-independent pricing, defined as the ratio $\optSC(\instance)/\optU(\instance)$, measures the revenue gain from allowing the seller to tailor the price to the realized advertising signal, relative to posting a single quality-independent price. As motivated earlier, posting a single price is common in many applications due to its operational simplicity, transparency, and implementability. From a theoretical perspective, however, one might worry that a single price is overly restrictive. \Cref{thm:voa bound} shows that this concern is mild: once the seller jointly optimizes the advertising strategy and single posted price, she can achieve a $2$-approximation to the signal-contingent-pricing benchmark. This provides a theoretical justification for focusing on simple quality-independent pricing in the main model.


\xhdr{Proof sketch of the theorem}
Here we sketch the high-level proof idea behind \Cref{thm:voa bound}; the formal analysis is deferred to \Cref{apx:voa proof}. The proof proceeds in three steps. First, we show that full-information advertising is optimal for the signal-contingent-pricing benchmark. Under the affine-in-type valuation class, the optimal conditional revenue given a signal is convex in the posterior averages; by Jensen's inequality, splitting a posterior into more informative signals can only improve revenue, so full-information advertising is optimal. Second, we upper-bound the signal-contingent-pricing benchmark by decomposing revenue into two components: one from the type-dependent part and one from the baseline-quality part. We then show that the no-information benchmark can capture each component separately with an appropriate quality-independent price. This yields $\optSC\leq 2\optNI$, which implies all three upper bounds. Third, we prove tightness. The tightness of $\optU/\optNI$ and $\optSC/\optNI$ follows from \Cref{example:voa lb 1}, which features binary quality and an additive valuation with a bounded equal-revenue type distribution. The tightness of $\optSC/\optU$ is established in \Cref{example:voa lb 2}, in which signal-contingent pricing exploits the low- and high-quality states separately, whereas any quality-independent price must compromise between them.


\xhdr{Necessity of joint optimization}
Our benchmark $\optU(\instance)$ is achieved by jointly optimizing the single quality-independent price and the advertising strategy, and this joint optimization is essential for the factor-of-two guarantee. \Cref{example:unbounded-adver-fixed-price} below demonstrates that if the quality-independent price is held fixed, there exist instances in which the revenue gap between optimal advertising and no-information advertising is unbounded. Consequently, the ratio $\optU(\instance)/\optNI(\instance)$---which isolates the value of advertising---is meaningful only when both benchmarks are permitted to optimize the quality-independent price. Similarly, \Cref{example:unbounded-sc-pricing-fixed-adver} below demonstrates that if the advertising strategy is held fixed, signal-contingent pricing can outperform quality-independent pricing by an unbounded factor. Consequently, the ratio $\optSC(\instance)/\optU(\instance)$---which isolates the impact of quality-independent price---is meaningful only when both benchmarks are permitted to optimize the advertising strategy. Taken together, these results show that the boundedness of our characterization relies critically on optimizing both levers simultaneously: fixing either one while varying the other can lead to arbitrarily large revenue gaps.


\begin{example}
\label{example:unbounded-adver-fixed-price}
Consider an instance with additive valuation $\buyerUtility(\type, \quality) = \type + \quality$ and a degenerate type distribution at $\type=0$. Let the product quality be binary, $\quality\in\{0,1\}$, with prior $\prior(0)=\prior(1)=0.5$.

Suppose the seller fixes the quality-independent price at $\price=1$ rather than optimizing it jointly with advertising. Under the no-information advertising strategy $\noinfor$, the customer's posterior expected valuation equals the prior mean quality, $0.5$. Because this is strictly below the price, the customer never purchases, so $\Rev_\instance(1,\noinfor)=0$. Under the full-information advertising strategy $\fullinfor$, the customer perfectly observes the realized quality and purchases whenever $\quality=1$ (since the valuation then equals the price). The resulting revenue is $\Rev_\instance(1,\fullinfor) = 1\cdot \prior(1) = 0.5$. 
Consequently, the ratio $\sup_{\scheme}\Rev_\instance(1,\scheme)/\Rev_\instance(1,\noinfor)$ is unbounded, showing that the value of advertising can be arbitrarily large when the price is fixed.
\hfill\halmos
\end{example}

\begin{example}
\label{example:unbounded-sc-pricing-fixed-adver}
Consider an instance with additive valuation $\buyerUtility(\type, \quality) = \type + \quality$ and a degenerate type distribution at $\type=0$. Let the product quality distribution $\prior$ be continuous on $[\eps, 1]$ with CDF $\prob{\quality \le t} = 1 - \eps/t$ for $t \in [\eps, 1)$, and place an atom of mass $\eps$ at $\quality=1$.

Suppose the seller fixes the full-information advertising strategy $\fullinfor$ rather than optimizing it jointly with pricing. Under the optimal quality-independent price $\price = \eps$, the customer always purchases at this price $\price = \eps$, so $\Rev_\instance(\eps,\fullinfor)=\eps$. Under the signal-contingent price policy $\signalPrice$ where $\signalPrice(\signal) = \quality$ for every realized signal $\signal=\quality$, the customer perfectly observes the realized quality and purchases at price $\signalPrice(\signal) = \quality$. 
The resulting revenue is $\SCRev_\instance(\signalPrice,\fullinfor) = \expect{\quality}=\eps+\eps\cdot \ln(1/\eps)$. 
Consequently, the ratio $\SCRev_\instance(\signalPrice,\fullinfor)/\Rev_\instance(\price,\fullinfor)$ is unbounded as $\eps\to 0$, showing that the impact of quality-independent price can be arbitrarily bad when the advertising strategy is fixed.
\hfill\halmos
\end{example}

%% file: Paper/4-algo.tex
\revcolor{The preceding section analyzed the value of advertising and the impact of quality-independent pricing on the seller's expected revenue.
We observed that the optimal advertising strategy can double the seller's expected revenue compared to optimal pricing without advertising.
At the same time, as also illustrated in the preceding section, solving for the optimal advertising strategy and the corresponding optimal price requires accurate knowledge of the underlying demand function (namely, the customers' type distribution~$\typeCDF$).
This assumption hinders the practical application of the resulting pricing and advertising policies, since demand functions are typically unknown and must be learned from sales data.
In this section, we study the design of dynamic pricing and advertising policies that incorporate demand learning in the presence of demand uncertainty.\footnote{\revcolor{In \Cref{apx:alternative-benchmark-learning}, we also extend our algorithmic and analytical framework to design learning algorithms for two variants of the model: (i) the seller can only offer a
single quality-independent price without advertising; and (ii) the seller can jointly optimize advertising and a
price menu that varies with the realized signal. These two variants correspond to the benchmarks $\optNI$ and $\optSC$, respectively, studied in \Cref{thm:voa bound}.}}

In \Cref{subsec:equiv}, we present an equivalent reformulation that enables tractable advertising strategies.
In \Cref{subsec:challenges}, we discuss several important challenges that persist even after this simplification.
Finally, in \Cref{subsec:algorithm}, we present our algorithm; the regret guarantee and its analysis are given in the next section.}

\subsection{An Equivalent Reformulation for Tractable Advertising Strategies}
\label{subsec:equiv}
Recall that in every round, customer $t$ sees the offered price $\price_t$ and the advertising strategy $\scheme_t$, which specifies, for each possible value $\quality$ of the realized product quality, the distribution over signals $\scheme_t(\cdot|\quality)\in\Delta(\signalSpace)$ that the seller will send.
After the product quality $\quality_t$ is realized, customer $t$ sees a signal $\signal_t\sim \scheme_t(\cdot|\quality_t)$ drawn from the seller's declared advertising strategy. The customer uses this signal, together with the prior $\prior$, to form a Bayesian posterior $\posteriorDist_t(\cdot|\signal_t)\in \Delta(\qualitySpace)$ over the product quality. A Bayesian rational customer then takes an action $\decision_t\in\{0,1\}$ to maximize expected utility.
In particular, we have
\renewcommand{\arraystretch}{1.2}
\begin{equation*}
    \decision_t =
    \left\{\begin{array}{ll}
    1 & \text{if } \expect[\quality\sim\posteriorDist_t(\cdot|\signal_t)]{\buyerUtility(\type_t, \quality)} \ge \price_t\\
    0 & \text{ otherwise}
    \end{array}
    \right.
\end{equation*}
From the decision formula above, it is clear that the choice of advertising strategy affects customer $t$'s decision only through the realized posterior $\posteriorDist_t(\cdot|\signal_t)$. Consequently, the seller's choice of advertising strategy in round~$t$ can be reduced to selecting a distribution over posteriors $\posteriorDist_t$.
The seller's choice can in fact be simplified further when the customer's valuation function is linear in the product quality $\quality$ (\Cref{assump:valuation}). By the linearity of the valuation function,
\begin{align*}
\expect[\quality\sim\posteriorDist_t(\cdot|\signal_t)]{\buyerUtility(\type_t, \quality)} = \buyerUtility(\type_t, \expect[\quality\sim\posteriorDist_t(\cdot|\signal_t)]{\quality}) = \buyerUtility(\type_t, \posteriorMean_t)
\end{align*}
where $\posteriorMean_t\triangleq\expect[\quality\sim\posteriorDist_t(\cdot|\signal_t)]{\quality}$ is the realized posterior mean. Here, the posterior mean $\posteriorMean_t\in [0,1]$ since the product quality $\quality \in \qualitySpace \subseteq [0,1]$. Therefore, the customer purchases ($\decision_t=1$) if and only if $\buyerUtility(\type_t, \posteriorMean_t)\ge \price_t$.
As a result, we can reduce the seller's advertising decision in round $t$ to the choice of a distribution (PDF) $\distOfMean_t(\cdot) \in\Delta([0, 1])$ over posterior means.

However, the choice of posterior mean distribution $\distOfMean_t$ must be \emph{feasible}, that is, it must be inducible by an advertising scheme given the prior $\prior$.
For simplicity of exposition, we consider a discrete quality space $\qualitySpace = \{\qualityVal_1, \ldots, \qualityVal_m\} \subseteq [0, 1]$, where $0 = \qualityVal_1 < \qualityVal_2 < \ldots < \qualityVal_m = 1$ and $\stateNum = |\qualitySpace|$ is the cardinality of the quality space.
Recall that the distribution of posterior means is induced by the distribution of posteriors, whose feasibility is characterized by the Bayes-consistency condition \eqref{bayes consistency vanilla}.
Thus, we can deduce the feasibility of the posterior mean distribution by restating condition \eqref{bayes consistency vanilla} in terms of conditional means. Specifically, with a slight abuse of notation, for every $i\in[\stateNum]$ we denote by $\distOfMean_i\in\Delta([0, 1])$ the conditional distribution over posterior means when the realized state is $\qualityVal_i$.
We say that a posterior mean distribution $\distOfMean$ is feasible if one can construct a set of conditional distributions $(\distOfMean_i)_{i\in[\stateNum]}$ satisfying the following Bayes-consistency condition, and vice versa \citep{KG-11}:\footnote{For readers familiar with the information design literature, this condition essentially states that the distribution over posterior means $\distOfMean$ is feasible if and only if it is a mean-preserving contraction of the prior \citep{BG-79,AMS-95}.}
\begin{align}
    \label{bayes consistency}
    \tag{{\textsc{BC-PostMean}}}
    \frac{\sum_{i\in[\stateNum]} \prior_i \distOfMean_i(\posteriorMean)\qualityVal_i}{\sum_{i\in[\stateNum]} \prior_i \distOfMean_i(\posteriorMean)}
    = \posteriorMean, \quad
    \forall \posteriorMean\in\supp(\distOfMean)~.
\end{align}
where $\prior_i$ is the prior probability of the realized state $\qualityVal_i$.
Throughout this paper, we use a collection of distributions $(\distOfMean_i)_{i\in[\stateNum]}$ satisfying condition \eqref{bayes consistency} as a convenient way to construct feasible distributions over posterior means: $\distOfMean(\posteriorMean)=\sum_i \distOfMean_i(\posteriorMean) \prior_i$.

With the above observations, we can assume without loss of generality that the seller's advertising strategy directly chooses a distribution $\distOfMean_t$ over posterior means satisfying \eqref{bayes consistency}, without explicitly designing the underlying advertising scheme $\{\scheme, \signalSpace\}$.
 
\xhdr{Timeline} We summarize the timeline of the new equivalent game as follows. At each round $t \in[\timeHorizon]$:
\begin{enumerate}
    \item
    The seller decides a price\footnote{Starting from \Cref{sec:alg design}, we assume that the seller chooses a price $\price \in[\plower,\pupper]$, where $\pupper\in\reals_+$ is a constant upper bound. This follows the standard assumption in the dynamic pricing literature \citep{KL-03,KZ-14}.} $\price_t \in [\plower, \pupper]$ and an
    advertising strategy $\distOfMean_t =
    (\distOfMean_{i, t})_{i\in[\stateNum]}$ satisfying~\ref{bayes consistency};
    \item A customer $t$ with private type $\type_t\sim \typeCDF$ arrives;
    \item A product quality $\quality_t\sim\prior$ and a posterior mean
    $\posteriorMean_t\sim \distOfMean_t$ are realized;
    \item The customer observes the posterior mean $\posteriorMean_t$ and purchases the product ($\decision_t=1$), generating revenue $\price_t$ for the seller, if and only if $\buyerUtility(\type_t,\posteriorMean_t) \ge  \price_t$.
\end{enumerate}
Note that the seller knows the form of the customer's valuation function $\buyerUtility$.
Moreover, the seller observes the realized product quality $\realizedQuality_t$,
the realized posterior mean $\posteriorMean_t$,
and the customer's decision $\decision_t$,
but knows neither the type CDF $\typeCDF$ (i.e., the demand function $\demand$)
nor the realized customer type $\type_t$.

\xhdr{Revenue and regret} Given the new formulation, we can also rewrite the revenue and regret in terms of the chosen posterior mean distributions $\distOfMean_t$, $t=1,\ldots, \timeHorizon$.
We define the following function $\inverseVal(\price, \posteriorMean)$, which we refer to as the {\it critical type} for a given price $\price$ and posterior mean $\posteriorMean$.
\begin{definition}[Critical type]
\label{defn:critical type}
For any price $\price\in [\plower, \pupper]$ and posterior mean $\posteriorMean\in [0, 1]$,
define the critical type
$\inverseVal(\price, \posteriorMean)
\triangleq \inf\{\type\in\typeSpace: \buyerUtility(\type, \posteriorMean) \ge \price\}$ as the smallest type whose expected value is weakly greater than the price $\price$ and who therefore purchases.
\end{definition}
Under \Cref{assump:valuation}, by the monotonicity of the valuation function in the customer's type $\type$, for any price $\price$, posterior mean $\posteriorMean$, and type $\type$, we have
$\indicator{\buyerUtility(\type, \posteriorMean) \ge \price} = \indicator{\type \ge \inverseVal(\price,\posteriorMean)}$.
Therefore, customer $t$ purchases the product if and only if ${\type_t \ge \inverseVal(\price_t, \posteriorMean_t)}$.

Consequently, given the price $\price_t=\price$, the advertising strategy $\distOfMean_t=\distOfMean$, and the prior distribution $\prior$, the expected revenue in any round $t$ is\footnote{Here, we slightly abuse notation and redefine $\Rev(\cdot,\cdot)$ as a function of the price $\price$ and the posterior mean distribution $\distOfMean$, rather than the price $\price$ and the advertising strategy $\scheme$ used earlier.}
\begin{align}
\label{def:revpq}
    \Rev(\price, \distOfMean) = \expect[\quality \sim \prior, \type \sim \typeCDF, \posteriorMean\sim\distOfMean]{\price \cdot  \indicator{\type \ge \inverseVal(\price, \posteriorMean)}}
    =  \price \sum\nolimits_{i\in[\stateNum]} \prior_i \int_0^1 \distOfMean_i(\posteriorMean) \demand(\inverseVal(\price,  \posteriorMean))\cdot \mathrm{d}\posteriorMean~.
\end{align}
Suppose the seller's online policy offers a price $\price_t$ and an advertising strategy $\distOfMean_t$ in round $t$, where $\price_t$ and $\distOfMean_t$ may depend on the history of observations up to round $t$.
Then the expected regret defined in \eqref{eq:vanilla regret} can be equivalently written as
\begin{align*}
    \Reg{\timeHorizon} = \timeHorizon \Rev(\optPricing,\optAdver)
    - \sum\nolimits_{t\in[\timeHorizon]} \expect{\Rev(\price_t, \distOfMean_t)}~.
\end{align*}
Here, the expectation is taken with respect to any randomness in the algorithm's choice of $\price_t$ and $\distOfMean_t$; and
$\optPricing$ and $\optAdver=(\optAdver_i)_{i\in[\stateNum]}$ are the optimal price and advertising strategy for a given $\typeCDF$ (and the induced $\kappa(\cdot, \cdot)$, which is determined by $\buyerUtility$). Given the expression for $\Rev(\price, \distOfMean)$ derived above, the optimal price and advertising strategy $(\optPricing, \optAdver)$ can be characterized as the solution to the following program:
\begin{align}
    \label{program: benchmark}
    \tag{$\prog_\textsc{OPT}$}
    \begin{alignedat}{2}
    \sup\nolimits_{\price, \distOfMean} ~
    \Rev(\price,\distOfMean) \quad \text{s.t. }~~
    \distOfMean_i\in\Delta([0, 1]), ~
     i \in [\stateNum]; ~~
    \frac{\sum_{i\in[\stateNum]} \prior_i \distOfMean_i(\posteriorMean)\bar{\quality}_i}{\sum_{i\in[\stateNum]} \prior_i \distOfMean_i(\posteriorMean)}
    = \posteriorMean, ~
    \posteriorMean\in[0, 1]~;
    \end{alignedat}
\end{align}
where the second constraint is the Bayes-consistency condition \eqref{bayes consistency}.

\subsection{Algorithm Design: Challenges and Ideas}
\label{subsec:challenges}
\label{subsec:algorithmoverview}

\revcolor{In this section, we discuss the challenges in developing a regret-optimal learning algorithm, together with the key algorithmic ideas for overcoming them. In addition, we discuss how learning interacts with the choice of advertising strategy in our algorithm.}

\xhdr{Challenge one: high-dimensional continuous decision space} In the previous section, we obtained a considerable simplification of the problem by reducing the seller's advertising strategy in every round $t$ to
a \emph{distribution} $\distOfMean_t \in \Delta({[0,1]})$ over posterior means
satisfying the \eqref{bayes consistency} condition.
However, the decision space (a.k.a.\ the space of arms) remains high-dimensional, and thus a naive application of (uniform or adaptive) discretization-based bandit techniques, e.g., from the Lipschitz bandit literature \citep{KSU-08,S-11}, would not achieve the desired results.

\noindent\underline{\emph{Algorithmic idea: exploring over a one-dimensional type space.}} Our algorithm uses a \emph{model-based approach} instead. The key observation is that the customer's purchase decision depends on the price $\price_t$ and the posterior mean $\posteriorMean_t$ only through the scalar critical type $\inverseVal(\price_t, \posteriorMean_t)$; hence, learning the demand model reduces to estimating the one-dimensional function $\demand(\type)=1-\typeCDF(\type)$. Specifically, we use customer purchase responses to construct upper confidence bound estimates of the demand model $\demand(\type)=1-\typeCDF(\type)$ at the points of a \emph{discretized type space} $\discTyepSpace \subseteq \typeSpace$. We then use these estimates to build a piecewise-constant demand function that serves as an upper confidence bound (UCB) on the true demand function $\demand$.
In each round, we then select the price and advertising strategy by solving an optimization problem analogous to \ref{program: benchmark}, but with the UCB demand function in place of the true demand.

\xhdr{Challenge two: efficient discretization of the type space}
The next challenge is to design a discretization scheme for the type space that achieves
(a) efficient learning, i.e., the discretized space can be efficiently explored to accurately estimate the demand function at those points, and (b) a Lipschitz property, i.e., the optimal revenue under the UCB demand estimate is close to the true optimal revenue whenever the estimation error on the discretized space is small.

There are two main difficulties in achieving this:
(i) the lack of any smoothness/Lipschitz assumption on the CDF $\typeCDF$; and (ii) the sensitivity of the Bayes-consistency condition \eqref{bayes consistency}.
Difficulty (i) limits what can be learned from the estimates at the discretization points, while difficulty (ii) limits which discretizations are even feasible.
To see difficulty (i), recall that given a price $\price_t$ and a realized posterior mean $\posteriorMean_t$, customer $t$'s purchase decision is given by $\decision_t = \indicator{\type_t \ge \inverseVal(\price_t, \posteriorMean_t)}$; thus, the seller can obtain an estimate of the demand function at the point $\inverseVal(\price_t, \posteriorMean_t)$. Without any smoothness or Lipschitz property of the demand function, these estimates cannot be accurately extrapolated to other points. This means that when solving our revenue optimization problem (the estimated version of \ref{program: benchmark}) for a price and advertising strategy, we can only rely on demand estimates at the points of a discretized type space, say $\discTyepSpace \subseteq \typeSpace$. However, if we restrict attention to a discretized type space $\discTyepSpace$, then the support of the posterior mean distributions (a.k.a.\ advertising strategies) must be restricted to points $\posteriorMean$ whose corresponding critical types $\inverseVal(\price,\posteriorMean)$ lie in $\discTyepSpace$.

Difficulty (ii) arises because the \eqref{bayes consistency} condition makes the set of feasible advertising strategies very sensitive to their support. In particular, if we use a uniform-grid discretization (which is commonly used in the prior dynamic pricing literature, e.g., \citealp{KL-03,BDKS-15}), it is easy to construct examples of a prior distribution and a valuation function for which there are no, or very few, feasible advertising strategies with the restricted support.
\begin{example}
\label{ex:uniform-discre-failure}
Consider the additive valuation function, i.e., $\buyerUtility(\type, \quality) = \type + \quality$,
and thus $\inverseVal(\price, \posteriorMean) = \price - \posteriorMean$.
Consider the quality space
$\qualitySpace = \{\qualityVal_i\}_{i\in[3]}$.
Given a small $\varepsilon$,
consider a uniform-grid discretization of the type space
$\discTyepSpace$, i.e., $\discTyepSpace =
\{0, \varepsilon, 2\varepsilon, \ldots, 1\}$.
If we also restrict the price $\price$ to a
uniform-grid discretization of the price space,
i.e., $\price = k\varepsilon$ for some $k\in \integers^+$, then to ensure $\inverseVal(\price, \posteriorMean)\in\discTyepSpace$,
the support of the advertising strategy
(i.e., the distribution of posterior means)
must also be restricted to the set $\discTyepSpace$.
However, if the prior distribution $\prior$ places negligible probability
on qualities $\qualityVal_1$ and $\qualityVal_3$, and
quality $\qualityVal_2 \notin \discTyepSpace$,
then no posterior distribution with mean in $\discTyepSpace$ can be constructed.
Therefore, no feasible advertising strategy exists.

Note that this difficulty cannot be fixed by simply enlarging the discretized type space to $\discTyepSpace \cup \qualitySpace$: even then, we would need an advertising strategy with $\inverseVal(\price,\posteriorMean)=\price-\posteriorMean\in \discTyepSpace \cup \qualitySpace$ for all prices $\price$ on the grid, which requires the support of the strategy (i.e., the posterior means $\posteriorMean$) to be restricted to $\{k\varepsilon-\qualityVal_i\}_{k\in\integers^+, i\in[3]}$; such posterior means again may not be attainable.
\hfill\halmos
\end{example}

\noindent\underline{\emph{Algorithmic idea: a novel discretization scheme.}}
Our algorithm employs a carefully designed \emph{quality-and-price-dependent discretization} scheme. The above example shows that we cannot uniformly discretize the price and type spaces using $\varepsilon$-(uniform-)grids, as no feasible advertising strategy may exist under such a discretization. It further shows that this difficulty cannot be resolved by simply adding the $\stateNum$ quality points to the discretized type space $\discTyepSpace$.
Instead, in our discretization scheme, we first uniformly discretize the price space into an $\varepsilon$-grid $\discPrice$. Then, to construct the discretized type space $\discTyepSpace$, we include, in addition to the points on an $\varepsilon$-grid over $[0,1]$, all points
$\{\inverseVal(\price, \quality)\}$ for every price $\price\in\discPrice$ and quality $\quality\in\qualitySpace$. This yields a discretized type space $\discTyepSpace$ of size $\stateNum \pupper/\varepsilon$.
We prove that our construction guarantees the existence of a near-optimal price and advertising strategy whose support lies in $\discTyepSpace$.
Establishing this claim requires a careful rounding argument (Procedure \ref{algo_construction_adver_general}), which forms one of the main technical ingredients of our regret analysis in \Cref{sec:analysis}.

\revcolor{\xhdr{The interplay of advertising and learning}
The construction above reflects a key difference between our problem and the standard dynamic-pricing problem. When the type distribution $\typeCDF$ and thus demand model $\demand$ are known, the advertising policy is chosen only for its static revenue effect: it reshapes the distribution of posterior means $\posteriorMean$, and the seller evaluates each induced ``market segment'' through the demand term $\demand(\inverseVal(\price, \posteriorMean))$. When $\typeCDF$ is unknown, however, the same advertising policy also affects what the seller learns. Conditional on the realized posterior mean $\posteriorMean$, the customer's purchase response is a Bernoulli observation of the demand function at the critical type $\inverseVal(\price, \posteriorMean)$. Thus, advertising is not only a surplus-extraction device; it also serves as a sampling rule over critical types.
In the no-advertising benchmark, the posterior mean is fixed at the prior mean, so choosing a price is essentially equivalent to choosing a single critical type. The seller's problem then reduces to a standard one-dimensional dynamic-pricing problem, for which a uniform discretization of the price or induced type space suffices to obtain the usual $\timeHorizon^{2/3}$-type regret rate \citep{KL-03}.

Once advertising is allowed, a posted price must be combined with a distribution over posterior means, and therefore with a distribution over critical types. To learn and optimize such policies, the seller must estimate demand at the critical types generated by feasible posterior supports. This is the sense in which learning distorts the design of the information policy: the discretization we construct, and the advertising strategy built on top of it, must preserve \eqref{bayes consistency} while still delivering a good revenue guarantee. Specifically, every posterior mean used by the algorithm must induce a critical type at which demand can be statistically learned. The next subsection formalizes these ideas in our learning algorithm.}

\subsection{Details of the Proposed Algorithm}
\label{subsec:algorithm}

Our dynamic pricing and advertising algorithm (\Cref{algo:dynamic pricing and advertising}) jointly discretizes the price space and the type space using the following {\em quality-and-price-based discretization} scheme: given a parameter $\varepsilon$, we define the sets
\begin{equation}
    \begin{alignedat}{1}
        \label{discretized set defn}
        \discPrice
        & \triangleq \{\varepsilon, 2\varepsilon, \ldots, \pupper\}~;\\
        \discTyepSpace
        & \triangleq
        \{0, \varepsilon, \ldots, 1-\varepsilon, 1\} \cup \{(\inverseVal(\price, \quality) \wedge 1)\vee 0\}_{\price\in\discPrice, \quality\in\qualitySpace}~.
    \end{alignedat}
\end{equation}
In each round $t$, we restrict the price and advertising strategy $(\price_t,\distOfMean_t)$ to pairs $(\price,\distOfMean=(\distOfMean_i)_{i\in[\stateNum]})$ such that $\price\in \discPrice$ and, given the price $\price$, each conditional distribution
$\distOfMean_i$ has support restricted to the set $\discPostMeanSpace_\price$ defined as
\begin{align*}
    \discPostMeanSpace_\price \triangleq
    \{\posteriorMean: \inverseVal(\price, \posteriorMean) \in \discTyepSpace\}~.
\end{align*}
Given the price and advertising strategy $(\offeredPrice_t, \distOfMean_t)$, let the realized posterior mean in round $t$ be $\posteriorMean_t \sim \distOfMean_t$, and let the corresponding critical type be $x_t \triangleq \inverseVal(\offeredPrice_t, \posteriorMean_t) \in [0, 1]$.
Note that the above restrictions on the price and advertising strategy guarantee that $x_t \in \discTyepSpace$.

Next, to compute the offered price and advertising strategy in round $t$, we optimize an upper confidence bound on the revenue function, constructed from upper confidence bounds $\trueUCBDemand(x)$, $x\in\discTyepSpace$, on the demand function, which we compute as follows.
For every type $x\in\discTyepSpace$, let $\chosenSet_t(x)$ denote the set of rounds before round $t$ in which the induced \criticalValue\ is exactly $x$, and let $\chosenCounter_t(x) $ be the number of such rounds.
That is,
\begin{align*}
    \chosenSet_t(x)
    \triangleq \left\{\tau < t: \inverseVal(\offeredPrice_\tau, \posteriorMean_\tau)
    = x\right\}~;
    \quad
    \chosenCounter_t(x)
    \triangleq \left|\chosenSet_t(x)\right|~,
    \quad
    x \in \discTyepSpace~.
\end{align*}
Recall that the customer's purchase decision follows $\decision_\tau=\indicator{\type_\tau \ge \inverseVal(\price_\tau,\posteriorMean_\tau)}$.
We estimate the demand function at $x$
as $\averageDemand_t(x) \triangleq
    \frac{\sum_{\tau\in\chosenSet_t(x)} \decision_\tau}{\chosenCounter_t(x)}$.
We can now define the following UCB index:
\begin{align}
\label{defn:true UCB demand}
    \trueUCBDemand_t(x) =
    \min_{x'\in\discTyepSpace: x'\le x}
    \displaystyle\averageDemand_t(x') +
    \sqrt{\frac{16\log \timeHorizon}{\chosenCounter_t(x')}} + \frac{\sqrt{(1+\chosenCounter_t(x')) \ln (1+\chosenCounter_t(x'))}}{\chosenCounter_t(x')}
    \wedge 1, \quad
    x\in\discTyepSpace~.
\end{align}
Then, for any pair consisting of a discretized price $\price \in \discPrice$ and an advertising strategy $\distOfMean=(\distOfMean_i \in \Delta({\discPostMeanSpace_\price}), i\in[\stateNum] \}$ with discretized support for that price,
we define the following estimate of the seller's revenue:
\begin{align*}
    \UCBRev_t(\price, \distOfMean) \triangleq
    \price\sum\nolimits_{i\in[\stateNum]} \prior_i \int_0^1 \distOfMean_i(\posteriorMean)
    \trueUCBDemand_t(\inverseVal(\price, \posteriorMean))\cdot \mathrm{d}\posteriorMean~.
\end{align*}
This is well-defined since, by construction, $\inverseVal(\price, \posteriorMean) \in \discTyepSpace$ for every such pair $(\price, \posteriorMean) \in \discPrice \times \discPostMeanSpace_\price$.
Finally, we let $(\offeredPrice_t, \offeredAdver_t)$ be an optimal solution to the following optimization problem:
\begin{equation}
    \label{empricial opt}
    \tag{$\prog_t^{\cc{UCB}}$}
    \begin{alignedat}{2}
    \sup\nolimits_{\price\in\discPrice, \distOfMean} ~
    \UCBRev_t(\price, \distOfMean)\quad
    && \text{s.t. }  \distOfMean_i\in\Delta({\discPostMeanSpace_\price}), ~~
    i \in [\stateNum]; ~~
    \frac{\sum_{i\in[\stateNum]} \prior_i \distOfMean_i(\posteriorMean)\qualityVal_i}{\sum_{i\in[\stateNum]} \prior_i \distOfMean_i(\posteriorMean)}
    = \posteriorMean, ~~
    \posteriorMean\in \discPostMeanSpace_\price ~.
    \end{alignedat}
\end{equation}

\let\oldnl\nl
\newcommand{\nonl}{\renewcommand{\nl}{\let\nl\oldnl}}

\begin{algorithm2e}
\caption{Algorithm for Dynamic Pricing and Advertising with  Demand Learning.}
\label{algo:dynamic pricing and advertising}
\SetAlgoLined\DontPrintSemicolon
\textbf{Input:} Discretization parameter 
$\discrePrecision$. \\ 
For the first $\numDiscType$ rounds, for each $x\in\discTyepSpace$,
offer a price $\price$ with any no information advertising s.t.\ 
$\inverseVal(\price, \expect[\quality\sim\prior]{\quality}) = x$.\\
\nonl\algcomment{No information advertising provides
a completely uninformative signal -- the distribution $\phi(\cdot|\quality)$ of signals does not depend on the realized quality $\quality$}\\
\For{each round $t = \numDiscType+ 1, \numDiscType + 2, \ldots, T$}{
    For all $x\in \discTyepSpace$, compute 
    $\trueUCBDemand_{t}(x)$ as defined in \eqref{defn:true UCB demand}. \\
    Offer the price $\offeredPrice_t$ and an advertising 
    $\offeredAdver_t$ computed as the solution to
    program \ref{empricial opt}. \label{line:solved solution}\\
    \nonl\algcomment{$\offeredPrice_t$ and $ \offeredAdver_t$ satisfy
    $\inverseVal(\offeredPrice_t, \posteriorMean)\in\discTyepSpace$ 
    for every $\posteriorMean\in\supp(\offeredAdver_t)$.}\\
    Observe realized posterior mean 
    $\posteriorMean_t \sim \offeredAdver_t$ 
    and customer's purchase 
    decision $\decision_t\in\{0, 1\}$. \\
    Update $\left\{
    \chosenSet_{t+1}(x), \chosenCounter_{t+1}(x), \averageDemand_{t+1}(x)
    \right\}_{x\in\discTyepSpace}$.
}
\end{algorithm2e}
We summarize our algorithm in
\Cref{algo:dynamic pricing and advertising}.
The main computational bottleneck
of \Cref{algo:dynamic pricing and advertising}
is solving the high-dimensional program \ref{empricial opt}
in each round $t \ge \numDiscType+1$.
As we show in \Cref{prop:complexity of empricial opt},
this program can be solved efficiently and optimally.

\begin{restatable}{proposition}{complexityofempricialopt}
\label{prop:complexity of empricial opt}
Let $\varepsilon$ be the discretization
parameter for the set $\discPrice$ defined in \eqref{discretized set defn}.
There exists an algorithm with running time polynomial in $\sfrac{|\discTyepSpace|\pupper}{\varepsilon}$
that solves the program \ref{empricial opt}.
\end{restatable}
The proof of this result is adapted from \citet{ABSY-19,C-19}, and exploits the monotonicity of the function $\UCBDemand_t$; it is deferred to \Cref{apx:proof of complexity}.
We conclude this section with the following remark on the extension to unbounded type support.
\begin{remark}
\label{rmk:unbounded type}
Our algorithm and analysis can be extended to the case of unbounded type support (e.g., $\typeSpace=[0, \infty)$).
In particular, since the price is bounded in $[0, \pupper]$
and the quality is bounded in $[0, 1]$, the monotonicity of the valuation function implies that the critical type $\inverseVal(\price, \posteriorMean)$ induced by any $\price\in[0, \pupper]$ and $\posteriorMean\in[0, 1]$ lies in $[\inverseVal(0, 1), \inverseVal(\pupper, 0)]$. Thus, an instance with unbounded type support is equivalent to an instance with bounded type support $[\inverseVal(0, 1), \inverseVal(\pupper, 0)]$.
\end{remark}


%% file: Paper/5-analysis.tex
In this section, we present our main regret bound
for the dynamic pricing and advertising algorithm (\Cref{algo:dynamic pricing and advertising}) introduced in the preceding section.
\begin{theorem}[Regret bound]
\label{thm:regret upper bound}
For every instance $\instance$ in which the customers' valuation function satisfies \Cref{assump:valuation},
\Cref{algo:dynamic pricing and advertising} with parameter $\varepsilon = \Theta(({\stateNum\log \timeHorizon}/{\timeHorizon})^{\sfrac{1}{3}} )$
has expected regret $O(\timeHorizon^{\sfrac{2}{3}}
(\stateNum\log\timeHorizon)^{\sfrac{1}{3}})$, where $\timeHorizon$ is the time horizon and $\stateNum$ is the cardinality of the discrete quality space $\qualitySpace$. 

Moreover, for every online algorithm, there exists an instance satisfying \Cref{assump:valuation} and $\stateNum = 1$ on which its regret is $\Omega(\timeHorizon^{2/3})$.
\end{theorem}

For this result, we consider an arbitrary but discrete quality space $\qualitySpace$ of cardinality $\stateNum$. We do not assume any smoothness or Lipschitz properties of the type distribution $\typeCDF$. When the valuation function $\buyerUtility(\cdot, \quality)$ is $\lipschitzconst$-Lipschitz in the type $\type$,
our regret bound becomes
$O(\timeHorizon^{\sfrac{2}{3}} (\lipschitzconst\stateNum \log \timeHorizon)^{\sfrac{1}{3}})$.
\revcolor{In \Cref{subsec:improvement}, we establish improved regret bounds for the case of additive valuations and equally spaced quality spaces (see \Cref{additive and uniformly-spaced}), and we extend the result to arbitrarily large and continuous quality spaces (see \Cref{additive and arbitrary}).}

As stated in the theorem, our algorithm achieves the optimal regret dependence (up to logarithmic factors) on the time horizon $\timeHorizon$. The hardness result simply follows from the observation that our model recovers the well-studied dynamic pricing and learning problem with unlimited supply \citep{KL-03,BDKS-15} as a special case (by setting the product quality to be deterministic, i.e., $m=1$, in which case the advertising scheme reveals no information and thus has no impact on the customer's purchase decision). For the dynamic pricing problem (without advertising), a lower bound of $\Omega(\timeHorizon^{\sfrac{2}{3}})$ on the regret is known \citep{KL-03}. Therefore, the dependence on $\timeHorizon$ in our result cannot be improved.
In fact, our bound matches this lower bound in the case of a binary or constant-size quality space, which is common setting in the information design literature
\citep{KG-11,BDHN-22,AK-20,TH-21,BCVZ-22,FTX-22}.

\revcolor{In the remainder of this section, we outline the regret analysis in \Cref{subsec:proof-outline}. In \Cref{subsec:rounding}, we develop a rounding argument that complements our quality-and-price-dependent discretization scheme and bounds the discretization error; this is a key technical step of the analysis. Finally, in \Cref{subsec:optimism}, we characterize the estimation error and its impact on the regret over the course of the learning dynamics.}


\subsection{Proof Outline}
\label{subsec:proof-outline}
In this section, we outline the regret analysis of \Cref{algo:dynamic pricing and advertising}.
Recall that in every round $t$, \Cref{algo:dynamic pricing and advertising} sets the price $\offeredPrice_t$ and advertising strategy $\distOfMean_t$ to an optimal solution of the program \ref{empricial opt}, which approximates the benchmark \ref{program: benchmark} in two ways.
First, it restricts the price and the support of the advertising strategy to a discretized space $\discPrice \times \{\discPostMeanSpace_p, p\in\discPrice\}$. Second, it approximates the true demand function with an upper confidence bound $\UCBDemand_t$. Our proof consists of two main steps that bound the errors introduced by each of these approximations.
All missing proofs in this section are deferred to \Cref{apx:main proof}.

\xhdr{Step 1: bounding the discretization error via a rounding argument} To separate the discretization error from the error due to demand estimation, we consider an intermediate optimization problem \ref{program: intermediate} (in \Cref{subsec:rounding}) obtained by replacing the UCB demand function $\UCBDemand_t$ in program \ref{empricial opt} with the true demand function $\demand$ (while keeping the discretized space for $\price,\distOfMean$).
    Let $(\discretizedOptPrice, \discretizedOptAdver)$ be an optimal solution of program \ref{program: intermediate}.
    We show that the revenue
    $\Rev(\discretizedOptPrice, \discretizedOptAdver)$ is sufficiently
    close (within $2\varepsilon$) to the optimal revenue $\Rev(\optPrice, \optAdver)$.
    This bound is obtained through a careful {\em rounding} argument:
    we show that the optimal price $\optPrice$ and
    the optimal advertising strategy $\optAdver$ can be rounded
    to a new price $\newPrice$
    and a new advertising strategy $\newAdver$ that satisfy
    (i) feasibility ({\bf \Cref{lem:feasibility adver}}): $\newPrice\in\discPrice$ and $\supp(\newAdver)\subseteq\discPostMeanSpace_{\newPrice}$;
    and
    (ii) a revenue guarantee ({\bf \Cref{lem:revenue guarantee}}): $\Rev(\newPrice, \newAdver) \ge \Rev(\optPrice, \optAdver) - 2\varepsilon$.

\xhdr{Step 2: bounding the estimation error and establishing optimism}
    Next, we show that the UCB estimates of the demand function $\UCBDemand_t(x)$, $x\in \discTyepSpace$, concentrate around the true demand function $\demand$ with high probability, and we derive explicit bounds on the gap between the true and estimated functions ({\bf \Cref{lem:high-prob estimation error}}).
    This allows us to establish
    (i) revenue optimism ({\bf \Cref{lem:optimistic estimates}}): the algorithm's revenue estimates are (almost) optimistic, i.e.,
    $\UCBRev_t(\offeredPrice_t, \offeredAdver_t) \ge \Rev(\discretizedOptPrice, \discretizedOptAdver) \ge \Rev(\optPrice, \optAdver) - 2\varepsilon$;
    and
    (ii) revenue approximation ({\bf \Cref{lem:single-round regret}}): the optimistic revenue estimates are close to the true revenue in round $t$, with the gap between the two inversely proportional to the number of observations; in particular,
    $\UCBRev_t(\offeredPrice_t, \offeredAdver_t) -
    \Rev(\offeredPrice_t, \offeredAdver_t)
    \le
    5\offeredPrice_t \expect[\posteriorMean\sim \offeredAdver_t]{\sqrt{\log \timeHorizon/\chosenCounter_t(\inverseVal(\offeredPrice_t, \posteriorMean))}}$.

Putting all the pieces together yields the regret bound stated in \Cref{thm:regret upper bound}. Specifically, the above observations imply that the regret in each round can be bounded, roughly, by $2\varepsilon + 5\offeredPrice_t \expect[\posteriorMean\sim \offeredAdver_t]{\sqrt{\log \timeHorizon/\chosenCounter_t(\inverseVal(\offeredPrice_t, \posteriorMean))}}$. Then, using the constraint $\sum\nolimits_{x\in \discTyepSpace} \chosenCounter_T(x) \le T$, we show that, in the worst case, the total regret over $T$ rounds is bounded by $O(T\varepsilon + \sqrt{|\discTyepSpace| T\log \timeHorizon)}$. The theorem statement is then obtained by substituting $|\discTyepSpace| = O(\stateNum/\varepsilon)$ and optimizing over the parameter $\varepsilon$.

\subsection{A Rounding Procedure to Bound the Discretization Error}
\label{subsec:rounding}

\input{Paper/4-1-rounding}

\subsection{Estimation Error and Optimism}
\label{subsec:regret}
\label{subsec:optimism}
\input{Paper/4-2-regret-analysis}

%% file: Paper/4-1-rounding.tex
In this section, we bound the revenue loss due to discretization.
Specifically, let $(\discretizedOptPrice, \discretizedOptAdver)$
be an optimal solution of the following program:
\begin{equation}
\label{program: intermediate}
    \tag{$\widetilde\prog$}
    \begin{alignedat}{2}
    \max_{\price}\max_{\distOfMean} ~~
    & \price\sum_{i\in[\stateNum]} \prior_i \int_0^1 \distOfMean_i(\posteriorMean)
    \demand(\inverseVal(p, \posteriorMean))\cdot \mathrm{d}\posteriorMean
    \quad && \text{s.t. }\\
    &  p\in\discPrice; ~~ \distOfMean_i\in\Delta({\discPostMeanSpace_\price}), \quad
    &&  i \in [\stateNum]\\
    &\frac{\sum_{i\in[\stateNum]} \prior_i \distOfMean_i(\posteriorMean)\qualityVal_i}{\sum_{i\in[\stateNum]} \prior_i \distOfMean_i(\posteriorMean)}
    = \posteriorMean, \quad
    &&  \posteriorMean\in \discPostMeanSpace_p \\
    \end{alignedat}
\end{equation}
The main result of this section is the following:
\begin{restatable}{proposition}{discretizationerror}
\label{prop:discretization error}
For any type CDF $\typeCDF$, we have
$\Rev(\optPrice, \optAdver) - \Rev(\discretizedOptPrice, \discretizedOptAdver)
\le 2\varepsilon$.
\end{restatable}
In what follows, we outline the high-level steps of the proof.
All missing formal proofs can be found in \Cref{apx:missing procedure}.
Our argument builds on a novel rounding procedure (see Procedure~\ref{algo_construction_adver_general} in \Cref{apx:missing procedure})
that rounds the optimal price $\optPrice$ and
the optimal advertising strategy $\optAdver$
to a new price $\newPrice$
and a new advertising strategy $\newAdver$ satisfying:
(i) $\newPrice\in\discPrice$ and $\supp(\newAdver)\subseteq\discPostMeanSpace_{\newPrice}$;
and
(ii) a revenue loss of at most $2\varepsilon$, i.e., $\Rev(\optPrice, \optAdver) -
\Rev(\newPrice, \newAdver) \le 2\varepsilon$.
It is worth noting that
Procedure \ref{algo_construction_adver_general},
which we believe is of independent interest,
works for any valuation function that is linear in quality and satisfies \Cref{assump:valuation}. It
only requires knowledge of the critical-type function $\inverseVal(\cdot, \cdot)$
and the prior $\prior$;
in particular, Procedure \ref{algo_construction_adver_general}
does not rely on any knowledge or estimates
of the unknown demand function. Indeed, \Cref{prop:discretization error}
continues to hold if we replace the demand function $\demand$
in the revenue formulation \eqref{def:revpq}
with any non-increasing function.
A graphical illustration of Procedure~\ref{algo_construction_adver_general} is provided in \Cref{fig:rounding}.
\begin{figure}
    \centering
    \input{Paper/plots/rounding}
    \caption{Graphical illustration of Procedure \ref{algo_construction_adver_general}.
    Given the input price and advertising strategy $(p, \distOfMean)$,
    fix a posterior mean $\posteriorMean \in\supp(\distOfMean)$ with
    $\{i'\in[\stateNum]: \distOfMean_{i'}(\posteriorMean) > 0\} = \{i, j\}$ (shown as a black dashed line).
    The procedure first identifies $x = \inverseVal(p, \posteriorMean)$
    and $x^\dagger = \inverseVal(\newPrice, \posteriorMean)
    \in ((z-1)\varepsilon, z\varepsilon)$,
    where the constructed price $\newPrice$ is defined as in the procedure.
    It then finds two posterior means $\leftPosteriorMean$ and $\rightPosteriorMean$ (here,
    $\leftPosteriorMean \ge \qualityVal_i$ and $\rightPosteriorMean \le \qualityVal_j$)
    such that
    $\inverseVal(\newPrice, \leftPosteriorMean) = z\varepsilon$ and
    $\inverseVal(\newPrice, \rightPosteriorMean) = (z-1)\varepsilon$
    (shown as brown dashed lines), with $\inverseVal(\newPrice, \rightPosteriorMean) < \inverseVal(\newPrice, \leftPosteriorMean) < \inverseVal(p, \posteriorMean)$.
    \label{fig:rounding}
    }
\end{figure}

\xhdr{Details and guarantees of Procedure~\ref{algo_construction_adver_general}} Procedure \ref{algo_construction_adver_general} takes as input an advertising strategy $\distOfMean$ that satisfies $|\{i'\in[\stateNum]: \distOfMean_{i'}(\posteriorMean) > 0\}| \le 2$ for every posterior mean
$\posteriorMean\in\supp(\distOfMean)$.
This structural requirement says that
under the advertising strategy $\distOfMean$,
the realized signal either fully reveals the product quality
or leaves the customer uncertain between at most two product qualities.
Indeed, there exists an optimal advertising strategy for
program \ref{program: benchmark} that satisfies
this structural requirement:
\begin{lemma}[\citealp{FTX-22}]
\label{lem:binary support}
There exists an optimal advertising strategy $\optAdver$ such that
$|\{i\in[\stateNum]: \optAdver_i(\posteriorMean) > 0\}| \le 2$
for every $\posteriorMean \in \supp(\optAdver)$.
\end{lemma}
Intuitively, this result follows from the fact that
the extreme points of the set of distributions with a fixed expectation are binary-supported.
We can also deduce the following property of
the optimal price $\optPrice$ and the optimal advertising strategy $\optAdver$:
\begin{restatable}{lemma}{nobadposterior}
\label{lem:no bad posterior}
There exist an optimal price $\optPrice$ and
an optimal advertising strategy $\optAdver$
such that for every posterior mean
$\posteriorMean \in\supp(\optAdver)$ with
$\posteriorMean \notin \qualitySpace$, we have
$\optPrice \le \max_{\type\in\typeSpace}\buyerUtility(\type, \posteriorMean)$.
\end{restatable}
This property
follows from the observation that
if there exists a posterior mean $\posteriorMean\in\supp(\optAdver)$
with $\posteriorMean\notin \qualitySpace$, then, by \Cref{lem:binary support}, it must be that $\{i'\in[\stateNum]: \optAdver_{i'}(\posteriorMean) > 0\} = \{i, j\}$ for some $i < j$
with $\qualityVal_i < \posteriorMean < \qualityVal_j$.
Now, if $\optPrice > \max_{\type\in\typeSpace}\buyerUtility(\type, \posteriorMean)$,
then every type's valuation at this posterior mean is below the price $\optPrice$, so this posterior mean contributes nothing to the revenue. One can therefore redistribute the probability mass
$\optAdver(\posteriorMean)$ on this posterior mean to $\qualityVal_i$ and $\qualityVal_j$ without losing any revenue, which yields a pair $(\optPrice, \optAdver)$ with the desired property.

With \Cref{lem:binary support} and \Cref{lem:no bad posterior} in place,
we now formally state two guarantees on the price and advertising strategy
produced by Procedure \ref{algo_construction_adver_general}.
\begin{restatable}[Feasibility guarantee]{lemma}{feasibilityadver}
\label{lem:feasibility adver}
Given an input price and advertising strategy $(p, \distOfMean)$ satisfying the properties stated in \Cref{lem:binary support} and \Cref{lem:no bad posterior}, the output price $\newPrice$ and
advertising strategy $\newAdver$ of Procedure \ref{algo_construction_adver_general} satisfy:
$\newPrice\in\discPrice$,
$\newAdver$ is feasible, and
$\inverseVal(\newPrice, \posteriorMean)
\in \discTyepSpace$ for all $\posteriorMean\in\supp(\newAdver)$.
\end{restatable}
\begin{restatable}[Revenue guarantee]{lemma}{revenueguarantee}
\label{lem:revenue guarantee}
Fix a price $p \ge 2\varepsilon$ and
a feasible advertising strategy $\distOfMean$, and
let $\newPrice, \newAdver = \Round(p, \distOfMean)$ be the output
of Procedure~\ref{algo_construction_adver_general}. Then, $\Rev(p, \distOfMean) - \Rev(\newPrice, \newAdver) \le 2\varepsilon$.
\end{restatable}

At a high level, the argument is as follows. Given an input price $\price$,
Procedure~\ref{algo_construction_adver_general} outputs the closest price $\newPrice\in\discPrice$
satisfying $\price - 2\varepsilon \le \newPrice \le \price -\varepsilon$.
Given an input advertising strategy $\distOfMean$,
for every posterior mean $\posteriorMean \in \supp(\distOfMean)$ with $\posteriorMean\notin \discPostMeanSpace_{\newPrice}$,
by \Cref{lem:binary support},
there must exist two qualities $\qualityVal_i, \qualityVal_j$ with $i< j$ such that
$\{i'\in[\stateNum]:\distOfMean_{i'}(\posteriorMean) > 0\} = \{i, j\}$.
For such a posterior mean $\posteriorMean$,
Procedure~\ref{algo_construction_adver_general} first identifies the
critical types $x = \inverseVal(\price, \posteriorMean)$ and
$x^\dagger = \inverseVal(\newPrice, \posteriorMean)$, where
$x^\dagger$ lies in a grid interval $((z-1)\varepsilon, z\varepsilon)$ for some $z\in\N^+$.
Then, Procedure~\ref{algo_construction_adver_general} uses
the critical-type function $\inverseVal(\newPrice, \cdot)$ at the constructed price $\newPrice$
to find two posterior means
$\leftPosteriorMean$ and $\rightPosteriorMean$
satisfying
$\inverseVal(\newPrice, \leftPosteriorMean) = z\varepsilon$ and
$\inverseVal(\newPrice, \rightPosteriorMean) = (z-1)\varepsilon$.\footnote{To see that such $\leftPosteriorMean$ and $\rightPosteriorMean$ always exist, note that because the valuation function
is assumed to be increasing in the type $\theta$ (see Assumption \ref{assump:1a}), given any $\theta,
p,q$,
if $v(\theta,q)=p$, then $\kappa(\price,q)=\theta$. Therefore, $\leftPosteriorMean$ and $\rightPosteriorMean$ are the values of $q$ satisfying $v(z\varepsilon, q) =\newPrice$ and $v((z-1)\varepsilon, q) =\newPrice$, respectively. Moreover,
since $v(\theta, q)$ is linear, and hence continuous, in $q$ for any given $\theta$,
such solutions $\leftPosteriorMean$ and $\rightPosteriorMean$ always exist.}
To construct a feasible
advertising strategy, we then round $\leftPosteriorMean$ up to $\qualityVal_i$
whenever $\leftPosteriorMean < \qualityVal_i$,
and round $\rightPosteriorMean$ down to $\qualityVal_j$ whenever
$\rightPosteriorMean > \qualityVal_j$.
By \Cref{assump:valuation}, together with \Cref{lem:no bad posterior},
we can also show that the two constructed
posterior means $\leftPosteriorMean$ and $\rightPosteriorMean$
satisfy
(a) $\leftPosteriorMean < \posteriorMean < \rightPosteriorMean$;
and, moreover,
(b) $\inverseVal(\newPrice,\leftPosteriorMean) \leq
\inverseVal(\price, \posteriorMean)$ and
$\kappa(\newPrice,\rightPosteriorMean) \leq
\kappa(\price, \posteriorMean)$.
Relation (a) enables us to decompose the probability mass $\distOfMean(\posteriorMean)$ on this posterior mean
into probabilities on the two posterior means $\leftPosteriorMean$ and $\rightPosteriorMean$ while still
satisfying the \eqref{bayes consistency} condition.
Together with the monotonicity of the demand function $\demand$,
relation (b) guarantees that the revenue of the output of Procedure \ref{algo_construction_adver_general} is within $2\varepsilon$ of the revenue of the input.

The proof of \Cref{prop:discretization error} then follows immediately from \Cref{lem:feasibility adver} and \Cref{lem:revenue guarantee}.

%% file: Paper/plots/rounding.tex
\begin{tikzpicture}[scale=0.9, transform shape]
\begin{axis}[
axis lines = middle,
hide axis,
xtick style={draw=none},
ytick style={draw=none},
xticklabels=\empty,
yticklabels=\empty,
xmin=-12,xmax=12,ymin=0.32,ymax=1.1,
width=0.9\textwidth,
height=0.5\textwidth,
samples=50]

\addplot[mark=none, black, samples=2, line width=0.4mm] coordinates {(-12, 1) (12, 1)};
\addplot[mark=*,only marks, fill=white] coordinates {(-4, 1)} node[above, pos=1]{};
\addplot[mark=*,only marks, fill=white] coordinates {(4, 1)} node[above, pos=1]{};
\addplot[] coordinates {(-4, 1)} node[above, pos=1]{\Large$\qualityVal_i$};
\addplot[] coordinates {(4, 1)} node[above, pos=1]{\Large$\qualityVal_j$};

\addplot[mark=none, black, samples=2, line width=0.4mm] coordinates {(-12, 0.7) (12, 0.7)};
\addplot[mark=*,only marks, fill=white] coordinates {(0, 0.7)} node[above, pos=1]{};

\addplot[] coordinates {(0, 0.7)} node[above, pos=1, yshift=0.1cm]{\Large$\posteriorMean$};
\addplot[] coordinates {(-2, 0.7)} node[above, pos=1, yshift=0.1cm]{\Large$\leftPosteriorMean$};
\addplot[] coordinates {(2, 0.7)} node[above, pos=1, yshift=0.1cm]{\Large$\rightPosteriorMean$};
\addplot[mark=*,only marks, fill=white] coordinates {(-2, 0.7)} node[above, pos=1]{};
\addplot[mark=*,only marks, fill=white] coordinates {(2, 0.7)} node[above, pos=1]{};

\addplot[mark=none, black, samples=2, line width=0.4mm] coordinates {(-12, 0.4) (12, 0.4)};

\addplot[gray, thick] coordinates {(-11,0.4) (-11,0.415)};
\addplot[gray, thick] coordinates {(-8,0.4) (-8,0.415)};
\addplot[gray, thick] coordinates {(-5,0.4) (-5,0.415)};
\addplot[gray, thick] coordinates {(-2,0.4) (-2,0.415)};
\addplot[gray, thick] coordinates {(1,0.4) (1,0.415)};
\addplot[gray, thick] coordinates {(4,0.4) (4,0.415)};
\addplot[gray, thick] coordinates {(7,0.4) (7,0.415)};
\addplot[gray, thick] coordinates {(10,0.4) (10,0.415)};

\addplot[mark=star,only marks, fill=white, mark size = 3pt, thick] coordinates {(2.5, 0.4)} node[above, pos=1]{};
\addplot[] coordinates {(2.5, 0.4)} node[above, pos=1, yshift=0.1cm]{\Large$x$};

\addplot[gray, thick] coordinates {(-3.5,0.4) (-3.5, 0.415)};
\addplot[] coordinates {(-3.5, 0.4)} node[above, pos=1, yshift=0.1cm]{\Large$x^\dagger$};

\addplot[] coordinates {(-2, 0.4)} node[below, pos=1, yshift = -0.14cm]{$z\varepsilon$};
\addplot[] coordinates {(-5, 0.4)} node[below, pos=1, ]{$(z-1)\varepsilon$};
\addplot[mark=star,only marks, fill=white, mark size = 3pt, thick, color = brown] coordinates {(-2, 0.4)} node[above, pos=1]{};
\addplot[mark=star, only marks, fill=white, mark size = 3pt, thick, color = brown] coordinates {(-5, 0.4)} node[above, pos=1]{};

\draw [-stealth,very thick, dashed, color = brown, line width=0.2mm] (axis cs: -2, 0.68) -- (axis cs:-2, 0.43);
\draw [-stealth,very thick, dashed, color = brown, line width=0.2mm] (axis cs: 2, 0.68) -- (axis cs:-5, 0.43);

\draw [-stealth,very thick, dashed, color = black, line width=0.2mm] (axis cs: 0, 0.68) -- (axis cs:2.49, 0.47);

\draw [-stealth,very thick, dashed, color = black, line width=0.2mm] (axis cs: -4, 0.98) -- (axis cs:0, 0.78);
\draw [-stealth,very thick, dashed, color = black, line width=0.2mm] (axis cs: 4, 0.98) -- (axis cs:0, 0.78);

\draw [-stealth,very thick, dashed, color = brown, line width=0.2mm] (axis cs: -4, 0.98) -- (axis cs:-2, 0.78);
\draw [-stealth,very thick, dashed, color = brown, line width=0.2mm] (axis cs: 4, 0.98) -- (axis cs:-2, 0.78);
\draw [-stealth,very thick, dashed, color = brown, line width=0.2mm] (axis cs: -4, 0.98) -- (axis cs:2, 0.78);
\draw [-stealth,very thick, dashed, color = brown, line width=0.2mm] (axis cs: 4, 0.98) -- (axis cs:2, 0.78);

\draw [dashed, color = gray, line width=0.1mm] (axis cs: -4, 0.98) -- (axis cs:-4, 0.7);
\draw [dashed, color = gray, line width=0.1mm] (axis cs: 4, 0.98) -- (axis cs:4, 0.7);

\addplot[] coordinates {(-9.5, 0.98)} node[below, pos=1, ]{quality space $\qualitySpace$};
\addplot[] coordinates {(-7.95, 0.68)} node[below, pos=1, ]{posterior mean space $\discPostMeanSpace_p$};
\addplot[] coordinates {(-11, 0.38)} node[below, pos=1, ]{set $\discTyepSpace$};



\end{axis}

\end{tikzpicture}

%% file: Paper/4-2-regret-analysis.tex
In this section, we present our estimation error analysis. We begin by showing that $\UCBDemand_t(x)$ provides an upper confidence bound on the true demand function $\demand(x)$ for all $x\in \discTyepSpace$, and we derive a bound on how much larger it can be than $\demand(x)$.
All missing proofs in this subsection can be found in \Cref{apx:proofs of regret analysis}.
\begin{restatable}{lemma}{highprobestimationerror}
\label{lem:high-prob estimation error}
For every $t \ge \numDiscType +1$,
the following holds with probability at least $1 - \sfrac{1}{T^2}$:
\begin{alignat}{2}
    \UCBDemand_t(x) &\ge \demand(x),
    \quad && \forall x\in \discTyepSpace;
    \label{high-prob:UCB lower bound}\\
    \UCBDemand_t(x) - \demand(x)
    & \le  2 \sqrt{\frac{16\log T}{\chosenCounter_t(x)}} + \frac{2\sqrt{(1+\chosenCounter_t(x)) \ln (1+\chosenCounter_t(x))}}{\chosenCounter_t(x)},
    \quad && \forall x \in \discTyepSpace.
    \label{high-prob:UCB upper bound}
\end{alignat}
\end{restatable}
To prove these inequalities for all points $x\in \discTyepSpace$,
we first show that the empirical estimates
$\averageDemand_t(x)$, $x\in\discTyepSpace$,
concentrate around the true demand values $\demand(x)$ as $\chosenCounter_t(x)$ increases.
We establish this concentration bound using a uniform bound obtained from the scalar-valued version of the
self-normalized martingale tail inequality \citep{APS-12}.

We next analyze how close the seller's optimistic
revenue estimate, computed using the upper confidence bound $\UCBDemand_t$,
is to the true revenue.
In particular, we have the following result.
\begin{restatable}{lemma}{optimisticestimates}
\label{lem:optimistic estimates}
For every time
$t\ge \numDiscType+1$, with probability at least $1 - \sfrac{2}{T^2}$,
we have
$\Rev(\discretizedOptPrice, \discretizedOptAdver)
\le \UCBRev_t(\discretizedOptPrice, \discretizedOptAdver)
\le \UCBRev_t(\offeredPrice_t, \offeredAdver_t)$.
\end{restatable}
This result follows from the bounds in
\Cref{lem:high-prob estimation error},
which establish that $\UCBDemand_t(x) \ge \demand(x)$
with high probability.
Finally, we again apply \Cref{lem:high-prob estimation error} to upper-bound the gap $\UCBRev_t(\offeredPrice_t, \offeredAdver_t) - \Rev(\offeredPrice_t, \offeredAdver_t)$.
\begin{restatable}{lemma}{singleroundregret}
\label{lem:single-round regret}
For every time
$t\ge \numDiscType+1$, with probability at least $1 - \sfrac{2}{T^2}$,
we have $\UCBRev_t(\offeredPrice_t, \offeredAdver_t) -
\Rev(\offeredPrice_t, \offeredAdver_t)
\le
5\offeredPrice_t
\sum_{\posteriorMean\in\supp(\offeredAdver_t)}
\offeredAdver_t(\posteriorMean)
\sqrt{\frac{\log T}{\chosenCounter_t(\inverseVal(\offeredPrice_t, \posteriorMean))}}$.
\end{restatable}
Intuitively, the gap between the seller's estimated revenue $\UCBRev_t(\offeredPrice_t, \offeredAdver_t)$
and the true expected revenue $\Rev(\offeredPrice_t, \offeredAdver_t)$
is bounded by a weighted sum (i.e., with weights $\{\offeredAdver_t(\posteriorMean)\}_{\posteriorMean\in\supp(\offeredAdver_t)}$) of
the demand estimation errors at the points $\inverseVal(\offeredPrice_t, \posteriorMean)$, i.e.,
$|\UCBDemand_t(\inverseVal(\offeredPrice_t, \posteriorMean)) -\demand(\inverseVal(\offeredPrice_t, \posteriorMean))|$.

\newcommand{\remove}[1]{}
\remove{
\begin{align*}
    \eqref{eq:regret upper bound helper multiple states}
    & \le \numDiscType + 2\varepsilon T +
    5\sqrt{2}\pupper\expect{\sum_{x\in\discTyepSpace}
    \sum_{t=\numDiscType + 1}^T 
    \distOfThreshold_t(x)
    \sqrt{\frac{\log T}{\chosenCounter_{t+1}(x)}}} \\
    & \le 
    \numDiscType + 2\varepsilon T + 
    10\sqrt{2}\pupper\sum_{x\in\discTyepSpace}\expect{
    \sqrt{\sum_{t=\numDiscType + 1}^T \distOfThreshold_t(x) \log T}}\\
    & \overset{(a)}{\le}
    \numDiscType + 2\varepsilon T 
    + 10\sqrt{2}\pupper  {\sum_{x\in\discTyepSpace} \sqrt{\expect{\sum_{t=\numDiscType + 1}^T \distOfThreshold_t(x)}\log T}}
    \overset{(b)}{\le}
    \numDiscType  +2\varepsilon T 
    + 10\sqrt{2}\pupper  \sqrt{\numDiscType T\log T}
\end{align*}
where inequality (a) follows from Jensen's inequality,
and inequality (b) follows from the 
observation that $\sum_{x\in\discTyepSpace}\expect{\sum_{t=\numDiscType + 1}^T \distOfThreshold_t(x)} \le T$
and we maximize the term $\sum_{x\in\discTyepSpace}
\sqrt{\expect{\sum_{t=\numDiscType + 1}^T \distOfThreshold_t(x)}}$
subject to the constraint $\sum_{x\in\discTyepSpace}\expect{\sum_{t=\numDiscType + 1}^T \distOfThreshold_t(x)} \le T$
to obtain the worst case regret bound.
}


%% file: Paper/conclusions.tex
In this work, we use a foundational information design framework, Bayesian persuasion, to model the effect of an advertising strategy on customers' beliefs about product quality and consequently their purchase decisions. 
This formulation allows us to quantify the tradeoffs between the design of the pricing and advertising strategies and their combined impact on the revenue outcomes. 
We first characterize the value of advertising and how it can help improve a seller's revenue. 
We show that for a broad class of customer valuation
functions, advertising can at most double the seller’s expected revenue, and this upper bound is tight.
We then study the seller's dynamic pricing and advertising with demand learning problem, where the demand function is apriori unknown.
We provide an efficient learning algorithm that has regret $O(T^{2/3}(m \log T)^{1/3})$ for a finite quality space with cardinality $m$.  
This result implies that when the number of qualities $m$ is a constant, there is almost no additional learning cost for the seller to additionally learn the optimal advertising compared to the standard dynamic pricing (without advertising) with non-parametric demand learning problem.

There are interesting future directions from this work. 
\revcolor{First, although our $2$-factor revenue comparison extends beyond valuations that are linear in product quality, 
in particular, its upper bound continues to hold for any valuation of the form $\buyerUtility(\type,\quality)=\type g_1(\quality)+g_2(\quality)$, where $g_1$ and $g_2$ 
may depend nonlinearly on $\quality$, 
our demand-learning results rely critically on linearity in quality. 
Under this assumption, a customer's posterior expected valuation depends on the posterior belief only through its mean quality, 
allowing us to represent an advertising policy as a distribution over one-dimensional posterior means and to associate each price-posterior pair with a scalar critical type. 
Without linearity in quality, the posterior mean is generally no longer a sufficient statistic: two posterior beliefs with the same mean may induce different expected valuations 
and purchase decisions. 
The seller would then need to optimize and learn over full posterior beliefs, or a higher-dimensional collection of posterior statistics, 
while simultaneously preserving Bayes consistency. 
Moreover, our quality-and-price-dependent discretization and rounding argument, which controls the approximation error by moving posterior means while preserving the induced critical types, does not directly extend to this higher-dimensional setting. 
It would therefore be interesting to develop new discretization and learning techniques for nonlinear-in-quality valuations and determine whether an efficient algorithm can still achieve the optimal $\widetilde O(\timeHorizon^{\sfrac{2}{3}})$ regret rate.}
Second, in our setting, the customers with i.i.d.\ private types arrive at each time round, and leave the market no matter whether they purchase the product or not. 
However, in practice, customers may strategize their purchase time for a more favorable price \citep{CF-18,CS-19}. 
How to model the advertising effect with strategic customers and achieve efficient demand learning is also another interesting future direction.

%% file: Paper/apx-example.tex
In this section, we briefly illustrate how the real-world applications can be captured by our problem.

\xhdr{Opaque product}
Consider a recurring rescued-produce box service, such as an Oddbox-like fruit-and-vegetable box service.\footnote{\revcolor{
Examples of recurring fruit-and-vegetable box services include 
\href{https://www.oddbox.co.uk/home-box}{Oddbox}, 
\href{https://hungryharvest.net/harvest-options-2}{Hungry Harvest}, 
\href{https://www.riverford.co.uk/shop/categories/seasonal-organic-boxes/veg-boxes}{Riverford Organic Farmers}, 
and \href{https://www.abelandcole.co.uk/small-mixed-fruit-veg-box}{Abel \& Cole}.}} 
The product in each purchase period is the upcoming produce box, rather than the long-term subscription contract or any individual item contained in the box. 
Customers purchase a box of a given size or category, and the posted price is attached to that box type or subscription tier, while the precise products in that box may vary across delivery periods.

The product-quality state represents value-relevant characteristics of the upcoming box, such as its composition, variety, freshness, scarcity, or aggregate retail value. 
The quality varies across periods because the produce available for rescue depends on the realized supply from growers and distributors.
Once the produce for an upcoming box has been sourced or assembled, the service (which is the seller) 
observes its composition and other value-relevant attributes, while prospective customers may not yet fully observe them.
The advertising signal represents the information that the service discloses about the box. 
This information may include a rescue list, selected featured items, broad product categories, an expected-value range, or an aggregate quality score. 
The customer uses this disclosed information to assess the box at the posted or subscription-tier price.
This application captures the central information asymmetry in our model: the seller observes value-relevant information about the product offered in the current purchase period and decides how much of that information to communicate to the customer. 

\xhdr{On-demand service}
Consider an on-demand service platform (e.g., Uber) that intermediates transactions between customers and service providers. The product in each purchase period is a specific service transaction, such as a ride, delivery, or home-service appointment. 
The platform posts an upfront price for the requested service before the precise provider or service configuration is assigned, and this realized configuration may vary across transactions.

The product-quality state captures value-relevant characteristics of the realized service match, such as the provider's reliability, rating, expected waiting time, vehicle or equipment characteristics, or overall service quality. 
This quality varies across transactions because different providers and service configurations may be assigned to different requests. 
Once a match has been formed, the platform observes information about these attributes that the customer may not yet fully observe.

The advertising signal captures the information that the platform discloses about the realized match before the customer decides whether to proceed with the transaction. 
This information may include selected provider attributes, a rating category, an estimated waiting time, vehicle or equipment information, or an aggregate quality score. 
The customer uses the disclosed information to assess the service at the previously quoted price.
Accordingly, this application captures the same central information asymmetry: the platform observes value-relevant information about the service offered in the current transaction and chooses how much of this information to communicate before the customer's final decision. 

%% file: Paper/7-numerical.tex
\newcommand{\normal}{\cc{Normal}}
\newcommand{\ETC}{\cc{ETC}}
In this section, we conduct numerical experiments to compare the empirical performance of our proposed algorithm with some baseline algorithms on synthetically generated data.
\revcolor{For each experimental setting and each algorithm, we simulate a selling horizon of $\timeHorizon = 8000$ rounds and conduct $10$ independent simulation runs. 
At every round $t$, the solid curve reports the sample mean of the empirical cumulative regret across these ten runs, while the shaded region reports the corresponding mean $\pm 0.25$ sample standard deviation.}

\begin{figure}[H]
    \centering
    \begin{minipage}{\textwidth}
        \centering
        \includegraphics[width=\linewidth]{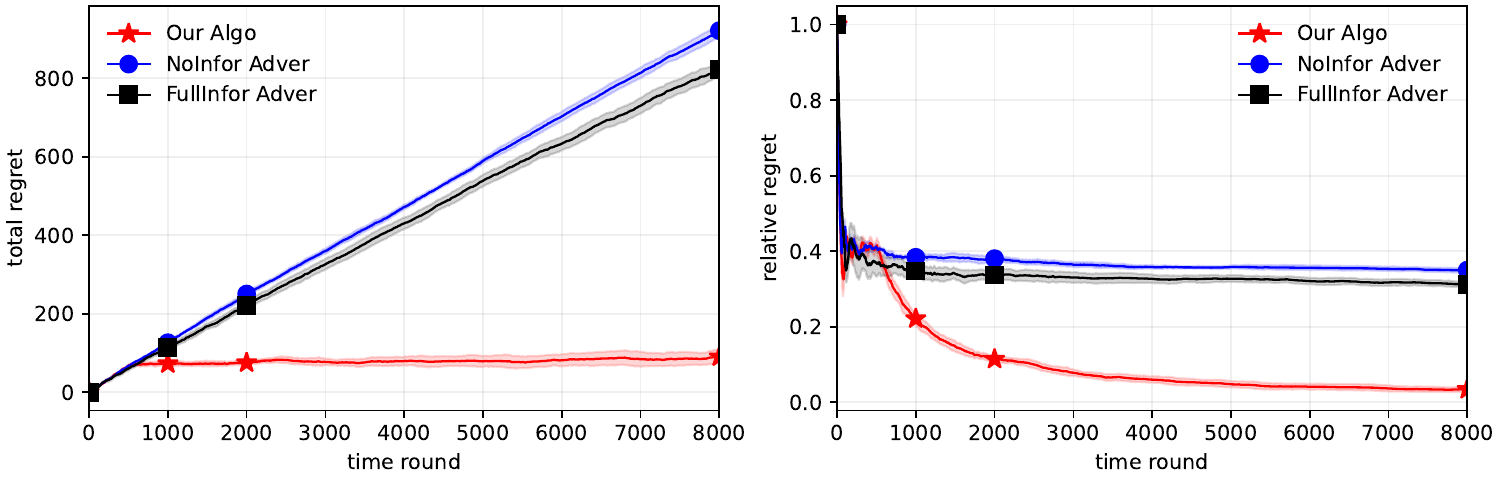}
    \end{minipage}%
    \caption{Setting 1: compared to no-information advertising strategy and full-information advertising strategy}
    \label{fig:exp diff}
\end{figure}
\begin{figure}[H]
    \centering
    \begin{minipage}{\textwidth}
        \centering
        \includegraphics[width=\linewidth]{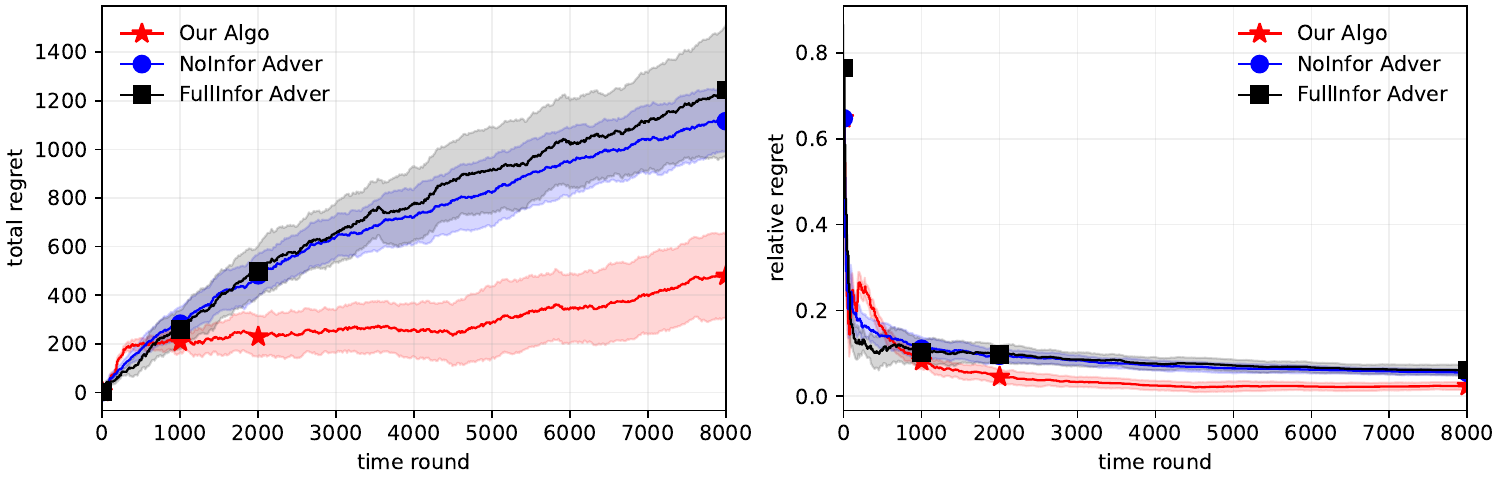}
    \end{minipage}%
    \caption{Setting 2: compared to no-information advertising strategy and full-information advertising strategy}
    \label{fig:normal diff}
\end{figure}

We consider two experiment settings with an additive valuation function i.e., $\buyerUtility(\type, \quality) = \type + \quality$: 
(1) The first setting is the binary-quality, continuous type distribution introduced in \Cref{subsec:example} (\Cref{fig:example-binary-sqrt-total}).
In this setting, the optimal price is $0.600$ and optimal advertising generates two possible posterior means $\{0, 0.600\}$, leading to an optimal revenue $0.329$.
(2) The second setting has a type $\type\sim \normal(1, 4)$ where $\normal(1, 4)$ is a normal distribution with mean value $1$ and variance $4$; the quality space is $\{1, 10\}$ with the mean value $4$.
In this setting, the optimal price is $5.666$ and optimal advertising has two possible posterior means $\{1, 6.5\}$, leading to a revenue $2.554$.

We first compare our algorithm with two baseline algorithms that always use a simple advertising strategy: the one that ignores the advertising (i.e., it simply keeps using the no-information advertising strategy), and the one that keeps using the full-information advertising strategy. 
Notice that when the advertising strategy is fixed, the seller's problem becomes a standard dynamic pricing problem.
Thus, in these two baselines, we optimize and learn the optimal price using the standard dynamic pricing algorithm (i.e., the UCB algorithm) developed in  \cite{KL-03} for the stochastic unknown demand.

In the first setting, the optimal price and optimal revenue for keeping using the no-information strategy are $0.250, 0.250$, respectively; and the optimal price and optimal revenue for keeping using the full-information strategy are $0.790, 0.263$, respectively. 
In the second setting, the optimal price and optimal revenue for keeping using the no-information strategy are $5.001, 2.500$, respectively; and the optimal price and optimal revenue for keeping using the full-information strategy are $6.952, 2.456$, respectively. 
The results of these two settings are reported in \Cref{fig:exp diff} and \Cref{fig:normal diff}, respectively. 
In both figures, we report the cumulative regret (defined as in Eqn.~\eqref{eq:vanilla regret}) and the relative regret (defined as $\frac{\Reg{T}}{T\cdot \sup_{p,  \phi} \Rev(p,\phi)}$). 
As we can see from these figures, in both settings, our proposed algorithm significantly outperforms these baselines that do not adaptively optimize the advertising strategies. 
When the time horizon $T$ is small, the revenue gap between our algorithm and the baseline algorithms is small. 
This is expected as our algorithm uses the initial rounds to maintain an estimated type CDF, and then optimizes an advertising strategy based on this estimated type CDF, thus leading to almost linear regret in initial rounds. 
When the time horizon $T$ becomes large, our algorithm then outperforms the baselines as our algorithm can use the (more accurate) estimated type CDF to optimize the advertising strategy, which is closer to the optimal advertising strategy.

We then compare our algorithm with two other commonly-studied  explore-then-commit (denoted by ETC) baseline algorithms: the first one uses $T_0 = T^{\sfrac{2}{3}}$ pure exploration rounds (henceforth denoted by ETC-$T^{\sfrac{2}{3}}$) and the second one uses $T_0 = T^{\sfrac{1}{2}}$ pure exploration rounds (henceforth denoted by ETC-$T^{\sfrac{1}{2}}$).
We implement the ETC algorithm as follows:
We use the same discretized type space $\discTypeSpace$ for both the ETC algorithm and our proposed algorithm. 
In the first $T_0$ rounds, we keep using the no-information strategy.
We construct the discretized price space $\discPrice$ for the ETC algorithm such that whenever a price $p\in\discPrice$ is chosen from this discretized price space, the induced critical type $\inverseVal(p, \expect[\quality\sim\prior]{\quality})\in \discTypeSpace$ must be in the discretized type space $\discTypeSpace$. Thus, for each discretized type $\type\in \discTypeSpace$, we can choose a certain price for $T_0/|\discTypeSpace|$ rounds to explore the value of the type CDF $\typeCDF(\type)$ at this discretized type $\type$.
After $T_0$ rounds, we can maintain an estimated type CDF $\typeCDF^{\ETC}$; we then compute the (empirical) optimal price and optimal advertising based on this estimated $\typeCDF^{\ETC}$ and implement this price and advertising for the remaining rounds.
\begin{figure}[H]
    \centering
    \begin{minipage}{\textwidth}
        \centering
        \includegraphics[width=\linewidth]{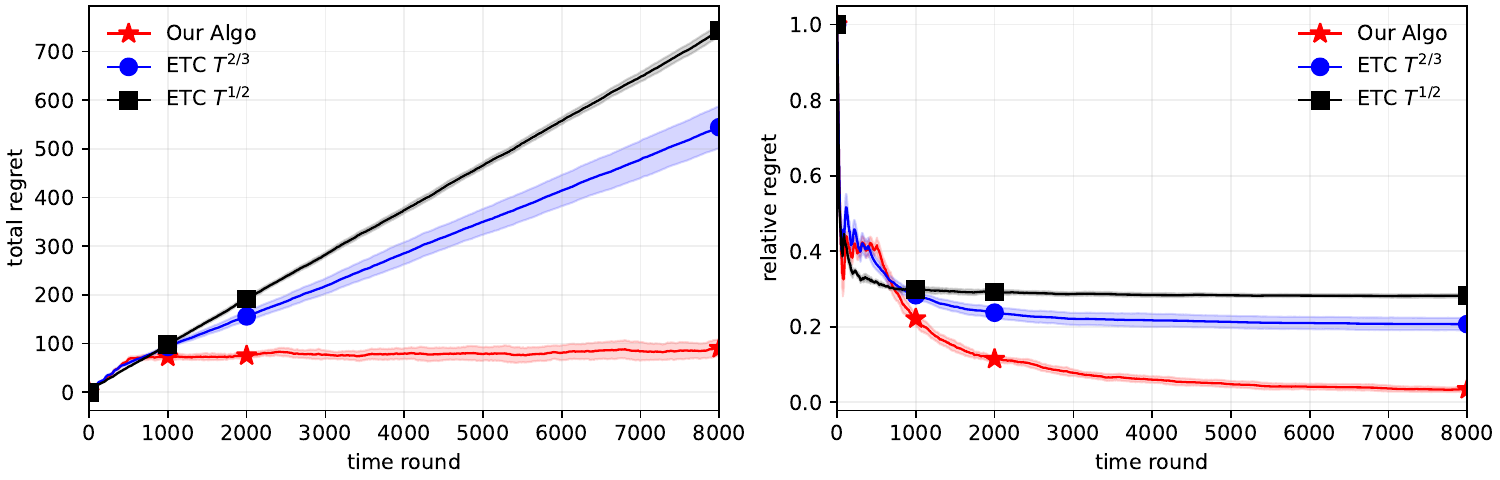}
    \end{minipage}%
    \caption{Setting 1: compared to ETC algorithm with different pure exploration rounds}
    \label{fig:exp ETC}
\end{figure}
\begin{figure}[H]
    \centering
    \begin{minipage}{\textwidth}
        \centering
        \includegraphics[width=\linewidth]{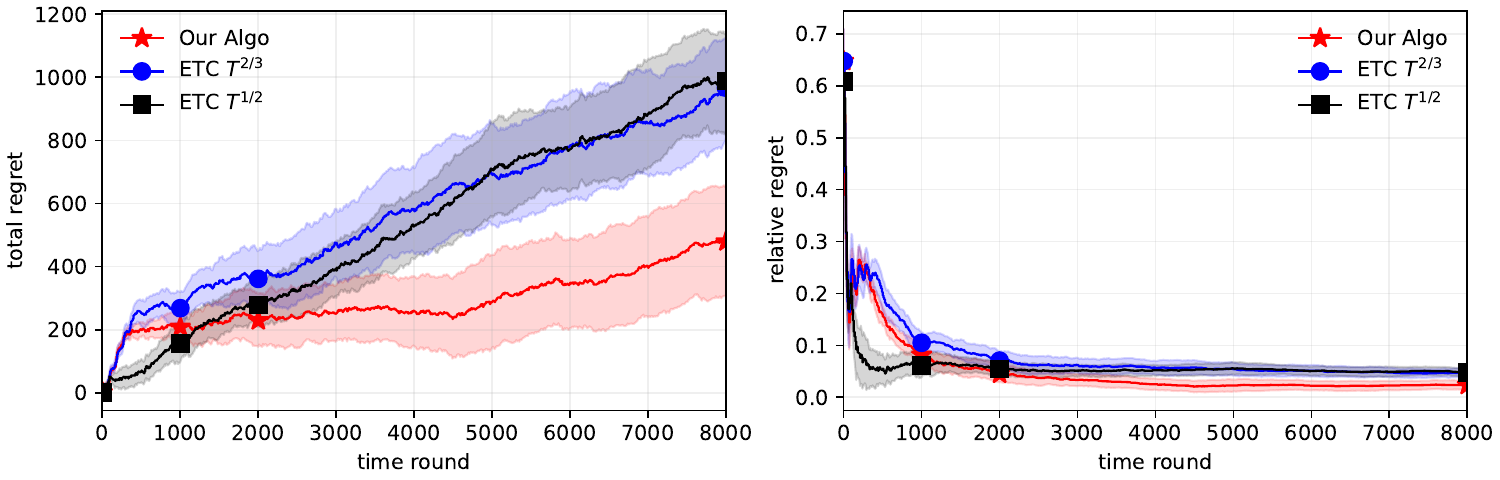}
        \label{fig:normal ETC adver cumu regret}
    \end{minipage}%
    \caption{Setting 2: compared to ETC algorithm with different pure exploration rounds}
    \label{fig:normal ETC}
\end{figure}
The results of two settings are reported in \Cref{fig:exp ETC} and \Cref{fig:normal ETC}, respectively. 
As we can see from these figures, in both settings, our proposed algorithm still significantly outperforms these baseline algorithms.  
We can also observe that even though the ETC-$T^{\sfrac{1}{2}}$ algorithm initially outperforms the ETC-$T^{\sfrac{2}{3}}$ algorithm, the ETC-$T^{\sfrac{2}{3}}$ algorithm then gradually outperforms the ETC-$T^{\sfrac{1}{2}}$ algorithm.
This is because the ETC-$T^{\sfrac{2}{3}}$  algorithm has a longer pure exploration length than that of ETC-$T^{\sfrac{1}{2}}$, thus has larger regret. But after the pure exploration phase, the ETC-$T^{\sfrac{2}{3}}$  algorithm has a better estimate for the unknown type CDF $\typeCDF$, and thus is able to compute a price and an advertising strategy that has revenue closer to the optimal one.
Consequently, the incurred per-round  regret of ETC-$T^{\sfrac{2}{3}}$ is smaller than that of ETC-$T^{\sfrac{1}{2}}$.

%% file: Paper/apx-rev-gap-miss.tex

In this section, we further illustrate the revenue loss that arises when the underlying true type distribution is exponential, but the seller misspecifies it as a Gaussian distribution with the same mean.

We consider the same binary-quality and additive-valuation environment as in \Cref{subsec:example}.
The product quality is $\quality\in\{0,1\}$ with $\prior(1)=\highQualityProb=\frac{1}{2}$ and $\prior(0)=1-\highQualityProb=\frac{1}{2}$.
The customers have an additive valuation function $\buyerUtility(\type,\quality)=\type+\quality$.
An advertising strategy induces a posterior mean $\posteriorMean=\expect{\quality\mid \signal}\in[0,1]$.
By Bayes consistency, in this binary-quality example, the seller's advertising strategy is equivalently described by a random posterior mean $\posteriorMean\in[0,1]$ satisfying $\expect{\posteriorMean}=\frac{1}{2}$.
Given a price $\price$ and posterior mean $\posteriorMean$, a type-$\type$ customer purchases if and only if
$\type+\posteriorMean\geq \price$, or equivalently $\type\geq \price-\posteriorMean$.
Therefore, for any customer type distribution with demand function $\demand(x)=1-\typeCDF(x)$, the seller's expected revenue can be simplified as follows
\begin{align*}
    \price \cdot \expect{\demand(\price-\posteriorMean)}
    \quad
    \text{s.t. }
    \posteriorMean\in[0,1]\text{ and }\expect{\posteriorMean}=\frac{1}{2}.
\end{align*}
In this example, the true type distribution is exponential with rate $\expRate=1$, denoted by $\trueTypeCDF(\type)=1-\exp(-\type)$.
Thus, the true demand function is 
$\trueDemand(x) = 1$ for $x \le 0$ and 
$\trueDemand(x) = \exp(-x)$ for $x > 0$.
However, suppose the seller incorrectly believes that the type distribution is Gaussian with the same mean $1$, but with variance $\gaussVar$ where $\gaussStd$ denotes the standard deviation of the misspecified Gaussian distribution.
The seller's misspecified type distribution is
$\gaussTypeCDF(\type)
=
\normalCDF\left(\frac{\type-1}{\gaussStd}\right)$
where $\normalCDF$ is the standard normal CDF.
The corresponding misspecified demand function is
$\gaussDemand(x)
=
1-\normalCDF\left(\frac{x-1}{\gaussStd}\right)$.
The seller first solves the static optimal pricing-and-advertising problem under the misspecified Gaussian distribution, and then implements the resulting price and advertising strategy in the true exponential environment.\footnote{This example uses an unbounded type space for transparency. If one wants to maintain a bounded type space, both the exponential and Gaussian distributions can be truncated on a sufficiently large interval and then normalized. The displayed revenue gap changes by an arbitrarily small amount when the truncation bound is sufficiently large.}

\xhdr{True optimal benchmark under the exponential distribution}
We first compute the optimal pricing-and-advertising policy under the true exponential distribution.
For any fixed price $\price$, define the optimal true revenue from price $\price$ as
$\truePriceRevenue(\price)
\triangleq
\sup\nolimits_{\posteriorMean:\, \expect{\posteriorMean}=1/2}
\price\cdot \expect{\trueDemand(\price-\posteriorMean)}$.
Equivalently, $\truePriceRevenue(\price)$ is obtained by evaluating the concave envelope of the function $\posteriorMean\mapsto \trueDemand(\price-\posteriorMean)$ at the prior mean $\frac{1}{2}$, and then multiplying by the price $\price$.
We now compute $\truePriceRevenue(\price)$ explicitly.
\begin{itemize}
    \item 
    First, when $0\leq \price\leq \frac{1}{2}$, the seller can choose the no-information advertising strategy that leads to $\posteriorMean\equiv \frac{1}{2}$.
    Then $\price-\posteriorMean\leq 0$, and thus all customers purchase.
    Since no policy can generate revenue larger than the posted price $\price$, we obtain $\truePriceRevenue(\price)=\price$ for $\price\in[0, \frac{1}{2}]$.
    \item 
    Second, when $\frac{1}{2}<\price\leq 1$, the function $\posteriorMean\mapsto \trueDemand(\price-\posteriorMean)$ is equal to $\exp(-(\price-\posteriorMean))$ on $[0,\price)$ and is equal to $1$ on $[\price,1]$.
    The concave envelope at the prior mean $\frac{1}{2}$ is given by the chord connecting the two points $(0,\exp(-\price))$ and $(\price,1)$.
    We thus have for $\frac{1}{2}<\price\leq 1$:
    \begin{align*}
        \truePriceRevenue(\price)
        =
        \price\left[
        \exp(-\price)
        +
        \frac{1}{2\price}\left(1-\exp(-\price)\right)
        \right]
        =
        \frac{1}{2}+\left(\price-\frac{1}{2}\right)\exp(-\price)~.
    \end{align*}
    \item 
    Third, when $\price>1$, the function $\posteriorMean\mapsto \trueDemand(\price-\posteriorMean)$ is equal to $\exp(-(\price-\posteriorMean))$ on the entire interval $[0,1]$.
    Since this function is convex in $\posteriorMean$, its concave envelope at $\frac{1}{2}$ is given by the chord connecting the two endpoints $\posteriorMean=0$ and $\posteriorMean=1$.
    We thus have for $ \price>1$:
    \begin{align*}
        \truePriceRevenue(\price)
        =
        \price\cdot \frac{1}{2}\left(\exp(-\price)+\exp(-(\price-1))\right)
        =
        \frac{1+\mathrm{e}}{2}\price\exp(-\price)~.
    \end{align*}
\end{itemize}
By calculation, we can deduce that the true optimal revenue is 
$\trueOptRev
=
\frac{1}{2}+\frac{1}{2\mathrm{e}}
\approx 0.684$.

\xhdr{Gaussian-optimized policy under the misspecified distribution}
We next describe how the misspecified seller computes her pricing-and-advertising policy.
Under the Gaussian belief with variance $\gaussVar$, for any fixed price $\price$, the seller solves
\begin{align*}
    \sup\nolimits_{\posteriorMean:\, \expect{\posteriorMean}=1/2}
    \expect{\gaussDemand(\price-\posteriorMean)}.
\end{align*}
Because this is a one-dimensional concavification problem with a single mean constraint, it is without loss of optimality to consider distributions supported on at most two posterior means.
Therefore, for each fixed price $\price$, we can optimize over two posterior means $0\leq \leftPosteriorMean\leq \frac{1}{2}\leq \rightPosteriorMean\leq 1$.
The case $\leftPosteriorMean=\rightPosteriorMean=\frac{1}{2}$ is interpreted as the degenerate no-information advertising strategy.
The Gaussian-optimized pricing-and-advertising policy is then characterized by the following program:
\begin{align*}
    \left(\gaussOptPrice,\gaussLeftPostMean,\gaussRightPostMean\right)
    \in
    \argmax_{\price\geq 0,\ 0\leq \leftPosteriorMean\leq \frac{1}{2}\leq \rightPosteriorMean\leq 1}
    \price
    \left[
    \left(1-\frac{\frac{1}{2}-\leftPosteriorMean}{\rightPosteriorMean-\leftPosteriorMean}\right)\gaussDemand(\price-\leftPosteriorMean)
    +
    \frac{\frac{1}{2}-\leftPosteriorMean}{\rightPosteriorMean-\leftPosteriorMean}\gaussDemand(\price-\rightPosteriorMean)
    \right],
\end{align*}
Let the probability assigned by this Gaussian-optimized advertising strategy to the high posterior mean $\gaussRightPostMean$ be
$\gaussPostWeight
\triangleq
\frac{\frac{1}{2}-\gaussLeftPostMean}{\gaussRightPostMean-\gaussLeftPostMean}$.
When the seller implements this Gaussian-optimized policy in the true exponential environment, the true revenue becomes
\begin{align*}
    \misspecifiedRev
    =
    \gaussOptPrice
    \left[
    (1-\gaussPostWeight)\trueDemand(\gaussOptPrice-\gaussLeftPostMean)
    +
    \gaussPostWeight\trueDemand(\gaussOptPrice-\gaussRightPostMean)
    \right].
\end{align*}
Therefore, the absolute revenue loss due to the misspecified type distribution is 
$\misspecifiedGap
\triangleq
\trueOptRev-\misspecifiedRev$
and the relative revenue loss is
$\relativeMisspecifiedGap
\triangleq
\frac{\misspecifiedGap}{\trueOptRev}$.

\begin{figure}[t]
    \centering
    \begin{tikzpicture}
    \begin{axis}[
        width=0.64\textwidth,
        height=0.40\textwidth,
        xlabel={Misspecified Gaussian variance $\gaussVar$},
        ylabel={Relative revenue loss (\%)},
        xmin=0.2, xmax=4.05,
        ymin=10, ymax=32,
        xtick={0.25,1,2,3,4},
        ytick={10,15,20,25,30},
        grid=major,
        legend style={at={(0.03,0.97)}, anchor=north west, draw=none, fill=none}
    ]
    \addplot+[mark=*, thick] coordinates {
        (0.25,12.433)
        (0.50,13.375)
        (0.75,14.749)
        (1.00,14.977)
        (1.25,14.693)
        (1.50,14.556)
        (1.75,14.504)
        (2.00,15.621)
        (2.25,17.530)
        (2.50,19.401)
        (2.75,21.230)
        (3.00,23.013)
        (3.25,24.750)
        (3.50,26.440)
        (3.75,28.082)
        (4.00,29.679)
    };
    \addlegendentry{$100\cdot \relativeMisspecifiedGap$}
    \addplot[dashed, thick] coordinates {(1,10) (1,32)};
    \end{axis}
    \end{tikzpicture}
    \caption{\revcolor{Revenue loss from misspecifying the customer type distribution in \Cref{apx:rev-gap}}}
    \label{fig:misspecified-type-distribution}
\end{figure}

\Cref{fig:misspecified-type-distribution} 
shows that matching the mean of the type distribution is far from sufficient.
Even when the Gaussian distribution also matches the variance of the true exponential distribution, namely when $\gaussVar=1$, the Gaussian-optimized policy still loses about $15\%$ of the true optimal revenue.
In this case, the Gaussian-optimized policy posts price $\gaussOptPrice\approx 1.397$ and uses a partial-pooling advertising strategy with posterior means $\gaussLeftPostMean=0$ and $\gaussRightPostMean\approx 0.594$.
The high posterior mean is realized with probability approximately $0.842$.
When this policy is implemented in the true exponential environment, it generates revenue of approximately $0.582$, compared to the true optimal revenue $\trueOptRev\approx 0.684$.

The intuition is as follows.
The type distribution affects not only the optimal price, but also how the seller should pool or separate product qualities through advertising.
Under the true exponential distribution, the optimal policy fully discloses quality and posts price $1$.
In contrast, under the misspecified Gaussian distribution, the seller may either pool qualities or choose a substantially higher price.
For example, when the misspecified variance is large, the Gaussian belief assigns relatively high probability to large willingness-to-pay realizations around the relevant price thresholds.
The seller then chooses a higher price that appears attractive under the Gaussian belief.
However, in the true exponential environment, this higher price substantially reduces demand and leads to a large revenue loss.

This example therefore highlights why demand learning is relevant even when customer types are i.i.d.
Although the type distribution $\typeCDF$ suffices to characterize the static optimal pricing-and-advertising policy, the seller still needs to learn the shape of $\typeCDF$, not merely its mean or variance, in order to compute the right joint pricing-and-advertising policy.

%% file: Paper/apx-comparisons.tex
Several closely related studies with known customer type or valuation distributions examine information disclosure jointly with monopoly pricing or richer mechanisms. In this section, we provide a detailed discussion of them.

\xhdr{Comparisons with \citet{RS-17}}
It is useful to compare our static model with the model of \citet{RS-17}. 
In their setting, a customer has an uncertain valuation for the selling item, the customer observes a signal about this valuation, and the seller then posts a take-it-or-leave-it price. 
Their main focus is customer-optimal information policy, but their analysis also identifies seller-optimal information structures. 
In particular, when the seller's cost is $\sellerCost$, a seller-optimal information structure pools all values above $\sellerCost$: the customer only learns whether his valuation is above $\sellerCost$, and conditional on this event, the seller posts the corresponding posterior mean valuation as the price.

Their framework shares several structural similarities with ours. Both models combine an information policy with a single quality-independent price, and consider use quasilinear payoffs with binary purchase decisions.
In addition, \citet{RS-17} also consider that a customer's belief matters only through its posterior expected valuation, which is the same as in our setting as the customer's valuation function linearly depending on the product quality, and thus the advertising affects customer demand through the posterior mean quality.
The zero-seller-cost case in \citet{RS-17} has a particularly simple analogue in our framework with additive valuation function. 
When the seller's cost is $\sellerCost=0$ and valuations are nonnegative, the seller-optimal information structure reduces to no information: the customer only learns the prior mean valuation, and the seller posts this mean as the price. 
The following lemma formalizes the corresponding result in our model.

\begin{lemma}[No-information optimality under degenerate customer types]
\label{lem:degenerate-type-no-info}
Suppose the valuation function is additive, i.e., $\buyerUtility(\type,\quality)=\type+\quality$, and the customer type distribution is degenerate at $\degType$, i.e., $\type=\degType$ with probability one. Then no-information advertising strategy $\noinfor$ is optimal. 
\end{lemma}

\Cref{lem:degenerate-type-no-info} shows that the zero-cost seller-optimal benchmark in \citet{RS-17} maps cleanly to our model when customer types are degenerate. In that case, all valuation uncertainty comes from product quality, and the seller can extract the entire expected valuation by pooling all quality information and posting the prior mean valuation as the price.
The proof of \Cref{lem:degenerate-type-no-info} is provided in \Cref{apx:no-info-optimality}, and we actually prove a stronger result (see \Cref{prop:no-info-sufficient}) which establishes the sufficient conditions for optimality of the no-information advertising.

Our model can therefore be viewed as extending this benchmark by allowing part of the customer's valuation to be intrinsic and not controlled by advertising. In our formulation, $\quality$ is the product-quality component whose perceived value can be affected by the seller's advertising signal, while $\type$ captures the customer's idiosyncratic preference or willingness-to-pay component that is known to the customer but not observed or controlled by the seller. When $\type$ is nondegenerate, no-information advertising no longer pools the entire valuation uncertainty; it only pools uncertainty about the product-quality component. The remaining heterogeneity in $\type$ affects how each posterior quality belief translates into demand at a given price. Consequently, informative advertising can become valuable for the seller, because it can reshape the distribution of posterior product valuations to better match the underlying type distribution.
This distinction also drives our learning problem. 
Whereas \citet{RS-17} characterize the optimal information policy under a known valuation distribution, our seller must learn the unknown type distribution $\typeCDF$: each price–posterior-mean pair induces a critical type, and the resulting purchase decision reveals demand at that type. 
Advertising therefore affects not only revenue but also which parts of demand are sampled as the seller jointly learns and optimizes pricing and advertising over time.


\xhdr{Comparison with \citet{ES-07} and \citet{BHM-26}} 
\citet{ES-07} study a mechanism-design problem in which a seller can disclose additional information that refines bidders' knowledge of their private valuations. Their main result shows that full disclosure is revenue maximizing and can be implemented through a handicap auction. This conclusion is closely related to our signal-contingent-pricing benchmark $\optSC(\instance)$: once prices can depend on the realized signal, further refining information cannot reduce revenue, and full-information advertising is optimal by a convexity/Jensen argument.

The key difference is that our main benchmark restricts the seller to a single quality-independent price. 
Under this constraint, full disclosure need not be optimal because the seller cannot tailor prices across posterior segments. 
Partial disclosure can therefore outperform both full disclosure and no disclosure by balancing revenues across customer groups. 
Our goal is to characterize the value of advertising in this simpler posted-price environment, 
show that it remains within a factor of two of the stronger signal-contingent benchmark, and study the additional challenge of learning when the customer type distribution is unknown.

\citet{BHM-26} study a setting in which the seller jointly designs buyers' information and a nonlinear pricing mechanism. Their optimal mechanism combines persuasion and screening, using value pooling and reduced product variety to mitigate informational rents. Like our paper, their analysis treats information design and pricing as jointly determined instruments.
The main difference is that their model centers on screening through menus of products and payments, whereas our main benchmark uses only a single quality-independent price and advertising about product quality. As a result, pooling serves different purposes: in their setting it reduces informational rents, while in ours it shapes the distribution of posterior qualities faced by a common price. 

\xhdr{Comparison with \cite{BCW-22}}
It is worth comparing the value of advertising, i.e., the ratio $\optU(\instance)/\optNI(\instance)$,
established in \Cref{thm:voa bound} to the recent work \citep{BCW-22}, which considers a setting where a seller uses information about customers' characteristics to offer different prices to different market segments.
They show that the best possible revenue from perfectly discriminating across the different segments in the market (i.e., the revenue of third-degree price policy) is also twice the revenue from a single price across all customer segments.
Our setting differs from theirs as the seller in our setting is segmenting the market through advertising the product characteristics, not by using customer information, and the seller is offering a {\em single quality-independent} price to all segments -- thus, there is no price discrimination.
In addition, the results obtained in \cite{BCW-22} depend on two key assumptions: 
that the revenue function is concave (with respect to price) for each market segment and that the market segments share a common support.
They establish a multiplicative revenue gap of 2 using a geometric approach that utilizes these two assumptions.
However, these two assumptions are not necessarily satisfied in our setting, so our proof follows a completely different approach.\footnote{\revcolor{To see this, take the additive valuation $\buyerUtility(\type, \quality) = \type+\quality$ as an example.  
Each signal realization $\sigma$ realized from the seller's advertising strategy induces a corresponding market segment where the revenue function (denoted by $\Rev(p\mid \sigma)$) in this segment is $\Rev(p\mid \sigma)\triangleq p\cdot (1 - \typeCDF(p - \expect[\posteriorDist(\cdot\mid \sigma)]{\quality}))$.
As we do not make any assumptions about the customers' type distribution $\typeCDF$, the function $\Rev(p\mid \sigma)$ is not necessarily concave. Moreover, the support of this segment is $[\expect[\posteriorDist(\cdot\mid \sigma)]{\quality}, 1]$, which varies with different signals.}}

The above comparisons also provide the following managerial insights for the seller using advertising as a potential lever to improve their revenue. 
We recall that one common way to improve the seller's revenue is price discrimination -- like the one adopted in \citet{BCW-22} and many other works \citep[e.g.,][]{BBM-15,CEL-22,BMSW-24} -- the seller increases revenue by tailoring prices to specific customer segments based on their characteristics. 
However this approach may bear two important issues: (1) it requires the seller to have access to detailed customer characteristics so that they can accurately personalize the market segment; (2) it may raise fairness/privacy concerns.
In contrast, we show that by focusing on product advertising or differentiation, one can guide customers' perceptions and indirectly influence customer segmentation  based on product attributes rather than personal characteristics, which may be more feasible when customer data is limited or fairness/privacy concerns restrict the use of such data.

%% file: Paper/apx-sufficient-conditions-noinfor.tex
To prove \Cref{lem:degenerate-type-no-info}, we establish the following stronger result which establishes the sufficient conditions for optimality of the no-information advertising.
\begin{proposition}[Sufficient conditions for no-information optimality]
\label{prop:no-info-sufficient}
Suppose the valuation function satisfies \Cref{assump:valuation}.
Let $\priorMean\triangleq \expect[\quality\sim\prior]{\quality}$ be the prior mean quality, and let $\inverseVal(\price,\posteriorMean)$ be the critical type induced by price $\price$ and posterior mean quality $\posteriorMean$.
Then the no-information advertising strategy $\noinfor$ is optimal under quality-independent pricing, namely 
$\sup\nolimits_{\price,\scheme}\Rev(\price,\scheme)
=
\sup\nolimits_{\price}\Rev(\price,\noinfor)$,
if either of the following conditions holds:
\begin{enumerate}[label=(\roman*)]
    \item 
    There exists a price $\postedMeanPrice\geq 0$ such that
    $\buyerUtility(\type,\priorMean)
    \in
    \{0,\postedMeanPrice\}$ almost surely.

    \item 
    For every price $\price\geq 0$, the function
    $\posteriorMean
    \mapsto
    \demand(\inverseVal(\price,\posteriorMean))$
    is concave on the convex hull of the quality space.
\end{enumerate}
\end{proposition}

For the additive valuation function $\buyerUtility(\type,\quality)=\type+\quality$, we have $\inverseVal(\price,\posteriorMean)=\price-\posteriorMean$, with the usual boundary convention when the threshold lies outside the type support. Hence, $\demand(\inverseVal(\price,\posteriorMean))=\demand(\price-\posteriorMean)$. Therefore, if $\demand$ is concave on the relevant type range, then condition~(ii) holds.

If $\type=\degType$ almost surely, then under no-information advertising, $\buyerUtility(\type,\priorMean)=\buyerUtility(\degType,\priorMean)$ almost surely. 
Thus, condition~(i) holds with $\postedMeanPrice=\buyerUtility(\degType,\priorMean)$, and no-information advertising is optimal. This recovers \Cref{lem:degenerate-type-no-info} and corresponds to the zero-cost seller-optimal case in \citet{RS-17}.

\begin{proof}[Proof of \Cref{prop:no-info-sufficient}]
Fix an arbitrary advertising strategy $\scheme$. For every realized signal $\signal$, let
$\postMeanSignal\triangleq \expect[\quality\sim\posteriorDist(\cdot\mid\signal)]{\quality}$ denote the posterior mean quality induced by $\signal$. By Eqn.~\eqref{bayes consistency vanilla}, we have
$\expect{\postMeanSignal}
= \priorMean$.
Since $\buyerUtility$ is linear in quality, the posterior expected valuation of a type-$\type$ customer after observing signal $\signal$ is
$\expect[\quality\sim\posteriorDist(\cdot\mid\signal)]{\buyerUtility(\type,\quality)}
=
\buyerUtility(\type,\postMeanSignal)$.
Moreover, by \Cref{assump:valuation}, $\buyerUtility(\type,\posteriorMean)$ is increasing in $\type$. Thus, given price $\price$ and posterior mean quality $\posteriorMean$, the customer purchases if and only if $\type\geq \inverseVal(\price,\posteriorMean)$. Therefore, the revenue from price $\price$ and advertising strategy $\scheme$ can be written as
$\Rev(\price,\scheme)
=
\price\cdot
\expect{\demand(\inverseVal(\price,\postMeanSignal))}$.
Under the no-information advertising strategy $\noinfor$, the posterior mean quality is always $\priorMean$, and thus, we have
$\Rev(\price,\noinfor)
=
\price\cdot \demand(\inverseVal(\price,\priorMean))$.

We first prove the claim under condition~(i). Suppose there exists $\postedMeanPrice\geq 0$ such that $\buyerUtility(\type,\priorMean)\in\{0,\postedMeanPrice\}$ almost surely. Under no-information advertising, if the seller posts price $\postedMeanPrice$, then every customer with no-information valuation $\postedMeanPrice$ purchases, while customers with no-information valuation $0$ do not contribute positive revenue. 
Thus, under no-information advertising, if the seller posts price $\postedMeanPrice$, 
\begin{align*}
    \Rev(\postedMeanPrice,\noinfor)
    =
    \postedMeanPrice\cdot
    \prob{\buyerUtility(\type,\priorMean)\geq \postedMeanPrice}
    =
    \postedMeanPrice\cdot
    \prob{\buyerUtility(\type,\priorMean)=\postedMeanPrice}
    =
    \expect[\type\sim\typeCDF]{\buyerUtility(\type,\priorMean)}
    =
    \expect[\type\sim\typeCDF,\quality\sim\prior]{\buyerUtility(\type,\quality)}~.
\end{align*}
The first equality follows from the revenue definition under $\noinfor$. The second and third equalities follow from the two-point-support condition $\buyerUtility(\type,\priorMean)\in\{0,\postedMeanPrice\}$ almost surely. The last equality follows from the linearity of $\buyerUtility$ in quality and the independence between customer type and product quality. Thus, no-information advertising with price $\postedMeanPrice$ extracts the full ex ante surplus.

It remains to show that no pricing-and-advertising strategy can generate more than the full ex ante surplus. For any price $\price$ and advertising strategy $\scheme$, we have
\begin{align*}
    \Rev(\price,\scheme)
    & =
    \expect[\type,\signal]{\price\cdot
    \indicator{\buyerUtility(\type,\postMeanSignal)\geq \price}}\\
    & \leq
    \expect[\type,\signal]{\buyerUtility(\type,\postMeanSignal)\cdot
    \indicator{\buyerUtility(\type,\postMeanSignal)\geq \price}}
    \tag{By linearity in quality, and independence between $\type$ and $\signal$}
    \\
    & \leq
    \expect[\type,\signal]{\buyerUtility(\type,\postMeanSignal)}
    \tag{By $\buyerUtility(\type,\postMeanSignal) \ge 0$}\\
    & =
    \expect[\type\sim\typeCDF]{\buyerUtility(\type,\expect{\postMeanSignal})}\\
    & =
    \expect[\type\sim\typeCDF]{\buyerUtility(\type,\priorMean)}
    \tag{By Eqn.~\eqref{bayes consistency vanilla}}
    \\
    & =
    \expect[\type\sim\typeCDF,\quality\sim\prior]{\buyerUtility(\type,\quality)} 
    \tag{By linearity in quality}
\end{align*}
The first inequality follows because, on the purchase event, the posted price is no larger than the customer's posterior expected valuation. 
Thus,
\begin{align*}
    \sup\nolimits_{\price,\scheme}\Rev(\price,\scheme)
    \leq
    \expect[\type\sim\typeCDF,\quality\sim\prior]{\buyerUtility(\type,\quality)}
    =
    \Rev(\postedMeanPrice,\noinfor)
    \leq
    \sup\nolimits_{\price}\Rev(\price,\noinfor)
    \leq
    \sup\nolimits_{\price,\scheme}\Rev(\price,\scheme)~.
\end{align*}
Thus, we have
$\sup\nolimits_{\price,\scheme}\Rev(\price,\scheme)=\sup\nolimits_{\price}\Rev(\price,\noinfor)$ under condition~(i).

We next prove the claim under condition~(ii). 
Fix any price $\price$ and any advertising strategy $\scheme$. Since the function $\posteriorMean\mapsto \demand(\inverseVal(\price,\posteriorMean))$ is concave and $\expect{\postMeanSignal}=\priorMean$, we have 
\begin{align*}
    \Rev(\price,\scheme)
    =
    \price\cdot
    \expect{\demand(\inverseVal(\price,\postMeanSignal))}
    \le
    \price\cdot \demand\left(\inverseVal\left(\price,\expect{\postMeanSignal}\right)\right)
    =
    \price\cdot
    \demand(\inverseVal(\price,\priorMean))
    =
    \Rev(\price,\noinfor)~,
\end{align*}
where the second inequality is due to Jensen's inequality.

Since this inequality holds for every price $\price$ and advertising strategy $\scheme$, we have
$\sup\nolimits_{\price,\scheme}\Rev(\price,\scheme)
\leq
\sup\nolimits_{\price}\Rev(\price,\noinfor)$.
The reverse inequality holds because $\noinfor$ is a feasible advertising strategy. 
Thus, no-information advertising is optimal under condition~(ii).
\end{proof}

%% file: Paper/apx-proof-voa-new.tex
In this section, we prove \Cref{thm:voa bound} with the following stronger statement.
\begin{restatable}[Revenue comparisons]{theorem}{voaboundstronger}
\label{thm:voa bound stronger}
{For every instance $\instance$
where the customers' valuation function 
satisfies \Cref{assump:valuation} and} admits the affine-in-type representation $\buyerUtility(\type,\quality)
=
\type\cdot \auxfunc_1(\quality)+\auxfunc_2(\quality)$
and $\auxfunc_1,\auxfunc_2:\qualitySpace\to\reals_+$ are measurable functions, 
we have
\begin{align*}
    \optNI(\instance)
    \leq
    \optU(\instance)
    \leq
    \optSC(\instance)
    \leq
    2\optNI(\instance)~.
\end{align*}
Moreover, all three factors are tight:
\begin{align*}
    \sup\nolimits_{\instance}
    \frac{\optU(\instance)}{\optNI(\instance)}
    =
    2~,\quad
    \sup\nolimits_{\instance}
    \frac{\optSC(\instance)}{\optNI(\instance)}
    =
    2~,\quad
    \sup\nolimits_{\instance}
    \frac{\optSC(\instance)}{\optU(\instance)}
    =
    2~,
\end{align*}
where the supremum is over all instances where customers' valuation function admits the above affine-in-type representation.
\end{restatable}
We first prove \Cref{thm:voa bound stronger}, and then use \Cref{thm:voa bound stronger} to prove \Cref{thm:voa bound}. 
The tightness proofs use the following two instances.
\begin{example}
\label{example:voa lb 1}
In this instance, the product has binary quality $\quality\in\{0,1\}$ with $\prior(1)=\eps$ and $\prior(0)=1-\eps$. The valuation function is additive, $\buyerUtility(\type,\quality)=\type+\quality$. The customer type distribution is supported on $[0,1]$ and has the CDF 
$\typeCDF_{\eps}(\type)=1-\frac{\eps}{\type+\eps}$ for $\type\in[0,1)$, with an atom of size $\frac{\eps}{1+\eps}$ at $\type=1$.
We use $\instance_\eps$ to denote this instance.
\end{example}

\begin{example}
\label{example:voa lb 2}
In this instance, we fix $\eps\in(0,1)$ and define the following quantities
$\tightAlpha=\sqrt{\eps},
\tightDelta=\sqrt{\eps},
\tightTau=\eps$.
The product has binary quality $\quality\in\{0,1\}$ with 
$\prior(1) = \tightAlpha$.
The customer type is binary with $\prob{\type=1}=\tightDelta$ and $\prob{\type=0}=1-\tightDelta$.
The valuation function is
$\tightVal(\type,\quality)
=
\eps+(1-\eps)\type\left(\tightTau+(1-\tightTau)\quality\right)$.
We use $\instance_\eps\primed$ to denote this instance.
\end{example}

\begin{proof}[Proof of \Cref{thm:voa bound stronger}]
For notational simplicity, we write the proof for a finite signal space. The same argument applies to a general measurable signal space by replacing summations over signals with integrals.
We proceed with the proof in four steps.

\xhdr{Step 1: full-information advertising is optimal under signal-contingent pricing}
Fix an arbitrary advertising strategy $\scheme$. Let $\scheme(\signal)$ denote the ex ante probability that signal $\signal$ is realized, and recall that $\posteriorDist(\cdot\mid \signal)$ denotes the posterior belief induced by signal $\signal$. 
For every signal $\signal\in\signalSpace$, we define the following two posterior averages
\begin{align*}
    \posteriorAuxOne_{\signal}
    \triangleq
    \expect[\posteriorDist(\cdot\mid\signal)]{\auxfunc_1(\quality)}~,
    \quad
    \posteriorAuxTwo_{\signal}
    \triangleq
    \expect[\posteriorDist(\cdot\mid\signal)]{\auxfunc_2(\quality)}~.
\end{align*}
Given signal $\signal$, a type-$\type$ customer's posterior expected valuation is
$\expect[\posteriorDist(\cdot\mid\signal)]{\buyerUtility(\type,\quality)}
=
\type\cdot \posteriorAuxOne_{\signal}+\posteriorAuxTwo_{\signal}$.
Recall that $\demand(x)\triangleq\prob[\type\sim\typeCDF]{\type\geq x}$ denotes the demand function. 
For any pair $(A,B)\in\reals_+^2$, we define the optimal conditional revenue
\begin{align*}
    \condRev(A,B)
    \triangleq
    \sup\nolimits_{x\geq 0}(A x+B)\cdot \demand(x)~.
\end{align*}
This expression represents the optimal revenue from selling to customers whose posterior expected valuation is $\type A+B$. 
Indeed, choosing threshold $x$ corresponds to posting price $A x+B$; the case of prices below $B$ is dominated by choosing $x=0$.

The function $\condRev(A,B)$ is convex in $(A,B)$ because it is the supremum of affine functions $(A,B)\mapsto (Ax+B)\demand(x)$. 
Thus, for each signal $\signal$, Jensen's inequality implies
\begin{align*}
    \condRev(\posteriorAuxOne_{\signal},\posteriorAuxTwo_{\signal})
    =
    \condRev\left(
    \expect[\posteriorDist(\cdot\mid\signal)]{\auxfunc_1(\quality)},
    \expect[\posteriorDist(\cdot\mid\signal)]{\auxfunc_2(\quality)}
    \right)
    \leq
    \expect[\posteriorDist(\cdot\mid\signal)]
    {\condRev(\auxfunc_1(\quality),\auxfunc_2(\quality))}~.
\end{align*}
Taking expectation over signals and using Eqn.~\eqref{bayes consistency vanilla}, we obtain
\begin{align*}
    \sup\nolimits_{\signalPrice}\SCRev_{\instance}(\signalPrice,\scheme)
    =
    \sum\nolimits_{\signal\in\signalSpace}
    \scheme(\signal)\cdot
    \condRev(\posteriorAuxOne_{\signal},\posteriorAuxTwo_{\signal})
    & \leq
    \sum\nolimits_{\signal\in\signalSpace}
    \scheme(\signal)\cdot
    \expect[\posteriorDist(\cdot\mid\signal)]
    {\condRev(\auxfunc_1(\quality),\auxfunc_2(\quality))}\\
    & =
    \expect[\quality\sim\prior]
    {\condRev(\auxfunc_1(\quality),\auxfunc_2(\quality))}.
\end{align*}
The last expression is exactly the revenue achieved by full-information advertising with the optimal signal-contingent price after each realized quality. 
Thus, full-information advertising $\fullinfor$ is optimal to achieve $\optSC(\instance)$.

\xhdr{Step 2: upper bounding $\optSC(\instance)$ by $2\optNI(\instance)$}
We define the monopoly revenue associated with the type distribution alone as
$\typeRevenue
\triangleq
\sup\nolimits_{x\geq 0}x\cdot \demand(x)$.
For any $(A,B)\in\reals_+^2$, we know that 
\begin{align*}
    \condRev(A,B)
    =
    \sup\nolimits_{x\geq 0}(A x+B)\demand(x)
    \leq
    A\typeRevenue+B~,
\end{align*}
where the inequality follows from the definition of $\typeRevenue$ and $\demand(x)\leq 1$.

For notational simplicity, we define the following prior expectations of functions $\auxfunc_1, \auxfunc_2$,
$\priorAuxOne
\triangleq
\expect[\quality\sim\prior]{\auxfunc_1(\quality)},
\priorAuxTwo
\triangleq
\expect[\quality\sim\prior]{\auxfunc_2(\quality)}$.
Using the bound above and the full-information characterization from Step 1, we have
\begin{align}
    \label{ineq:full-infor-UB}
    \optSC(\instance)
    =
    \expect[\quality\sim\prior]
    {\condRev(\auxfunc_1(\quality),\auxfunc_2(\quality))}
    \leq
    \expect[\quality\sim\prior]
    {\auxfunc_1(\quality)\typeRevenue+\auxfunc_2(\quality)}
    =
    \priorAuxOne\cdot \typeRevenue+\priorAuxTwo~.
\end{align}

We next show that the no-information benchmark can obtain each term on the right-hand side of Eqn.~\eqref{ineq:full-infor-UB}. Under no-information advertising $\noinfor$, a type-$\type$ customer's expected valuation is $\type\cdot \priorAuxOne+\priorAuxTwo$. First, for any threshold $x\geq 0$, posting price $\priorAuxOne x$ sells to every customer with type at least $x$. Thus, we have
\begin{align}
    \label{ineq:lower-bound-1}
    \optNI(\instance)
    =
    \sup\nolimits_{\price}\Rev_{\instance}(\price,\noinfor)
    \geq
    \sup\nolimits_{x\geq 0}\priorAuxOne x\cdot \demand(x)
    =
    \priorAuxOne\cdot \typeRevenue~.
\end{align}
Second, posting price $\priorAuxTwo$ sells to all customers, because $\type\cdot\priorAuxOne+\priorAuxTwo\geq \priorAuxTwo$ for every type $\type$. Thus, we have
$\optNI(\instance)
\geq
\priorAuxTwo$.
Combining Eqn.~\eqref{ineq:full-infor-UB} and Eqn.~\eqref{ineq:lower-bound-1}, we obtain
\begin{align*}
    \optSC(\instance)
    \leq
    \priorAuxOne\cdot \typeRevenue+\priorAuxTwo
    \leq
    2\max\{\priorAuxOne\cdot \typeRevenue,\priorAuxTwo\}
    \leq
    2\optNI(\instance)~.
\end{align*}
Finally, $\optNI(\instance)\leq\optU(\instance)$ because $\noinfor$ is a feasible advertising strategy, and $\optU(\instance)\leq\optSC(\instance)$ because quality-independent pricing is a special case of signal-contingent pricing. 
Thus,
\begin{align*}
    \optNI(\instance)
    \leq
    \optU(\instance)
    \leq
    \optSC(\instance)
    \leq
    2\optNI(\instance)~.
\end{align*}
This proves all three upper bounds.

\xhdr{Step 3: tightness of $\optU/\optNI$ and $\optSC/\optNI$}
We analyze \Cref{example:voa lb 1}.
We first compute the optimal
revenue under the no-information advertising strategy.
Under no-information advertising, the posterior mean quality is $\expect[\quality\sim\prior]{\quality}=\eps$, so a type-$\type$ customer's expected valuation is $\type+\eps$. Given price $\price$, the customer purchases if and only if $\type\geq \price-\eps$. Thus,
\begin{align*}
    \Rev(\price,\noinfor)
    =
    \price\cdot (1 - \typeCDF_{\eps}(\price-\eps))
    =
    \left\{
    \begin{array}{ll}
        \price, & 0\leq \price\leq \eps,\\
        \eps, & \eps<\price\leq 1+\eps,\\
        0, & \price>1+\eps.
    \end{array}
    \right.
\end{align*}
Thus, we can deduce that the optimal revenue under the no-information advertising strategy is $\optNI(\instance_{\eps})=\eps$.

Now consider full-information advertising $\fullinfor$ and quality-independent price $\price=1$. If $\quality=1$, then every customer purchases because $\type+1\geq 1$. If $\quality=0$, then the customer purchases if and only if $\type\geq 1$, which occurs with probability $\boundedDemand(1)=\frac{\eps}{1+\eps}$. Thus,
\begin{align*}
    \Rev(1,\fullinfor)
    =
    1\cdot
    \left(\prior(1)\cdot 1+\prior(0)\cdot (1 - \typeCDF_\eps(1))\right)
    =
    \eps+(1-\eps)\cdot \frac{\eps}{1+\eps}
    =
    \frac{2\eps}{1+\eps}~.
\end{align*}
Since $\optSC(\instance_{\eps})\geq \optU(\instance_{\eps})\geq \Rev(1,\fullinfor)$, we have
\begin{align*}
    \frac{\optU(\instance_{\eps})}{\optNI(\instance_{\eps})}
    \geq
    \frac{2}{1+\eps}~,
    \quad
    \frac{\optSC(\instance_{\eps})}{\optNI(\instance_{\eps})}
    \geq
    \frac{2}{1+\eps}~.
\end{align*}
Letting $\eps\downarrow 0$, together with the upper bound in Step 2, proves
\begin{align*}
    \sup\nolimits_{\instance}
    \frac{\optU(\instance)}{\optNI(\instance)}
    =
    2~,
    \quad
    \sup\nolimits_{\instance}
    \frac{\optSC(\instance)}{\optNI(\instance)}
    =
    2~.
\end{align*}

\xhdr{Step 4: tightness of $\optSC/\optU$}
We analyze \Cref{example:voa lb 2}.
We first lower bound $\optSC$. Consider full-information advertising. If $\quality=0$, posting signal-contingent price $\eps$ sells to all customers and generates revenue $\eps$ in the low-quality state. If $\quality=1$, posting signal-contingent price $1$ sells to high-type customers and generates revenue $\tightDelta$ in the high-quality state. Thus,
\begin{align}
    \label{ineq:full-infor-LB}
    \optSC(\instance_\eps\primed)
    \geq
    (1-\tightAlpha)\eps+\tightAlpha\tightDelta
    =
    (1-\sqrt{\eps})\eps+\eps
    =
    (2-\sqrt{\eps})\eps~.
\end{align}

We next upper bound $\optU$. Fix any advertising strategy $\scheme$, and let $\tightSignalMean\triangleq \expect[\quality\sim\posteriorDist(\cdot\mid\signal)]{\quality}$ be the posterior mean quality after signal $\signal$. 
By Eqn.~\eqref{bayes consistency vanilla}, we have $\expect{\tightSignalMean}=\tightAlpha$. 
For the low type $\type=0$, the posterior expected valuation is always $\eps$. For the high type $\type=1$, the posterior expected valuation after signal $\signal$ is
$\tightVal(1,\tightSignalMean)
=
\eps+(1-\eps)\left(\tightTau+(1-\tightTau)\tightSignalMean\right)$.
Fix any quality-independent price $\price$. If $\price\leq \eps$, then the revenue is at most $\eps$. Now suppose $\price>\eps$. Then low-type customers do not purchase. 
We define the following quantities to simplify the notation.
\begin{align*}
    \tightM
    \triangleq
    \tightTau+(1-\tightTau)\tightSignalMean~,
    \quad
    \tightMBar
    \triangleq
    \expect{\tightM}
    =
    \tightTau+(1-\tightTau)\tightAlpha~,
    \quad 
    \tightThreshold
    \triangleq
    \frac{\price-\eps}{1-\eps}~.
\end{align*}
A high-type customer purchases if and only if $\tightM\geq \tightThreshold$. By Markov's inequality,
we have 
$\prob{\tightM\geq \tightThreshold}
\leq
\min\left\{1,\frac{\tightMBar}{\tightThreshold}\right\}$.
Since $\price=\eps+(1-\eps)\tightThreshold$, we have
\begin{align}
    \label{ineq:no-infor-UB2}
    \Rev(\price,\scheme)
    =
    \price\cdot \tightDelta\cdot
    \prob{\tightM\geq \tightThreshold}
    \leq
    \tightDelta\left(\eps+(1-\eps)\tightThreshold\right)
    \min\left\{1,\frac{\tightMBar}{\tightThreshold}\right\}
    \leq
    \tightDelta\left(\eps+(1-\eps)\tightMBar\right)~.
\end{align}
The last inequality holds by considering two cases: if $\tightThreshold\leq \tightMBar$, then $\eps+(1-\eps)\tightThreshold\leq \eps+(1-\eps)\tightMBar$; if $\tightThreshold>\tightMBar$, then
\begin{align*}
    \left(\eps+(1-\eps)\tightThreshold\right)\frac{\tightMBar}{\tightThreshold}
    =
    \tightMBar\left(\frac{\eps}{\tightThreshold}+1-\eps\right)
    \leq
    \eps+(1-\eps)\tightMBar~.
\end{align*}
Substituting $\tightAlpha=\sqrt{\eps}$, $\tightDelta=\sqrt{\eps}$, and $\tightTau=\eps$ into Eqn.~\eqref{ineq:no-infor-UB2}, we obtain
\begin{align*}
    \tightDelta\left(\eps+(1-\eps)\tightMBar\right)
    =
    \sqrt{\eps}
    \left[
    \eps+(1-\eps)\left(\eps+(1-\eps)\sqrt{\eps}\right)
    \right]
    =
    \eps+2\eps^{3/2}-2\eps^2-\eps^{5/2}+\eps^3
    \leq
    \eps+2\eps^{3/2}~.
\end{align*}
Combining the cases $\price\leq\eps$ and $\price>\eps$, we conclude that for every quality-independent price $\price$ and every advertising strategy $\scheme$,
we have 
$\Rev(\price,\scheme)
\leq
\eps+2\eps^{3/2}$, which implies that
$\optU(\instance_\eps\primed)
\leq
\eps(1+2\sqrt{\eps})$.
Combining Eqn.~\eqref{ineq:full-infor-LB}, we have
\begin{align*}
    \frac{\optSC(\instance_\eps\primed)}{\optU(\instance_\eps\primed)}
    \geq
    \frac{(2-\sqrt{\eps})\eps}{\eps(1+2\sqrt{\eps})}
    =
    \frac{2-\sqrt{\eps}}{1+2\sqrt{\eps}}~.
\end{align*}
Letting $\eps\rightarrow 0$, together with the upper bound in Step 2, proves 
$\sup\nolimits_{\instance}
\frac{\optSC(\instance)}{\optU(\instance)}
=
2$.
This completes the proof of \Cref{thm:voa bound stronger}.
\end{proof}

\begin{proof}[Proof of \Cref{thm:voa bound}]
We first show that every valuation function satisfying \Cref{assump:valuation} belongs to the affine-in-type class considered in \Cref{thm:voa bound stronger}. 
Fix any valuation function $\buyerUtility$ satisfying \Cref{assump:valuation}. Since $\buyerUtility(\cdot,\quality)$ is linear in type for every quality $\quality$, we can define
\begin{align*}
    \auxfunc_2(\quality)
    \triangleq
    \buyerUtility(0,\quality)~,
    \quad
    \auxfunc_1(\quality)
    \triangleq
    \buyerUtility(1,\quality)-\buyerUtility(0,\quality)~.
\end{align*}
Then, for every $\type\in[0,1]$ and $\quality\in\qualitySpace$, we have
$\buyerUtility(\type,\quality)
=
\type\cdot \auxfunc_1(\quality)+\auxfunc_2(\quality)$.
Moreover, since $\buyerUtility(\theta,\cdot)$ is linear in quality for every $\theta$, both $\auxfunc_1$ and $\auxfunc_2$ are linear, and hence measurable, functions of $\quality$. Since $\buyerUtility(\cdot,\quality)$ is increasing in type, we have $\auxfunc_1(\quality)\geq 0$ for every $\quality$. In addition, as $\buyerUtility(\cdot,\cdot)\ge 0$ always, we have $\auxfunc_2(\quality)=\buyerUtility(0,\quality)\geq 0$ for every $\quality$. 
Thus, the function $\buyerUtility$ admits the affine-in-type representation required by \Cref{thm:voa bound stronger}.

Applying \Cref{thm:voa bound stronger} to any instance whose valuation function satisfies \Cref{assump:valuation}, we obtain
$\optNI(\instance)
\leq
\optU(\instance)
\leq
\optSC(\instance)
\leq
2\optNI(\instance)$.
This proves the upper-bound part of \Cref{thm:voa bound}.

It remains to verify the tightness statements in \Cref{thm:voa bound}. The tightness examples used in the proof of \Cref{thm:voa bound stronger} all satisfy \Cref{assump:valuation}. Specifically, the examples proving the tightness of $\optU/\optNI$ and $\optSC/\optNI$ use the additive valuation function $\buyerUtility(\type,\quality)=\type+\quality$, which is linear and nondecreasing in quality, and is increasing, linear, and $1$-Lipschitz in type. The example proving the tightness of $\optSC/\optU$ uses the valuation function
$\tightVal(\type,\quality)
=
\eps+(1-\eps)\type\left(\tightTau+(1-\tightTau)\quality\right)$, where $\tightTau=\eps$. 
This function is linear in quality and linear in type. It is nondecreasing in quality because
$\frac{\partial \tightVal(\type,\quality)}{\partial \quality}
=
(1-\eps)\type(1-\tightTau)
\geq 0$,
and it is increasing and $1$-Lipschitz in type because
$\frac{\partial \tightVal(\type,\quality)}{\partial \type}
=
(1-\eps)\left(\tightTau+(1-\tightTau)\quality\right)
\in [0,1]$.
Thus, the same sequences of instances are feasible under \Cref{assump:valuation}. Consequently, each of the three suprema in \Cref{thm:voa bound} is at least $2$. Since the upper-bound part has already shown that each ratio is at most $2$, all three tightness statements follow.
\end{proof}

%% file: Paper/apx-alternative-benchmark-learning.tex
In this section, we design learning algorithms for two variants of the model: (i) the seller can only offer a single quality-independent price without advertising; and (ii) the seller can jointly optimize advertising and a price menu that varies with the realized signal. The optimal revenues of these two variants correspond to the pricing-only benchmark $\optNI$ and the signal-contingent-pricing benchmark $\optSC$, respectively, introduced and analyzed in \Cref{subsec:voa} (see \Cref{thm:voa bound}).

\subsection{Dynamic Learning against the Pricing-Only Benchmark}
\label{subsec:po-benchmark}
In this section, we consider the variant in which the seller can only offer a single quality-independent price without advertising.

In this variant, given any price $\price$, a customer with type $\type$ purchases the product if her expected valuation under the prior belief over the quality, $\expect[\quality \sim \prior]{\buyerUtility(\type, \quality)}$, is at least the price $\price$. Thus, the seller's pricing problem reduces to the classic dynamic pricing problem in which the customer's willingness to pay, $\expect[\quality \sim \prior]{\buyerUtility(\type, \quality)}$, is drawn from an unknown distribution. Invoking the results of \citet{KL-03}, we obtain the following theorem. For completeness, we also include a brief description of the algorithm and its regret analysis.

\begin{theorem}
\label{thm:po-benchmark:regret upper bound}
For every instance $\instance$,
there exists a UCB-type algorithm that implements quality-independent pricing without advertising and
has expected regret $O(\timeHorizon^{\sfrac{2}{3}}
(\log\timeHorizon)^{\sfrac{1}{3}})$ against the pricing-only benchmark $\optNI$, where $\timeHorizon$ is the time horizon.

Moreover, for every online algorithm, there exists an instance satisfying \Cref{assump:valuation} and $\stateNum = 1$ on which its regret is $\Omega(\timeHorizon^{2/3})$.
\end{theorem}

We remark that combining \Cref{thm:po-benchmark:regret upper bound} with \Cref{thm:voa bound} implies that the same UCB-type algorithm also achieves expected regret $O(\timeHorizon^{\sfrac{2}{3}}
(\log\timeHorizon)^{\sfrac{1}{3}})$ against half of our main advertising-and-pricing benchmark $\optU$, and even against half of the stronger signal-contingent-pricing benchmark $\optSC$. This echoes the managerial insight from \Cref{subsec:voa} that advertising may not be necessary when a multiplicative factor-of-$2$ loss is acceptable, even in the dynamic learning setting with uncertain demand.

\begin{proof}[Proof of \Cref{thm:po-benchmark:regret upper bound}]
Consider the following UCB-type algorithm. The seller fixes the no-information advertising strategy $\noinfor$ in all rounds. Given a discretization parameter $\discrePrecision>0$, the price grid $\simplePriceGrid$ is defined as
\begin{align*}
    \simplePriceGrid
    \triangleq
    \left\{0,\discrePrecision,2\discrePrecision,\ldots,
    \left\lfloor \frac{\pupper}{\discrePrecision}\right\rfloor\discrePrecision,\pupper\right\}~.
\end{align*}
The seller then runs a standard finite-armed \UCB\ algorithm over the price arms in $\simplePriceGrid$, where the realized reward from price $\price$ in a round is $\price$ times the purchase indicator. For every fixed price arm $\price\in\simplePriceGrid$, this reward is supported on $[0,\pupper]$ and has mean $\Rev(\price, \noinfor)$.

We first bound the discretization error. For an arbitrary price $\price\in[0,\pupper]$, let the rounded price $\roundedPrice$ be the largest grid point in $\simplePriceGrid$ that is weakly below $\price$.
The grid construction implies $\roundedPrice\ge \price-\discrePrecision$.
Since the critical type $\inverseVal(\price,\qualityVal)$ is weakly increasing in the price $\price$ and the demand function $\demand(\cdot)$ is weakly decreasing, we have $\demand(\inverseVal(\roundedPrice,\qualityVal))\ge \demand(\inverseVal(\price,\qualityVal))$.
Thus,
\begin{align*}
    \Rev(\roundedPrice, \noinfor)
    =
    \roundedPrice\demand(\inverseVal(\roundedPrice,\qualityVal))
    \ge
    (\price-\discrePrecision)\demand(\inverseVal(\price,\qualityVal))\ge
    \Rev(\price,\noinfor)-\discrePrecision~.
\end{align*}
Since this inequality holds for every price $\price\in[0,\pupper]$, the best grid price satisfies
$\max\nolimits_{\price\in\simplePriceGrid}\Rev(\price, \noinfor)
\ge
\optNI(\instance)-\discrePrecision$.

Next, the standard gap-independent guarantee for finite-armed \UCB\ with rewards supported on $[0,\pupper]$ implies that there exists a universal constant $C>0$ such that
\begin{align*}
    \timeHorizon \max\nolimits_{\price\in\simplePriceGrid}\Rev(\price, \noinfor)
    -
    \expect{\sum\nolimits_{t\in[\timeHorizon]}\Rev(\price_t,\noinfor)}
    \le
    C\pupper\sqrt{|\simplePriceGrid|\timeHorizon\log \timeHorizon}~.
\end{align*}
The price-grid cardinality satisfies $|\simplePriceGrid|
\le
2+\frac{\pupper}{\discrePrecision}$.
The preceding inequalities together imply
\begin{align*}
    \expect{\sum\nolimits_{t\in[\timeHorizon]}\Rev(\price_t,\noinfor)}
    \ge
    \timeHorizon\optNI(\instance)
    -
    \timeHorizon\discrePrecision
    -
    C\pupper\sqrt{\left(2+\frac{\pupper}{\discrePrecision}\right)\timeHorizon\log \timeHorizon} \ge
    \timeHorizon\optNI(\instance)
    -
    O\left(
    \timeHorizon\discrePrecision
    +
    \sqrt{\frac{\timeHorizon\log \timeHorizon}{\discrePrecision}}
    \right)~.
\end{align*}
Choosing the discretization precision
$\discrePrecision=\targetPrecision$ yields
\begin{align*}
    \expect{\sum\nolimits_{t\in[\timeHorizon]}\Rev(\price_t,\noinfor)}
    \ge
    \timeHorizon\optNI(\instance)
    -
    O\left(\timeHorizon^{\sfrac{2}{3}}(\log \timeHorizon)^{\sfrac{1}{3}}\right)~.
\end{align*}
This completes the proof of the regret upper bound as desired. The regret lower bound can be found in \citet{KL-03}.
\end{proof}

\newcommand{\discTyepSpaceSC}{\discTyepSpace^{\SC}}
\newcommand{\discPriceSC}{\discPrice^{\SC}}
\newcommand{\discPriceSCq}{\discPriceSC_{\quality}}
\newcommand{\discretizedOptPriceSC}{\discretizedOptPrice}

\subsection{Dynamic Learning against the Signal-Contingent-Pricing Benchmark}
\label{subsec:sc-benchmark}
In this section, we consider the variant in which the seller can jointly optimize advertising and a price menu that varies with the realized signal.

In this variant, as we show in the proof of \Cref{thm:voa bound} in \Cref{apx:voa proof}, under \Cref{assump:valuation}, the full-information advertising strategy $\fullinfor$ is optimal when the seller can use signal-contingent pricing, in which the price depends on the signal realization. It follows that, conditional on each realized quality, the seller's optimal pricing problem reduces to the classic dynamic pricing problem (e.g., see \Cref{thm:po-benchmark:regret upper bound} and \citealp{KL-03}). Thus, by running the UCB-type dynamic pricing algorithm separately for each realized quality, we obtain a meta-learning algorithm for the advertising and signal-contingent-pricing environment with regret $O\left(\timeHorizon^{\sfrac{2}{3}}(\stateNum\log \timeHorizon)^{\sfrac{1}{3}}\right)$\footnote{\revcolor{Note that the regret bound of the naive meta-learning algorithm can be expressed as $\sum\nolimits_{\quality\in\qualitySpace}O\left(\cdot \timeHorizon_\quality^{\sfrac{2}{3}}(\log \timeHorizon_\quality)^{\sfrac{1}{3}}\right)$ where $\timeHorizon_\quality$ is the time horizon in which quality $\quality$ is realized. Since $\sum\nolimits_{\quality\in\qualitySpace}\timeHorizon_\quality = \timeHorizon$, it is upper bounded by $O\left(\cdot \timeHorizon^{\sfrac{2}{3}}(\stateNum\log \timeHorizon)^{\sfrac{1}{3}}\right)$.}} against the signal-contingent-pricing benchmark $\optSC$. Below, however, we show that our model-based algorithmic framework and regret analysis yield a better learning algorithm, whose regret does not depend on the cardinality $\stateNum$ of the discrete quality space $\qualitySpace$---eliminating the dependence of the naive approach entirely.

\xhdr{Algorithm description} As in the model-based approach for our main model, we consider an algorithm that maintains a UCB estimate of the demand function over a discretized type space and uses it to compute an optimal signal-contingent pricing strategy over a discretized price space. (Recall that, due to the optimality of the full-information advertising policy, there is no need to optimize over the advertising strategy space.) The main difference between this algorithm and \Cref{algo:dynamic pricing and advertising} for the main model is that the present algorithm uses a uniform-grid discretization over the type space and a \emph{quality-dependent non-uniform discretization over the price space}, whereas \Cref{algo:dynamic pricing and advertising} uses a non-uniform discretization over the type space and a uniform discretization over the price space. Specifically, given a discretization parameter $\eps > 0$, we define the discretized type grid $\discTyepSpaceSC$ as
\begin{align*}
    \discTyepSpaceSC \triangleq \{\eps, 2\eps, \dots, 1\}~;
\end{align*}
and for each quality realization $\quality\in\qualitySpace$, we define the quality-dependent price grid $\discPriceSCq$ as
\begin{align*}
    \discPriceSCq \triangleq \{\buyerUtility(\type, \quality)\}_{\type\in\discTyepSpaceSC}~.
\end{align*}
As a sanity check, by construction, for every realized quality $\quality\in\qualitySpace$, the grid $\discPriceSCq$ has cardinality $1/\eps$, and for each discretized price $\price\in \discPriceSCq$, its critical type $\inverseVal(\price, \quality)$ belongs to the discretized grid $\discTyepSpaceSC$.

For each customer $t$, given the realized quality $\quality_t$, the algorithm computes the best estimated revenue using the UCB estimate. Specifically, with slight abuse of notation, we define
$\UCBRev_t(\price\mid \quality) \triangleq
\price\cdot \trueUCBDemand_t(\inverseVal(\price, \quality))$
and the algorithm posts the discretized price $\price_t$ given by
\begin{align*}
    \price_t \in \argmax\nolimits_{\price\in\discPriceSC_{\quality_t}}
    ~
    \UCBRev_t(\price\mid \quality_t)
\end{align*}
where $\trueUCBDemand_t(\cdot)$ follows the same definition and updating rule as in Eqn.~\eqref{defn:true UCB demand} in \Cref{subsec:algorithm}. We summarize our algorithm in \Cref{algo:sc dynamic pricing}. Its running time is polynomial in $1/\eps$.

\begin{algorithm2e}
\revcolor{
\caption{Algorithm for Dynamic Signal-Contingent Pricing with Full-Information Advertising and Demand Learning.}
\label{algo:sc dynamic pricing}
\SetAlgoLined\DontPrintSemicolon
\textbf{Input:} Discretization parameter
$\eps$. \\
For the first $1/\eps$ rounds, for each $x\in\discTyepSpaceSC$,
offer the full-information advertising strategy with the price menu
$\signalPrice(\quality) = \buyerUtility(x, \quality)$, so that the induced critical type satisfies
$\inverseVal(\signalPrice(\quality), \quality) = x$ for every realized quality $\quality$.\\
\nonl\algcomment{Full-information advertising fully reveals the realized quality -- the signal is the quality itself, and the price may depend on it.}\\
\For{each round $t = |\discTyepSpaceSC| + 1, |\discTyepSpaceSC| + 2, \ldots, T$}{
    For all $x\in \discTyepSpaceSC$, compute
    $\trueUCBDemand_{t}(x)$ as defined in Eqn.~\eqref{defn:true UCB demand}. \\
    Offer the full-information advertising strategy $\fullinfor$,
    observe the realized quality $\quality_t$, and post the price
    $\price_t \in \argmax_{\price\in\discPriceSC_{\quality_t}} \UCBRev_t(\price\mid \quality_t)$. \label{line:sc solved solution}\\
    \nonl\algcomment{By construction, $\inverseVal(\price_t, \quality_t)\in\discTyepSpaceSC$ for every $\price_t\in\discPriceSC_{\quality_t}$.}\\
    Observe the customer's purchase
    decision $\decision_t\in\{0, 1\}$. \\
    Update $\left\{
    \chosenSet_{t+1}(x), \chosenCounter_{t+1}(x), \averageDemand_{t+1}(x)
    \right\}_{x\in\discTyepSpaceSC}$.
}}
\end{algorithm2e}

\xhdr{Regret guarantee} Below, we present the regret guarantee of \Cref{algo:sc dynamic pricing}, which is optimal in its dependence on the time horizon. Its analysis follows the same framework as that of \Cref{thm:regret upper bound} for the main model.

\begin{theorem}
\label{thm:sc-benchmark:regret upper bound}
For every instance $\instance$ in which the customers' valuation function satisfies \Cref{assump:valuation},
\Cref{algo:sc dynamic pricing}
has expected regret $O(\timeHorizon^{\sfrac{2}{3}}
(\log\timeHorizon)^{\sfrac{1}{3}})$ against the signal-contingent-pricing benchmark $\optSC$, where $\timeHorizon$ is the time horizon.

Moreover, for every online algorithm, there exists an instance satisfying \Cref{assump:valuation} and $\stateNum = 1$ on which its regret is $\Omega(\timeHorizon^{2/3})$.
\end{theorem}
\begin{proof}
Following the analysis in \Cref{thm:voa bound} (\Cref{apx:voa proof}), the full-information advertising strategy is optimal in the signal-contingent-pricing variant under \Cref{assump:valuation}. The proof of the upper bound proceeds in three steps: we first bound the discretization error, then establish revenue optimism and a single-round revenue gap bound, and finally assemble the regret bound.

\xhdr{Step 1: discretization error}
Fix any instance $\instance$ satisfying \Cref{assump:valuation} and any quality realization $\quality\in\qualitySpace$. We claim that for every price $\price \ge 0$, there exists a grid price $\roundedPrice\in\discPriceSCq$ such that
\begin{align*}
    \Rev(\roundedPrice \mid \quality) \;\ge\; \Rev(\price \mid \quality) - \eps~,
\end{align*}
where $\Rev(\price \mid \quality) \triangleq \price \cdot \demand(\inverseVal(\price, \quality))$ denotes the expected revenue from posting price $\price$ when the realized quality is $\quality$.
Consequently,
\begin{align*}
    \max\nolimits_{\price\in\discPriceSCq} \Rev(\price \mid \quality)
    \;\ge\;
    \sup\nolimits_{\price \ge 0} \Rev(\price \mid \quality) - \eps~,
    \quad
    \forall \quality\in\qualitySpace~,
\end{align*}
and, by the optimality of full-information advertising for the signal-contingent-pricing benchmark, taking expectations over $\quality\sim\prior$ yields
\begin{align}
    \label{eq:sc discretization error}
    \expect[\quality\sim\prior]{\max\nolimits_{\price\in\discPriceSCq} \Rev(\price \mid \quality)}
    \;\ge\;
    \optSC(\instance) - \eps~.
\end{align}
To prove the claim, fix a quality realization $\quality\in\qualitySpace$ and an arbitrary price $\price \ge 0$.
If $\price > \buyerUtility(1, \quality)$, then no customer type purchases (since $\buyerUtility(\type, \quality) \le \buyerUtility(1, \quality) < \price$ for all $\type\in\typeSpace=[0,1]$ by monotonicity of $\buyerUtility(\cdot, \quality)$ in the type), so $\Rev(\price \mid \quality) = 0$ and any grid price $\roundedPrice\in\discPriceSCq$ trivially satisfies the claim.
Henceforth, assume $\price \le \buyerUtility(1, \quality)$.
Let $x \triangleq \inverseVal(\price, \quality) = \min\{\type\in\typeSpace: \buyerUtility(\type, \quality) \ge \price\}$ be the critical type of price $\price$ at quality $\quality$.
Since $\buyerUtility(\cdot, \quality)$ is affine, and hence continuous, in the type, the minimum is attained and
\begin{align}
    \label{eq:sc disc:critical valuation}
    \buyerUtility(x, \quality) \;\ge\; \price~.
\end{align}
Round the critical type down to the grid: let
$
    x^\flat \;\triangleq\; \eps\left\lfloor \frac{x}{\eps} \right\rfloor
    \;\in\; \{0\} \cup \discTyepSpaceSC$,
so that $x^\flat \le x \le x^\flat + \eps$.
Consider the grid price $
    \roundedPrice \;\triangleq\; \buyerUtility(x^\flat, \quality) \;\in\; \discPriceSCq$.
Since $\buyerUtility(\cdot, \quality)$ is increasing in the type, we have $\buyerUtility(\type, \quality) \ge \buyerUtility(x^\flat, \quality)$ if and only if $\type \ge x^\flat$, which means the critical type of the rounded price is exactly the grid point:
\begin{align}
    \label{eq:sc disc:rounded critical type}
    \inverseVal(\roundedPrice, \quality) \;=\; x^\flat~.
\end{align}
Moreover, since $\buyerUtility(\cdot, \quality)$ is $1$-Lipschitz in the type and $x - x^\flat \le \eps$,
\begin{align}
    \label{eq:sc disc:rounded price lower bound}
    \roundedPrice
    \;=\;
    \buyerUtility(x^\flat, \quality)
    \;\ge\;
    \buyerUtility(x, \quality) - (x - x^\flat)
    \;\ge\;
    \price - \eps~,
\end{align}
where the last inequality uses Eqn.~\eqref{eq:sc disc:critical valuation}.
Finally, since the demand function $\demand(\cdot)$ is non-increasing and $x^\flat \le x$, we have $\demand(x^\flat) \ge \demand(x)$. Combining this with Eqn.~\eqref{eq:sc disc:rounded critical type} and Eqn.~\eqref{eq:sc disc:rounded price lower bound},
\begin{align*}
    \Rev(\roundedPrice \mid \quality)
    \;=\;
    \roundedPrice \cdot \demand\big(\inverseVal(\roundedPrice, \quality)\big)
    \;=\;
    \roundedPrice \cdot \demand(x^\flat)
    \;\ge\;
    (\price - \eps)\cdot \demand(x)
    \;=\;
    \Rev(\price \mid \quality) - \eps \cdot \demand(x)
    \;\ge\;
    \Rev(\price \mid \quality) - \eps~.
\end{align*}
Since $\price$ was arbitrary, taking the supremum over $\price \ge 0$ on the right-hand side (and noting the maximum over the finite grid $\discPriceSCq$ on the left-hand side is attained) gives the per-quality statement, and Eqn.~\eqref{eq:sc discretization error} follows by taking expectations over $\quality \sim \prior$ and using $\optSC(\instance) = \expect[\quality\sim\prior]{\sup\nolimits_{\price\ge 0} \Rev(\price \mid \quality)}$.

\xhdr{Step 2: revenue optimism and single-round revenue gap}
We next establish the counterparts of \Cref{lem:optimistic estimates} and \Cref{lem:single-round regret} for \Cref{algo:sc dynamic pricing}. Throughout, we invoke \Cref{lem:high-prob estimation error} with the grid $\discTyepSpaceSC$ in place of $\discTyepSpace$; this is valid because the UCB index $\trueUCBDemand_t(\cdot)$ is defined and updated identically (\eqref{defn:true UCB demand}), and the exploration phase of \Cref{algo:sc dynamic pricing} guarantees $\chosenCounter_t(x) \ge 1$ for every $x\in\discTyepSpaceSC$ and every $t \ge |\discTyepSpaceSC| + 1$.
For each quality $\quality\in\qualitySpace$, let
\begin{align*}
    \discretizedOptPriceSC(\quality)
    \triangleq
    \argmax\nolimits_{\price\in\discPriceSCq} \Rev(\price\mid \quality)
\end{align*}
denote the optimal discretized price at quality $\quality$.
We claim that for every time $t \ge |\discTyepSpaceSC|+1$, with probability at least $1 - \sfrac{1}{T^2}$,
\begin{align}
    \label{eq:sc optimism}
    \Rev(\discretizedOptPriceSC(\quality_t)\mid \quality_t)
    \;\le\;
    \UCBRev_t(\discretizedOptPriceSC(\quality_t)\mid \quality_t)
    \;\le\;
    \UCBRev_t(\price_t\mid \quality_t)~.
\end{align}
To see this, condition on the high-probability event of \Cref{lem:high-prob estimation error} (with grid $\discTyepSpaceSC$), under which $\trueUCBDemand_t(x) \ge \demand(x)$ for every $x\in\discTyepSpaceSC$.
For the first inequality, note that the critical type induced by the discretized optimal price lies on the grid, i.e., $\inverseVal(\discretizedOptPriceSC(\quality_t), \quality_t)\in\discTyepSpaceSC$ by the construction of $\discPriceSC_{\quality_t}$. Therefore,
\begin{align*}
    \UCBRev_t(\discretizedOptPriceSC(\quality_t)\mid \quality_t)
    \;=\;
    \discretizedOptPriceSC(\quality_t)\cdot \trueUCBDemand_t\big(\inverseVal(\discretizedOptPriceSC(\quality_t), \quality_t)\big)
    \;\ge\;
    \discretizedOptPriceSC(\quality_t)\cdot \demand\big(\inverseVal(\discretizedOptPriceSC(\quality_t), \quality_t)\big)
    \;=\;
    \Rev(\discretizedOptPriceSC(\quality_t)\mid \quality_t)~.
\end{align*}
For the second inequality, recall that $\price_t \in \argmax_{\price\in\discPriceSC_{\quality_t}} \UCBRev_t(\price\mid \quality_t)$ maximizes the estimated revenue over the quality-dependent price grid. Since $\discretizedOptPriceSC(\quality_t)\in\discPriceSC_{\quality_t}$, the inequality follows directly from the optimality of~$\price_t$.

We further claim that for every time $t \ge |\discTyepSpaceSC|+1$, with probability at least $1 - \sfrac{1}{T^2}$,
letting $x_t \triangleq \inverseVal(\price_t, \quality_t)\in\discTyepSpaceSC$ denote the critical type induced in round $t$,
\begin{align}
    \label{eq:sc single-round gap}
    \UCBRev_t(\price_t\mid \quality_t) - \Rev(\price_t\mid \quality_t)
    \;\le\;
    \price_t\left(
    2\sqrt{\frac{16\log \timeHorizon}{\chosenCounter_t(x_t)}}
    +
    \frac{2\sqrt{(1+\chosenCounter_t(x_t)) \ln (1+\chosenCounter_t(x_t))}}{\chosenCounter_t(x_t)}
    \right)
    \;\le\;
    12\,\price_t \sqrt{\frac{\log \timeHorizon}{\chosenCounter_t(x_t)}}~.
\end{align}
To see this, condition again on the high-probability event of \Cref{lem:high-prob estimation error} (with grid $\discTyepSpaceSC$). Since $x_t = \inverseVal(\price_t, \quality_t)\in\discTyepSpaceSC$ by the construction of $\discPriceSC_{\quality_t}$, the estimation error bound Eqn.~\eqref{high-prob:UCB upper bound} applies at $x_t$, and hence
\begin{align*}
    \UCBRev_t(\price_t\mid \quality_t) - \Rev(\price_t\mid \quality_t)
    \;=\;
    \price_t\big(\trueUCBDemand_t(x_t) - \demand(x_t)\big)
    \;\le\;
    \price_t\left(
    2\sqrt{\frac{16\log \timeHorizon}{\chosenCounter_t(x_t)}}
    +
    \frac{2\sqrt{(1+\chosenCounter_t(x_t)) \ln (1+\chosenCounter_t(x_t))}}{\chosenCounter_t(x_t)}
    \right)~.
\end{align*}
For the simplified bound, recall that the exploration phase guarantees $\chosenCounter_t(x_t) \ge 1$. Thus, $1 + \chosenCounter_t(x_t) \le 2\chosenCounter_t(x_t)$ and $\ln(1 + \chosenCounter_t(x_t)) \le \ln(2\timeHorizon) \le 2\log \timeHorizon$, which gives
\begin{align*}
    \frac{2\sqrt{(1+\chosenCounter_t(x_t)) \ln (1+\chosenCounter_t(x_t))}}{\chosenCounter_t(x_t)}
    \;\le\;
    \frac{2\sqrt{2\chosenCounter_t(x_t)\cdot 2\log \timeHorizon}}{\chosenCounter_t(x_t)}
    \;=\;
    4\sqrt{\frac{\log \timeHorizon}{\chosenCounter_t(x_t)}}~.
\end{align*}
Combining this with $2\sqrt{16\log \timeHorizon / \chosenCounter_t(x_t)} = 8\sqrt{\log \timeHorizon / \chosenCounter_t(x_t)}$ yields Eqn.~\eqref{eq:sc single-round gap} with constant $12$.

\xhdr{Step 3: assembling the regret bound}
The regret of \Cref{algo:sc dynamic pricing} against the signal-contingent-pricing benchmark is
\begin{align*}
    \Reg{\timeHorizon}
    =
    \timeHorizon\cdot\optSC(\instance)
    -
    \expect{\sum\nolimits_{t\in[\timeHorizon]} \Rev(\price_t \mid \quality_t)}~.
\end{align*}
The exploration phase lasts $|\discTyepSpaceSC|$ rounds, each contributing at most $\pupper$ to the regret. For the remaining rounds, we have
\begin{align}
    \Reg{\timeHorizon}
    & \le \pupper|\discTyepSpaceSC| + \expect{\sum\nolimits_{t=|\discTyepSpaceSC|+1}^{\timeHorizon} \optSC(\instance) - \Rev(\price_t \mid \quality_t)} \nonumber \\
    & \overset{(a)}{\le} \pupper|\discTyepSpaceSC| + \varepsilon \timeHorizon +  \expect{\sum\nolimits_{t=|\discTyepSpaceSC|+1}^{\timeHorizon} \UCBRev_t(\price_t \mid \quality_t) - \Rev(\price_t \mid \quality_t)} \nonumber\\
    & \overset{(b)}{\le}  \pupper|\discTyepSpaceSC| + \varepsilon \timeHorizon +
    12\pupper\expect{\sum\nolimits_{t=|\discTyepSpaceSC| + 1}^{\timeHorizon}
    \sqrt{\frac{\log \timeHorizon}{\chosenCounter_t(\inverseVal(\price_t, \quality_t))}}} \nonumber \\
    & \overset{(c)}{\le}  \pupper|\discTyepSpaceSC| + \varepsilon \timeHorizon +
    12\pupper\,\expect{\sum\nolimits_{x\in\discTyepSpaceSC}
    \sum\nolimits_{n=1}^{\chosenCounter_{\timeHorizon+1}(x)-1}
    \sqrt{\frac{\log \timeHorizon}{n}}} ~,
    \label{eq:sc regret upper bound helper}
\end{align}
where the first inequality follows from the definition of regret and the bound $\pupper$ on the per-round revenue loss during the exploration phase. Inequality (a) follows from Eqn.~\eqref{eq:sc discretization error} and Eqn.~\eqref{eq:sc optimism}: taking expectations over $\quality_t\sim\prior$,
\begin{align*}
    \optSC(\instance)
    \;\le\;
    \expect[\quality\sim\prior]{\Rev(\discretizedOptPriceSC(\quality)\mid \quality)} + \varepsilon
    \;\le\;
    \expect{\UCBRev_t(\price_t \mid \quality_t)} + \varepsilon~,
\end{align*}
where the first step is the discretization error bound and the second step holds on the high-probability event of Eqn.~\eqref{eq:sc optimism} (the failure events across all rounds contribute at most $O(1/\timeHorizon)$ to the expected regret and are absorbed in the $O(\cdot)$ notation).
Inequality (b) follows from Eqn.~\eqref{eq:sc single-round gap} along with the upper bound $\pupper$ on the prices $\price_t$ in all rounds.
For inequality (c), we use that, by construction, the induced critical type $x_t = \inverseVal(\price_t, \quality_t)$ belongs to $\discTyepSpaceSC$ in every round, and that each round increments exactly one counter: $\chosenCounter_{t+1}(x) = \chosenCounter_t(x) + \indicator{x_t = x}$. Consequently, the per-round terms telescope deterministically across rounds: for each grid point $x\in\discTyepSpaceSC$, the summands $\sqrt{\log \timeHorizon/\chosenCounter_t(x_t)}$ over the rounds in which $x_t = x$ are precisely $\sqrt{\log \timeHorizon/n}$ for $n = 1, 2, \ldots, \chosenCounter_{\timeHorizon+1}(x)-1$.

We next bound the third term on the RHS of Eqn.~\eqref{eq:sc regret upper bound helper}. Using $\sum\nolimits_{n\in[N]} n^{-1/2} \le 2\sqrt{N}$ and the Cauchy--Schwarz inequality together with $\sum\nolimits_{x\in\discTyepSpaceSC} \chosenCounter_{\timeHorizon+1}(x) \le \timeHorizon$,
\begin{align*}
    \expect{\sum\nolimits_{x\in\discTyepSpaceSC}
    \sum\nolimits_{n=1}^{\chosenCounter_{\timeHorizon+1}(x)-1}
    \sqrt{\frac{\log \timeHorizon}{n}}}
    & \le
    2\,\expect{\sum\nolimits_{x\in\discTyepSpaceSC}
    \sqrt{\chosenCounter_{\timeHorizon+1}(x)\log \timeHorizon}}\\
    & \le
    2\sqrt{|\discTyepSpaceSC|\cdot \sum\nolimits_{x\in\discTyepSpaceSC}\chosenCounter_{\timeHorizon+1}(x)\cdot \log \timeHorizon}\\
    & \le
    2\sqrt{|\discTyepSpaceSC|\, \timeHorizon \log \timeHorizon}~.
\end{align*}
Substituting this bound into Eqn.~\eqref{eq:sc regret upper bound helper}, we obtain
\begin{align*}
    \Reg{\timeHorizon}
    \;\le\;
    \pupper|\discTyepSpaceSC|
    + \varepsilon \timeHorizon
    + 24\pupper\sqrt{|\discTyepSpaceSC|\, \timeHorizon \log \timeHorizon}
    + O(1)~,
\end{align*}
where the additive $O(1)$ term accounts for the low-probability failure events.
Now, by construction, the grid $\discTyepSpaceSC$ has cardinality $O(\sfrac{1}{\varepsilon})$.
Choosing $\varepsilon = \Theta\big((\sfrac{\log \timeHorizon}{\timeHorizon})^{\sfrac{1}{3}}\big)$ in the above regret bound, all terms are of order $O\big(\timeHorizon^{\sfrac{2}{3}}(\log \timeHorizon)^{\sfrac{1}{3}}\big)$ or smaller, and hence
\begin{align*}
    \Reg{\timeHorizon}
    \;\le\;
    O\left(\timeHorizon^{\sfrac{2}{3}}(\log \timeHorizon)^{\sfrac{1}{3}}\right)~.
\end{align*}

The lower bound follows from the same reduction as in the main model: with $\stateNum = 1$ (i.e., deterministic product quality), advertising reveals no information and signal-contingent pricing coincides with posted pricing, so our model recovers the classic dynamic pricing and learning problem, for which a regret lower bound of $\Omega(\timeHorizon^{\sfrac{2}{3}})$ holds \citep{KL-03}.
\end{proof}

%% file: Paper/apx-proof-complexity.tex
In this section, we prove \Cref{prop:complexity of empricial opt} in \Cref{sec:alg design}.

\complexityofempricialopt*
\begin{proof} 
Since the function $\UCBDemand_t$ is monotone with 
discontinuities at the points in the set $\discTyepSpace$, 
when we fix a price $p\in\discPrice$,
the function $\UCBDemand_t(\inverseVal(p, \cdot))$
is also monotone with discontinuities at the points in $\discPostMeanSpace_p = 
\{\posteriorMean: \inverseVal(p, \posteriorMean) = x\}_{x\in\discTyepSpace}$. 
Given a prior $\prior$, 
optimizing a monotone function with discontinuities over all feasible 
advertising strategies induced from the prior $\prior$ 
subject to the constraint that the support of advertising strategies
must be in the set $\discPostMeanSpace_p$ has been studied in
\cite{ABSY-19,C-19}.
It has been shown that there exists a polynomial (w.r.t.\ the
number of discontinuities) algorithm based on convex programming 
that can find an optimal advertising strategy 
(see Proposition 2 in \citealp{ABSY-19}). 
Thus, an exhaustive search over the discretized price space $\discPrice$
can lead to an optimal solution to the program \ref{empricial opt}. 
\end{proof}

%% file: Paper/apx-proof-analysis.tex
In this section, we present all technical discussion and proofs in \Cref{sec:analysis}.

\subsection{Missing Algorithm and Proofs in Section~{\ref{subsec:rounding}}}
\label{apx:missing procedure}

In this subsection, we provide the formal description of our rounding procedure in Procedure~\ref{algo_construction_adver_general} and the formal analysis of its theoretical guarantees.

\begin{myprocedure}
\caption{$\Round(p, \distOfMean)$: A critical-type guided procedure to round the strategy
$p, \distOfMean$}
\label{algo_construction_adver_general}
\SetAlgoLined\DontPrintSemicolon
\KwIn{$\varepsilon$, a price $p$ such that $p\ge 2\varepsilon$, 
and an advertising $\distOfMean$ 
such that $p, \distOfMean$ satisfy \Cref{lem:binary support} and \Cref{lem:no bad posterior}.}
\KwOut{A price $\newPrice\in\discPrice$,
an advertising $\newAdver$
satisfying $\supp(\newAdver)\subseteq \discPostMeanSpace_{\newPrice}$
}
\textbf{Initialization:} 
Let the set $\newSupp \leftarrow \emptyset$. \\
\nonl\algcomment{The set $\newSupp$ will be used to include the support of the advertising strategy $\newAdver$.}\\
Define price 
$\newPrice \leftarrow \max\{p'\in\discPrice: 
p - 2\varepsilon \le p' \le p - \varepsilon\}$.\\
\For{each posterior mean $\posteriorMean\in \supp(\distOfMean)$}{
\If( ~~
\nonl\algcomment{Namely, for this case
$\inverseVal(\newPrice, \posteriorMean) \in\discTyepSpace$.}){$\posteriorMean\in\discPostMeanSpace_{\newPrice}$}{
    $\newSupp \leftarrow \newSupp \cup\{\posteriorMean\}$,
    and let
    $\newAdver(\posteriorMean) = \distOfMean(\posteriorMean)$, 
    and let $\{i'\in[\stateNum]: \newAdver_{i'}(\posteriorMean) > 0 \} = \{i'\in[\stateNum]: \distOfMean_{i'}(\posteriorMean) > 0 \}$.
}
\Else{
    Suppose $\{i'\in[\stateNum]: \distOfMean_{i'}(\posteriorMean) > 0\} = \{i, j\}$ where $i< j$.\\
    Let $x \triangleq \inverseVal(p, \posteriorMean)$, 
    and let $x^\dagger \triangleq \inverseVal(\newPrice, \posteriorMean)
    \in ((z-1)\varepsilon, z\varepsilon)$ for some
    $z\in \N^+$. \\
    Let 
    $\leftPosteriorMean, \rightPosteriorMean$ satisfy
    $\inverseVal(\newPrice, \leftPosteriorMean) = z\varepsilon$, 
    $\inverseVal(\newPrice, \rightPosteriorMean) = (z-1)\varepsilon$. \\
    Let $\capLeftPosteriorMean\triangleq 
    \leftPosteriorMean \vee \qualityVal_i$, 
    and let 
    $\capRightPosteriorMean\triangleq 
    \rightPosteriorMean \wedge \qualityVal_j$. \\
    $\newSupp \leftarrow \newSupp \cup\{\capLeftPosteriorMean, \capRightPosteriorMean\}$.\\
    \nonl\algcomment{The conditional probabilities below are constructed to satisfy \eqref{bayes consistency}.}\\
    Let 
    $\newAdver_i(\capLeftPosteriorMean)
    = 
    \frac{\qualityVal_j - \capLeftPosteriorMean}{\qualityVal_j - \qualityVal_i}  \frac{1}{\prior_i}  \frac{\distOfMean(\posteriorMean) (\capRightPosteriorMean-\posteriorMean)}{\capRightPosteriorMean-\capLeftPosteriorMean}$ and 
    $\newAdver_i(\capRightPosteriorMean) 
    = 
    \distOfMean_i(\posteriorMean) - \newAdver_i(\capLeftPosteriorMean)$;
    $\newAdver_j(\capLeftPosteriorMean) 
    = 
    \frac{\capLeftPosteriorMean- \qualityVal_i}{\qualityVal_j - \qualityVal_i}  \frac{1}{\prior_j}  \frac{\distOfMean(\posteriorMean) (\capRightPosteriorMean-\posteriorMean)}{\capRightPosteriorMean-\capLeftPosteriorMean}$ 
    and 
    $\newAdver_j(\capRightPosteriorMean) 
    = 
    \distOfMean_j(\posteriorMean) - \newAdver_j(\capLeftPosteriorMean)$.
}
}
\end{myprocedure}

\nobadposterior*
\begin{proof}[Proof of \Cref{lem:no bad posterior}]
Let us fix the optimal price $\optPrice$ and 
optimal advertising $\optAdver$.
If there exists a posterior mean $\posteriorMean\in\supp(\optAdver)$
and $\posteriorMean\notin \qualitySpace$, then from \Cref{lem:binary support}, it must be the case $\{i'\in[\stateNum]: \optAdver_{i'}(\posteriorMean) > 0\} = \{i, j\}$ for some $i < j$
such that $\qualityVal_i < \posteriorMean < \qualityVal_j$.
If $\optPrice > \max_{\type\in\typeSpace} \inverseVal(\type, \posteriorMean)$, 
then it is easy to see that 
the revenue contributed from this posterior mean 
$\optPrice\sum\nolimits_{i}\prior_i\optAdver_i(\posteriorMean)\demand(\inverseVal(\optPrice, \posteriorMean))= 0$.
Thus, decoupling this posterior mean $\posteriorMean$ into the product qualities $\qualityVal_i$ and $\qualityVal_j$ will not lose any revenue.
\end{proof}

\feasibilityadver*
\begin{proof}[Proof of \Cref{lem:feasibility adver}]
$\newPrice\in\discPrice$ holds trivially by construction.
Below, we first show that the output $\newAdver$ is indeed a 
feasible advertising strategy,
and then prove that 
$\inverseVal(\newPrice, \posteriorMean')\in\discTyepSpace$
for every $\posteriorMean'\in\supp(\newAdver)$.
In the analysis below, let the price $p$ and the advertising strategy 
$\distOfMean$ be the input of Procedure \ref{algo_construction_adver_general}, 
and we will focus on an arbitrary posterior mean $\posteriorMean\in\supp(\distOfMean)$ and analyze 
the corresponding construction for $\newAdver$ from 
the posterior mean $\posteriorMean$.

\xhdr{Posterior mean distribution $\newAdver$ as a feasible advertising strategy}
Clearly, a feasible advertising strategy $\newAdver$ 
must satisfy $\newAdver\in\Delta([0, 1])$, i.e., $\newAdver$
is indeed a distribution over $[0, 1]$;
and the associated conditional
distributions $(\newAdver_i)_{i\in[\stateNum]}$ 
must be Bayes-consistent
as defined in \eqref{bayes consistency}.
In the analysis below, by \Cref{lem:no bad posterior}, we assume that 
$\optPrice \le \buyerUtility(1, \posteriorMean)$ for every 
$\posteriorMean \in \supp(\optAdver)$ and $\posteriorMean\notin\qualitySpace$.

We first prove that the constructed 
advertising strategy $\newAdver$ is indeed a feasible distribution.
We focus on the case where $\posteriorMean\notin\discPostMeanSpace_{\newPrice}$.
In this case, we must have $\newPrice \in (\buyerUtility(0, \posteriorMean), \buyerUtility(1, \posteriorMean))$, otherwise it either $\newPrice \le \buyerUtility(0, \posteriorMean)$ or $\newPrice = \buyerUtility(1, \posteriorMean)$, both of which fall into the scenario $\posteriorMean\in \discPostMeanSpace_{\newPrice}$. 
We first show the following claim: 
For any posterior mean
$\posteriorMean\in\supp(\distOfMean)$ with 
$\{i'\in[\stateNum]:\distOfMean_{i'}(\posteriorMean) > 0\} 
= \{i, j\}$, we have $\capLeftPosteriorMean \le \posteriorMean \le \capRightPosteriorMean$.
To see this, 
by definition, we have 
$\buyerUtility(x^\dagger, \posteriorMean) = \newPrice,
\buyerUtility(z\varepsilon, \leftPosteriorMean) = \newPrice,
\buyerUtility((z-1)\varepsilon, \rightPosteriorMean)  = \newPrice$, 
where $x^\dagger \in ((z-1)\varepsilon, z\varepsilon)$.
Under \Cref{assump:valuation} where customer's valuation 
$\buyerUtility(\cdot, \cdot)$ is monotone non-decreasing, 
we know $\leftPosteriorMean \le \posteriorMean \le \rightPosteriorMean$.
Now we show that $\newAdver_i(\capLeftPosteriorMean) \le \distOfMean_i(\posteriorMean)$ 
(a similar analysis can also show that
$\newAdver_j(\capLeftPosteriorMean)  \le \distOfMean_j(\posteriorMean)$). 
To see this, notice that from \Cref{lem:binary support}, 
we must have 
\begin{align*}
    \distOfMean(\posteriorMean)
    = \prior_i \distOfMean_i(\posteriorMean) + 
    \prior_j \distOfMean_j(\posteriorMean)~, \quad
    \frac{\prior_i \distOfMean_i(\posteriorMean)\qualityVal_i + 
    \prior_j \distOfMean_j(\posteriorMean)\qualityVal_j}{\distOfMean(\posteriorMean)}
    =\posteriorMean~.
\end{align*}
Thus, we must have 
$\distOfMean_i(\posteriorMean) = \frac{\distOfMean(\posteriorMean)\cdot (\qualityVal_j - \posteriorMean)}{\prior_i(\qualityVal_j -\qualityVal_i)}$.
Hence, 
\begin{align*}
    \distOfMean_i(\posteriorMean) - \newAdver_i(\capLeftPosteriorMean)
    & = \frac{\distOfMean(\posteriorMean)\cdot(\qualityVal_j - \posteriorMean)}{\prior_i(\qualityVal_j -\qualityVal_i)} 
    - 
    \frac{\qualityVal_j - \capLeftPosteriorMean}{\qualityVal_j - \qualityVal_i} \cdot \frac{1}{\prior_i} \cdot \frac{\distOfMean(\posteriorMean)\cdot(\capRightPosteriorMean-\posteriorMean)}{\capRightPosteriorMean-\capLeftPosteriorMean}\\
    & = \frac{\distOfMean(\posteriorMean)\cdot(\qualityVal_j-\capLeftPosteriorMean)}{\prior_i(\qualityVal_j-\qualityVal_i)}
    \cdot\left(\frac{\qualityVal_j - \posteriorMean}{\qualityVal_j-\capLeftPosteriorMean} - \frac{\distOfMean(\posteriorMean)\cdot(\capRightPosteriorMean-\posteriorMean)}{\capRightPosteriorMean-\capLeftPosteriorMean}\right) \ge 0~,
\end{align*}
where the last inequality follows from the
fact that 
$\qualityVal_i \le \capLeftPosteriorMean 
\le \posteriorMean \le \capRightPosteriorMean
\le \qualityVal_j$.
Together with the fact that $\distOfMean_i(\posteriorMean) \le 1$,
this shows that the value $\newAdver_i(\capLeftPosteriorMean)\in[0, 1]$.

We now argue that in the constructed advertising strategy $\newAdver$,
the summation of 
all conditional probabilities for realizing all  possible posterior means in $\newSupp$ indeed equals $1$.  
Notice that from Procedure \ref{algo_construction_adver_general}, 
for any posterior mean $\posteriorMean\in \supp(\distOfMean)$ with 
$\{i'\in[\stateNum]:\distOfMean_{i'}(\posteriorMean) > 0\} 
= \{\qualityVal_i, \qualityVal_j\}$,  
the constructed advertising strategy $\newAdver$ included 
two posterior means $\capLeftPosteriorMean, \capRightPosteriorMean$, and 
the probabilities for realizing 
posterior means $\capLeftPosteriorMean, \capRightPosteriorMean$ are
$\newAdver(\capLeftPosteriorMean)= \prior_i\newAdver_i(\capLeftPosteriorMean) + \prior_j\newAdver_j(\capRightPosteriorMean)$
(resp.\ $\newAdver(\capRightPosteriorMean)= \prior_i\newAdver_i(\capRightPosteriorMean) + \prior_j\newAdver_j(\capRightPosteriorMean)$). 
By construction, we have $\newAdver(\capLeftPosteriorMean)+\newAdver(\capRightPosteriorMean) = \distOfMean(\posteriorMean)$.
Hence, from
\begin{align*}
    \sum\nolimits_{\posteriorMean\in\supp(\distOfMean)}
    \newAdver(\capLeftPosteriorMean)+\newAdver(\capRightPosteriorMean)
    = \sum\nolimits_{\posteriorMean\in\supp(\distOfMean)} \distOfMean(\posteriorMean) = 1~,
\end{align*}
we know the constructed advertising strategy $\newAdver$ is indeed 
a feasible distribution.

We now show that the constructed advertising strategy $\newAdver$
indeed satisfies the condition \eqref{bayes consistency}. In other words, we want to prove that
for every $\posteriorMean'\in\supp(\newAdver)$, we have 
$\frac{\sum\nolimits_{i\in[\stateNum]} \prior_i \newAdver_i(\posteriorMean)\qualityVal_i}{\sum\nolimits_{i\in[\stateNum]} \prior_i \newAdver_i(\posteriorMean)}
= \posteriorMean'$.
Notice that when $\{i'\in[\stateNum]: \distOfMean_{i'}(\posteriorMean) > 0\} = \{i\}$,
the condition \eqref{bayes consistency} holds trivially.
When $\{i'\in[\stateNum]: \distOfMean_{i'}(\posteriorMean) > 0\} = \{i, j\}$, Procedure~\ref{algo_construction_adver_general} 
adds two posterior means $\capLeftPosteriorMean, \capRightPosteriorMean$
to the support of $\newAdver$. For the posterior mean $\capLeftPosteriorMean$:
\begin{align*}
    \frac{\prior_i\newAdver_i(\capLeftPosteriorMean)\qualityVal_i + \prior_j\newAdver_j(\capLeftPosteriorMean)\qualityVal_j}{\prior_i\newAdver_i(\capLeftPosteriorMean) + \prior_j\newAdver_j(\capLeftPosteriorMean)}
    & = \frac{\frac{\qualityVal_j - \capLeftPosteriorMean}{\qualityVal_j - \qualityVal_i} \cdot  \frac{\distOfMean(\posteriorMean)\cdot(\capRightPosteriorMean-\posteriorMean)}{\capRightPosteriorMean-\capLeftPosteriorMean}\cdot \qualityVal_i + 
    \frac{\capLeftPosteriorMean- \qualityVal_i}{\qualityVal_j - \qualityVal_i} \cdot \frac{\distOfMean(\posteriorMean)\cdot(\capRightPosteriorMean-\posteriorMean)}{\capRightPosteriorMean-\capLeftPosteriorMean}\cdot \qualityVal_j}{\frac{\qualityVal_j - \capLeftPosteriorMean}{\qualityVal_j - \qualityVal_i} \cdot  \frac{\distOfMean(\posteriorMean)\cdot(\capRightPosteriorMean-\posteriorMean)}{\capRightPosteriorMean-\capLeftPosteriorMean} + 
    \frac{\capLeftPosteriorMean- \qualityVal_i}{\qualityVal_j - \qualityVal_i} \cdot \frac{\distOfMean(\posteriorMean)\cdot(\capRightPosteriorMean-\posteriorMean)}{\capRightPosteriorMean-\capLeftPosteriorMean}} \\
    & = 
    \frac{\qualityVal_j - \capLeftPosteriorMean}{\qualityVal_j - \qualityVal_i} \cdot \qualityVal_i + 
    \frac{\capLeftPosteriorMean- \qualityVal_i}{\qualityVal_j - \qualityVal_i} \cdot \qualityVal_j
    = \capLeftPosteriorMean~.
\end{align*}
On the other hand, we observe  
\begin{align*}
    \newAdver(\capLeftPosteriorMean)\capLeftPosteriorMean + \newAdver(\capRightPosteriorMean) \capRightPosteriorMean 
    & = (\prior_i\newAdver_i(\capLeftPosteriorMean) + \prior_j\newAdver_j(\capLeftPosteriorMean)) \capLeftPosteriorMean + 
    (\prior_i(\distOfMean_i(\posteriorMean) - \newAdver_i(\capLeftPosteriorMean)) + \prior_j(\distOfMean_j(\posteriorMean) - \newAdver_j(\capLeftPosteriorMean))) \capRightPosteriorMean\\
    & = (\prior_i\distOfMean_i(\posteriorMean)+\prior_j\distOfMean_j(\posteriorMean))\capRightPosteriorMean
    - (\prior_i\newAdver_i(\capLeftPosteriorMean) + \prior_j\newAdver_j(\capLeftPosteriorMean))(\capRightPosteriorMean-\capLeftPosteriorMean)\\
    & = \distOfMean(\posteriorMean) \capRightPosteriorMean - \left(\frac{\qualityVal_j - \capLeftPosteriorMean}{\qualityVal_j - \qualityVal_i} \cdot  \frac{\distOfMean(\posteriorMean)\cdot(\capRightPosteriorMean-\posteriorMean)}{\capRightPosteriorMean-\capLeftPosteriorMean} + 
    \frac{\capLeftPosteriorMean- \qualityVal_i}{\qualityVal_j - \qualityVal_i} \cdot \frac{\distOfMean(\posteriorMean)\cdot(\capRightPosteriorMean-\posteriorMean)}{\capRightPosteriorMean-\capLeftPosteriorMean}\right) (\capRightPosteriorMean-\capLeftPosteriorMean) \\
    & = \distOfMean(\posteriorMean) \capRightPosteriorMean + \distOfMean(\posteriorMean)(\capRightPosteriorMean-\posteriorMean) = \distOfMean(\posteriorMean) \posteriorMean~.
\end{align*}
Thus, for the posterior mean $\capRightPosteriorMean$, we have
\begin{align*}
    \frac{\prior_i\newAdver_i(\capRightPosteriorMean)\qualityVal_i + \prior_j\newAdver_j(\capRightPosteriorMean)\qualityVal_j}{\prior_i\newAdver_i(\capRightPosteriorMean) + \prior_j\newAdver_j(\capRightPosteriorMean)}
    & = \frac{\prior_i(\distOfMean_i(\posteriorMean) - \newAdver_i(\capLeftPosteriorMean))\qualityVal_i + \prior_j(\distOfMean_j(\posteriorMean) - \newAdver_j(\capLeftPosteriorMean))\qualityVal_j}{\prior_i(\distOfMean_i(\posteriorMean) - \newAdver_i(\capLeftPosteriorMean)) + \prior_j(\distOfMean_j(\posteriorMean) - \newAdver_j(\capLeftPosteriorMean))}\\
    & = 
    \frac{\prior_i\distOfMean_i(\posteriorMean)\qualityVal_i + \prior_j\distOfMean_j(\posteriorMean) \qualityVal_j - \capLeftPosteriorMean \cdot (\prior_i\newAdver_i(\capLeftPosteriorMean) + \prior_j\newAdver_j(\capLeftPosteriorMean))}{\distOfMean(\posteriorMean) - (\prior_i\newAdver_i(\capLeftPosteriorMean) + \prior_j\newAdver_j(\capLeftPosteriorMean))} \\
    & = 
    \frac{\distOfMean(\posteriorMean) \posteriorMean - \newAdver(\capLeftPosteriorMean)\capLeftPosteriorMean }{\distOfMean(\posteriorMean) - 
    \newAdver(\capLeftPosteriorMean)}
    = \frac{\newAdver(\capRightPosteriorMean)\capRightPosteriorMean}{\newAdver(\capRightPosteriorMean)} = \capRightPosteriorMean~.
\end{align*}
We now have shown that the constructed advertising strategy $\newAdver$
indeed satisfies condition \eqref{bayes consistency}. 

\xhdr{Critical type $\inverseVal(\newPrice, \posteriorMean') \in \discTyepSpace$
for every $\posteriorMean'\in\supp(\newAdver)$:}
Fix a posterior mean $\posteriorMean \in \supp(\distOfMean)$.
We focus on the case $\posteriorMean\notin \discPostMeanSpace_{\newPrice}$ (the other case is trivial by construction).
We know that in Procedure \ref{algo_construction_adver_general}, the corresponding posterior means
$\capLeftPosteriorMean, \capRightPosteriorMean \in \supp(\newAdver)$.
Either $\capLeftPosteriorMean = \leftPosteriorMean$ or 
$\capLeftPosteriorMean = \qualityVal_i$, and 
either $\capRightPosteriorMean = \rightPosteriorMean$ or 
$\capRightPosteriorMean = \qualityVal_j$.
When $\capLeftPosteriorMean = \leftPosteriorMean$, we have
$\inverseVal(\newPrice, \capLeftPosteriorMean)
= \inverseVal(\newPrice, \leftPosteriorMean)= z\varepsilon \in \discTyepSpace$. 
When $\capLeftPosteriorMean = \qualityVal_i$,
we have
$\inverseVal(\newPrice, \capLeftPosteriorMean) 
= \inverseVal(\newPrice, \qualityVal_i)\in \discTyepSpace$
as $\newPrice\in\discPrice$.
A similar analysis also shows that $\inverseVal(\newPrice, \capRightPosteriorMean)\in\discTyepSpace$.
The proof completes.
\end{proof}

\revenueguarantee*
\begin{proof}[Proof of \Cref{lem:revenue guarantee}] 
We provide the proof when the input to Procedure 
\ref{algo_construction_adver_general} is $\optPrice, \optAdver$. 
The proof only utilizes the monotonicity of the function $\demand$.
In the analysis below, let $\newPrice, \newAdver$ 
be the advertising strategy output from 
Procedure \ref{algo_construction_adver_general} with the input
$\optPrice, \optAdver$.
We now fix a posterior mean $\posteriorMean\in\supp(\optAdver)$
and consider the following two cases: 

\noindent\textbf{Case 1: $\posteriorMean\in\discPostMeanSpace_{\newPrice}$.}
In this case, we have $\inverseVal(\newPrice, \posteriorMean)
\le \inverseVal(\optPrice, \posteriorMean)$. 

\noindent\textbf{Case 2: $\posteriorMean\notin\discPostMeanSpace_{\newPrice}$.}
Let $\{i'\in[\stateNum]:\optAdver_{i'}(\posteriorMean) > 0\} 
= \{i, j\}$.
Let $\capLeftPosteriorMean$, $\capRightPosteriorMean$
be the corresponding counterparts in the new advertising 
strategy $\newAdver$. 
We now show the following claim:
\begin{claim}
\label{claim:relation}
$\inverseVal(\newPrice, \capRightPosteriorMean)  
\le 
\inverseVal(\newPrice, \capLeftPosteriorMean) 
\le \inverseVal(\optPrice, \posteriorMean)$.
\end{claim}
To see the above claim, recall that in the construction, 
we have 
$\buyerUtility(x, \posteriorMean) \le \optPrice$,
$\buyerUtility(x^\dagger, \posteriorMean) = \newPrice $, and 
by construction, we have $\newPrice \le \optPrice -\varepsilon$.
Thus, by Assumption~\ref{assump:1a}, we have
$\varepsilon \le
\optPrice - \newPrice 
\le \buyerUtility(x, \posteriorMean) - \buyerUtility(x^\dagger, \posteriorMean)
\le x - x^\dagger$,
which implies that 
$z\varepsilon \le x^\dagger + \varepsilon \le x$.
Fix any price $p$. 
From Assumption~\ref{assump:1b}, we know that  the function $\inverseVal(p, \cdot)$ is monotone non-increasing. 
Recall that in previous analysis, we have shown
$\leftPosteriorMean \le \capLeftPosteriorMean \le  
\posteriorMean\le \capRightPosteriorMean\le \rightPosteriorMean$.
We thus have 
$\inverseVal(\newPrice, \capRightPosteriorMean)
\le 
\inverseVal(\newPrice, \capLeftPosteriorMean)
\le 
\inverseVal(\newPrice, \leftPosteriorMean) = z\varepsilon \le x
= \inverseVal(\optPrice, \posteriorMean)~.$

With the above observations, 
we have 
\begin{align*}
    &  \Rev(\optPrice, \optAdver) - 
    \Rev(\newPrice, \newAdver) \\
    = ~ & \optPrice \sum\nolimits_{i\in[\stateNum]} \prior_i \int_0^1 \optAdver_i(\posteriorMean)
    \demand(\inverseVal(\optPrice, \posteriorMean))\ \mathrm{d} \posteriorMean 
    - \newPrice \sum\nolimits_{i\in[\stateNum]} \prior_i \int_0^1 \newAdver_i(\posteriorMean)
    \demand(\inverseVal(\newPrice, \posteriorMean))\ \mathrm{d} \posteriorMean\\
    = ~ & \optPrice \int_0^1 \optAdver(\posteriorMean)
    \demand(\inverseVal(\optPrice, \posteriorMean))\ \mathrm{d} \posteriorMean 
    - \newPrice \int_0^1 \newAdver(\posteriorMean)
    \demand(\inverseVal(\newPrice, \posteriorMean))\ \mathrm{d} \posteriorMean\\
    \overset{(a)}{\le} ~ & 
    \optPrice \int_0^1 \optAdver(\posteriorMean)
    \demand(\inverseVal(\optPrice, \posteriorMean))\ \mathrm{d} \posteriorMean 
    - (\optPrice - 2\varepsilon)  \int_0^1 \newAdver(\posteriorMean)
    \demand(\inverseVal(\newPrice, \posteriorMean))\ \mathrm{d} \posteriorMean\\
    \overset{(b)}{=} ~ & 
    \optPrice \sum\nolimits_{\posteriorMean\in\supp(\optAdver)} \optAdver(\posteriorMean)
    \demand(\inverseVal(\optPrice, \posteriorMean)) - 
    \optPrice \sum\nolimits_{\posteriorMean\in\supp(\optAdver)}
    \left(\newAdver(\capLeftPosteriorMean)\demand(\newPrice, \capLeftPosteriorMean)
    + \newAdver(\capRightPosteriorMean) \demand(\newPrice, \capRightPosteriorMean)\right)
    + 2\varepsilon \\
    \overset{(c)}{=} ~ & 
    \optPrice \sum\nolimits_{\posteriorMean\in\supp(\optAdver)}\left(\optAdver(\posteriorMean) \demand\left(\inverseVal(\optPrice, \posteriorMean)\right) - \left(\newAdver(\capLeftPosteriorMean) \demand\left(\inverseVal(\newPrice, \capLeftPosteriorMean)\right) +
    \newAdver(\capRightPosteriorMean) \demand\left(\inverseVal(\newPrice, \capRightPosteriorMean)\right)\right) \right) + 2\varepsilon \\
    \overset{(d)}{\le}  ~ & 
    \sum\nolimits_{\posteriorMean\in\supp(\optAdver)}\optPrice\left(\optAdver(\posteriorMean) \demand(\inverseVal(\optPrice, \posteriorMean)) - \left(\newAdver(\capLeftPosteriorMean) \demand(\inverseVal(\optPrice, \posteriorMean)) +
    \newAdver(\capRightPosteriorMean) \demand(\inverseVal(\optPrice, \posteriorMean))\right) \right) + 2\varepsilon \\
    \overset{(e)}{=} ~ &
    \sum\nolimits_{\posteriorMean\in\supp(\optAdver)}\optPrice\left(\optAdver(\posteriorMean) \demand(\inverseVal(\optPrice, \posteriorMean)) - \optAdver(\posteriorMean) \demand(\inverseVal(\optPrice, \posteriorMean)) \right) + 2\varepsilon
    = 2\varepsilon~,
\end{align*}
where 
inequality (a) holds since $\newPrice \ge \optPrice-2\varepsilon$;
in equality (b), we, for simplicity, focus on the \textbf{else} case, the analysis for other scenarios is the same;
equality (c) holds by the construction of the 
strategy $\newAdver$,
inequality (d) holds by \Cref{claim:relation};
inequality (e) holds since by construction, we have 
for any $\posteriorMean\in\supp(\newAdver)$, we have 
$\newAdver(\capLeftPosteriorMean)+\newAdver(\capRightPosteriorMean) = \optAdver(\posteriorMean)$.
\end{proof}

\discretizationerror*
\begin{proof}[Proof of \Cref{prop:discretization error}]
Let $\newPrice, \newAdver = \Round(\optPrice, \optAdver)$. 
Then we have  $\Rev(\optPrice, \optAdver) - 
\Rev(\discretizedOptPrice, \discretizedOptAdver)
\le 
\Rev(\optPrice, \optAdver) - 
\Rev(\newPrice, \newAdver) \le 2\varepsilon$
where the first inequality follows from the feasibility guarantee
of price $\newPrice$ and advertising $\newAdver$ 
in \Cref{lem:feasibility adver}
and the definition of
$\discretizedOptPrice, \discretizedOptAdver$, 
and the second inequality follows from revenue guarantee 
in \Cref{lem:revenue guarantee}.
\end{proof}

\subsection{Missing Proofs of Section~\ref{subsec:optimism}}
\label{apx:proofs of regret analysis}

\highprobestimationerror*
To prove \Cref{lem:high-prob estimation error}, we use the following self-normalized martingale tail inequality
to prove the high-probability bounds. 
In particular, we use the following results about the uniform bound for self-normalized bound for martingales in \citet{APS-12}: 
\begin{lemma}[
\citealp{APS-12}]
\label{lem:uniform bound}
Let $\{\filtration_t\}_{t=1}^{\infty}$ be a filtration. 
Let $\{Z_t\}_{t=1}^\infty$ be a sequence of real-valued variables 
such that $Z_t$ is $\filtration_t$-measurable. 
Let $\{\eta_t\}_{t=1}^\infty$ be a sequence of real-valued random variables
such that $\eta_t$ is $\filtration_{t+1}$-measurable and 
is conditionally $R$-sub-Gaussian. 
Let $V > 0$ be deterministic. 
For any $\delta > 0$, 
with probability at least $1-\delta$, for all $t \ge 0$:
\begin{align}
    \label{ineq:uniform bound}
    \left|\sum\nolimits_{s\in[t]} \eta_s Z_s\right| \leq R \sqrt{2\left(V+\sum\nolimits_{s\in[t]} Z_s^2\right) \ln \left(\frac{\sqrt{V+\sum\nolimits_{s\in[t]} Z_s^2}}{\delta \sqrt{V}}\right)}~.
\end{align}
\end{lemma}

\begin{proof}[Proof of \Cref{lem:high-prob estimation error}]
To prove the results, we first show the following concentration
inequality for the empirical demand estimates of the points in 
the set $\discTyepSpace$:
the following holds with probability at least $1 - \sfrac{1}{T^2}$,
\begin{align}
    |\averageDemand_t(x) - \demand(x)| 
    \le 
    \sqrt{\frac{16\log T}{\chosenCounter_t(x)}} + \frac{\sqrt{(1+\chosenCounter_t(x)) \ln (1+\chosenCounter_t(x))}}{\chosenCounter_t(x)}, 
    \quad \forall x\in\discTyepSpace~.
    \label{high-prob:mean error}
\end{align}
To prove the above inequality,
we fix an arbitrary $x\in\discTyepSpace$. 
We define the random variable
$Z_t(x) = 
\indicator{\inverseVal(\offeredPrice_t, \posteriorMean_t) = x}$,
We also define random variable 
$\eta_t(x) = \decision_t(x) - \demand(x)$ if $Z_t(x) = 1$
at time step $t$. 
Then by definition, we know that 
the sequence $\{\sum\nolimits_{s\in[t]} \eta_t(x)Z_t(x)\}$ is a martingale
adapted to $\{\filtration_{t+1}\}_{t=0}^\infty$.
Moreover, the sequence of the variable $\{Z_t(x)\}_{t=1}^\infty$
is $\filtration_t$-measurable, and
the variable $\eta_t(x)$ is $1$-sub-Gaussian.
Now take $V = 1$ and substitute for $\eta_t(x) 
= \decision_t(x) - \demand(x)$, 
apply the uniform bound obtained 
in \Cref{lem:uniform bound}, we have for any $t \ge \numDiscType + 1$,
the following holds with probability at least $1 - \delta$,
\begin{align*}
    \left|\sum\nolimits_{s\in[t]} (\decision_s(x) - \demand(x)) Z_s(x)\right| 
    & \leq  \sqrt{2\left(1+\sum\nolimits_{s\in[t]} Z_s(x)^2\right) \ln \left(\frac{\sqrt{1+\sum\nolimits_{s\in[t]} Z_s(x)^2}}{\delta }\right)}~.
\end{align*}
Observe that in the above inequality, 
the term 
$\left|\sum\nolimits_{s\in[t]} (\decision_s(x) - \demand(x)) Z_s(x)\right|$
is exactly $|\sum\nolimits_{s\in\chosenSet_t(x)} \decision_s(x) - \chosenCounter_t(x)\demand(x)|$, 
and the term $\sum\nolimits_{s\in[t]} Z_s(x)^2$ exactly equals 
to $\chosenCounter_t(x)$.
Dividing both sides with $\chosenCounter_t(x)$, 
substituting for $\sum\nolimits_{s\in[t]} Z_s(x)^2 = \chosenCounter_t(x)$,
we obtain
\begin{align*}
    \left|\averageDemand_t(x) - \demand(x)\right|
    & \leq  
    \frac{1}{\chosenCounter_t(x)}\sqrt{2\left(1+ \chosenCounter_t(x)\right) \ln \left(\frac{\sqrt{1+ \chosenCounter_t(x)}}{\delta }\right)}  \\
    & = 
    \sqrt{\frac{2\left(1+ \chosenCounter_t(x)\right) \ln \frac{1}{\delta}+ \left(1+ \chosenCounter_t(x)\right) \ln \left(1+ \chosenCounter_t(x)\right)}{\chosenCounter_t(x)^2}}  \\
    & \le 
    \sqrt{\frac{4\ln \frac{1}{\delta}}{\chosenCounter_t(x)}}
    + \frac{\sqrt{(1+\chosenCounter_t(x)) \ln (1+\chosenCounter_t(x))}}{\chosenCounter_t(x)}~,
\end{align*}
where in last inequality we use the fact that 
$1 + \chosenCounter_t(x) \le 2\chosenCounter_t(x)$, 
and $\sqrt{u+v} \le \sqrt{u} + \sqrt{v}$ for any $u, v \ge 0$.
Setting $\delta=  T^{-5}$, 
we know that the above inequality 
holds with probability at least $1 - \sfrac{1}{T^5}$.
Taking the union bound over all choices of $t$ and over all
choices of $x\in\discTyepSpace$, 
we obtain that the first statement 
holds with probability at least $1 - \sfrac{1}{T^2}$
as long as $\numDiscType \le T$, which is the case for us.

For the Eqn.~\eqref{high-prob:UCB lower bound},
for notation simplicity, let $\CR_t(x) \triangleq 
\sqrt{\frac{16\log T}{\chosenCounter_t(x)}} + \frac{\sqrt{(1+\chosenCounter_t(x)) \ln (1+\chosenCounter_t(x))}}{\chosenCounter_t(x)}$ 
be the high-probability error, 
and we also write $\discTyepSpace = \{x^{(1)}, \ldots, x^{(\numDiscType)}\}$ where $x^{(i)} < x^{(j)}$ for any $i < j$. 
Now fix an arbitrary $x^{(i)}\in\discTyepSpace$, fix a time round 
$t \ge \numDiscType + 1$.
Denote the random variable 
$\smallIndex = \argmin_{i': i'\le i} 
\averageDemand_t(x^{(i')}) +
\CR_t(x^{(i')}) \wedge 1$. 
\begin{align*}
    \prob{\UCBDemand_t(x^{(i)}) \ge \demand(x^{(i)})}
    & =
    1 - \sum\nolimits_{j = 1}^{i}\prob{\smallIndex = j} \prob{\UCBDemand_t(x^{(i)}) < \demand(x^{(i)})\mid\smallIndex = j} \\
    & = 
    1 - \sum\nolimits_{j = 1}^{i}\prob{\smallIndex = j} \prob{\averageDemand_t(x^{(j)}) + \CR_t(x^{(j)}) < \demand(x^{(i)})\mid\smallIndex = j} \\
    & \overset{(a)}{\ge} 
    1 - \sum\nolimits_{j = 1}^{i}\prob{\smallIndex = j} \prob{\averageDemand_t(x^{(j)}) + \CR_t(x^{(j)}) < \demand(x^{(j)})\mid\smallIndex = j} \\
    & =
    1 - \sum\nolimits_{j = 1}^{i}\prob{\averageDemand_t(x^{(j)}) + \CR_t(x^{(j)}) < \demand(x^{(j)}), \smallIndex = j} \\
    & \ge
    1 - \sum\nolimits_{j = 1}^{i}\prob{\averageDemand_t(x^{(j)}) + \CR_t(x^{(j)}) < \demand(x^{(j)})} \\
    & \overset{(b)}{\ge}  1 - \numDiscType \delta \ge 1 - T^{-4}
\end{align*}
where inequality (a) holds since $\demand(x^{(j)}) \ge \demand(x^{(i)})$ 
for any $j \le i$, 
and inequality (b) holds follows from earlier analysis
where for a fixed $t$ and fixed $x\in\discTyepSpace$, we have
$\prob{\averageDemand_t(x) + \CR_t(x) < \demand(x)} 
\le \delta$.
Taking the union bound over all choices of $t$ and over all
choices of $x\in\discTyepSpace$ finishes the proof.


For the Eqn.~\eqref{high-prob:UCB upper bound}, 
from triangle inequality, 
we have 
\begin{align*}
    \left|\UCBDemand_t(x) - \demand(x)\right|
    & \le 
    \left|\UCBDemand_t(x) - \averageDemand_t(x)\right|+
    \left|\averageDemand_t(x) -  \demand(x)\right|\\
    & \le 
    \sqrt{\frac{16\log T}{\chosenCounter_t(x)}} + \frac{\sqrt{(1+\chosenCounter_t(x)) \ln (1+\chosenCounter_t(x))}}{\chosenCounter_t(x)}
    +
    \left|\averageDemand_t(x) -  \demand(x)\right| \\
    & \overset{(a)}{\le}
    2 \sqrt{\frac{16\log T}{\chosenCounter_t(x)}} + \frac{2\sqrt{(1+\chosenCounter_t(x)) \ln (1+\chosenCounter_t(x))}}{\chosenCounter_t(x)}~,
\end{align*}
where the inequality (a) holds with probability at least 
$1 - \sfrac{1}{T^2}$ according to the first statement we just proved.
\end{proof}

\optimisticestimates*
\begin{proof}[Proof of \Cref{lem:optimistic estimates}]
We begin our analysis by defining the following event.
For all $t = \numDiscType+1,  \ldots, \timeHorizon$, 
define events $\badEvent_t$
\begin{align*}
    \badEvent_t \triangleq 
    \bigcup_{x\in \discTyepSpace} 
    \left\{\UCBDemand_t(x) < \demand(x) \text{ or }
    \UCBDemand_t(x) > \demand(x) + 
    \sqrt{\frac{16\log T}{\chosenCounter_t(x)}} + \frac{\sqrt{(1+\chosenCounter_t(x)) \ln (1+\chosenCounter_t(x))}}{\chosenCounter_t(x)}
    \right\}~.
\end{align*}
From union bound, it follows that
\begin{align*}
    \prob{\badEvent_t} & \le 
    \sum\nolimits_{x\in \discTyepSpace}
    \prob{\UCBDemand_t(x) < \demand(x) \text{ or }
    \UCBDemand_t(x) > \demand(x) + 
    \sqrt{\frac{16\log T}{\chosenCounter_t(x)}} + \frac{\sqrt{(1+\chosenCounter_t(x)) \ln (1+\chosenCounter_t(x))}}{\chosenCounter_t(x)}
    }\\
    & \le  
    \sum\nolimits_{x\in \discTyepSpace}
    \prob{\UCBDemand_t(x) < \demand(x)}+ \\
    & \quad \sum\nolimits_{x\in \discTyepSpace} \prob{\UCBDemand_t(x) > \demand(x) + 
    \sqrt{\frac{16\log T}{\chosenCounter_t(x)}} + 
    \frac{\sqrt{(1+\chosenCounter_t(x)) \ln (1+\chosenCounter_t(x))}}{\chosenCounter_t(x)}
    }\\
    & \overset{(a)}{\le} 
    \frac{2}{T^2}~,
\end{align*}
where the inequality (a) follows from inequalities 
\eqref{high-prob:UCB lower bound}
and \eqref{high-prob:UCB upper bound} in
\Cref{lem:high-prob estimation error}.
Recall that whenever 
$\indicator{\badEventComple_{t}} = 1$, 
we have 
\begin{align*}
    & \Rev(\discretizedOptPrice, \discretizedOptAdver)
    - \UCBRev_t(\discretizedOptPrice, \discretizedOptAdver)\\
    = ~ & 
    \discretizedOptPrice\sum\nolimits_{i\in[\stateNum]} \prior_i \int_0^1 \discretizedOptAdver_i(\posteriorMean)
    \demand(\inverseVal(\discretizedOptPrice, \posteriorMean))\ \mathrm{d} \posteriorMean
    -
    \discretizedOptPrice\sum\nolimits_{i\in[\stateNum]} \prior_i \int_0^1 \discretizedOptAdver_i(\posteriorMean)
    \UCBDemand_t(\inverseVal(\discretizedOptPrice, \posteriorMean))\ \mathrm{d} \posteriorMean \\
    = ~ & \discretizedOptPrice\sum\nolimits_{i\in[\stateNum]} \prior_i \int_0^1 \discretizedOptAdver_i(\posteriorMean)
    \left(\demand(\inverseVal(\discretizedOptPrice, \posteriorMean)) - 
    \UCBDemand_t(\inverseVal(\discretizedOptPrice, \posteriorMean))\right)\ \mathrm{d} \posteriorMean \le 0~.
\end{align*}
Thus, whenever $\indicator{\badEventComple_{t}} = 1$, 
we have  
\begin{align*}
    \Rev(\discretizedOptPrice, \discretizedOptAdver)
    \le 
    \UCBRev_t(\discretizedOptPrice, \discretizedOptAdver)
    \overset{(a)}{\le} 
    \Rev(\offeredPrice_t, \distOfMean_t)~,
\end{align*}
where inequality (a) follows from our algorithm design. 
\end{proof}

\singleroundregret*
\begin{proof}[Proof of \Cref{lem:single-round regret}]
Follow from the definition of the event $\badEventComple_t$, 
when $\indicator{\badEventComple_t} = 1$, we have
\begin{align*}
    & \UCBRev_t(\offeredPrice_t, \offeredAdver_t)
    - \Rev(\offeredPrice_t, \offeredAdver_t) \\
    \le ~ &  
    \offeredPrice_t\sum\nolimits_{i\in[\stateNum]} \prior_i \int_0^1 \distOfMean_{i, t}(\posteriorMean)
    \left(\UCBDemand_t(\inverseVal(\offeredPrice_t, \posteriorMean)) - 
    \demand(\inverseVal(\offeredPrice_t, \posteriorMean)) \right)\ \mathrm{d} \posteriorMean \\
    \overset{(a)}{\le} ~ & 
    \offeredPrice_t\sum\nolimits_{i\in[\stateNum]} \prior_i \int_0^1 \distOfMean_{i, t}(\posteriorMean)
    \left(\sqrt{\frac{16\log T}{\chosenCounter_t(\inverseVal(\offeredPrice_t, \posteriorMean))}} + \frac{\sqrt{(1+\chosenCounter_t(\inverseVal(\offeredPrice_t, \posteriorMean))) \ln (1+\chosenCounter_t(\inverseVal(\offeredPrice_t, \posteriorMean)))}}{\chosenCounter_t(\inverseVal(\offeredPrice_t, \posteriorMean))}\right)\ \mathrm{d} \posteriorMean \\
    \overset{(b)}{\le} ~ & 
    5\offeredPrice_t\sum\nolimits_{i\in[\stateNum]} \prior_i \int_0^1 \distOfMean_{i, t}(\posteriorMean)
    \sqrt{\frac{\log T}{\chosenCounter_t(\inverseVal(\offeredPrice_t, \posteriorMean))}} \ \mathrm{d} \posteriorMean \\
    \overset{(c)}{=} ~ & 
    5\offeredPrice_t 
    \sum\nolimits_{\posteriorMean\in\supp(\offeredAdver_t)} 
    \offeredAdver_t(\posteriorMean)
    \sqrt{\frac{\log T}{\chosenCounter_t(\inverseVal(\offeredPrice_t, \posteriorMean))}}~,
\end{align*}
where inequality (a) follows from the definition of event $\badEventComple_t$, 
inequality (b) follows from the fact that 
$\chosenCounter_t(x) \le T, \forall t$ and thus, 
$\frac{\sqrt{(1+\chosenCounter_t(x)) \ln (1+\chosenCounter_t(x))}}{\chosenCounter_t(x)} \le \sqrt{\frac{\log T}{\chosenCounter_t(x)}}$,
and in last equality (c), we have $\distOfMean_t(\posteriorMean) 
= \sum\nolimits_{i\in[\stateNum]} \prior_i\distOfMean_{i, t}(\posteriorMean)$.
\end{proof}

\subsection{Putting it all together}
\label{subsec:finalproof}
We can now combine the above lemmas to prove 
\Cref{thm:regret upper bound}.
\begin{proof}[Proof of \Cref{thm:regret upper bound}]
We have that with probability at least $1-O(1/T)$,
\begin{align}
    \Reg{T}
    & \le \numDiscType + \expect{\sum\nolimits_{t=\numDiscType+1}^T \Rev(\optPrice, \optAdver) - \Rev(\offeredPrice_t, \offeredAdver_t)} \nonumber \\
    & \overset{(a)}{\le} \numDiscType + 2\varepsilon T +  \expect{\sum\nolimits_{t=\numDiscType+1}^T \UCBRev_t(\offeredPrice_t, \offeredAdver_t) - \Rev(\offeredPrice_t, \offeredAdver_t)} \nonumber\\
    & \overset{(b)}{\le}  \numDiscType + 2\varepsilon T +
    5\pupper\expect{\sum\nolimits_{t=\numDiscType + 1}^T
    \sum\nolimits_{\posteriorMean\in\supp(\offeredAdver_t)} 
    \offeredAdver_t(\posteriorMean)
    \sqrt{\frac{\log T}{\chosenCounter_t(\inverseVal(\offeredPrice_t, \posteriorMean))}}} \nonumber \\
    & \overset{(c)}{=}  \numDiscType + 2\varepsilon T +
    5\pupper\expect{\sum\nolimits_{x\in\discTyepSpace}
    \sum\nolimits_{t=\numDiscType + 1}^T 
    \distOfThreshold_t(x)
    \sqrt{\frac{\log T}{\chosenCounter_t(x)}}} ~,
    \label{eq:regret upper bound helper multiple states}
\end{align}
where the first inequality follows from the definition of regret. Inequality (a), follows from
\Cref{lem:optimistic estimates} and \Cref{prop:discretization error}, and
inequality (b) follows from \Cref{lem:single-round regret} along with upper bound $\pupper$ on prices $\offeredPrice_t$ in all rounds.
For inequality (c), we use that by construction 
$\inverseVal(\offeredPrice_t, \posteriorMean) \in \discTyepSpace$ for every posterior mean $\posteriorMean\in\supp(\offeredAdver_t)$.
and define distribution $\distOfThreshold_t\in\Delta(\discTyepSpace)$ 
over the set $\discTyepSpace$
as 
$$\distOfThreshold_t(x) = \sum\nolimits_{\posteriorMean\in\supp(\offeredAdver_t): \inverseVal(\offeredPrice_t, \posteriorMean) = x}\distOfMean_t(\posteriorMean), \quad  x\in \discTyepSpace~.$$ 
Define Bernoulli random variable $X_t(x) = \indicator{x_t = x}$,  where $x_t=\inverseVal(\offeredPrice_t, \posteriorMean_t)$. Then, from the definition of $\distOfThreshold_t(x)$ observe that $\prob{X_t(x)=1| \chosenCounter_t(x)}= \distOfThreshold_t(x)$. Also, by definition
$$\chosenCounter_{t+1}(x) = 1 + \sum\nolimits_{\ell = \numDiscType+1}^t X_\ell(x) \le 2 N_t(x)~.$$
We use these observations below to obtain a bound on the third term in the RHS of \eqref{eq:regret upper bound helper multiple states}:
\begin{align*}
    \expect{\sum\nolimits_{x\in\discTyepSpace}
    \sum\nolimits_{t=\numDiscType + 1}^T 
    \distOfThreshold_t(x)
    \sqrt{\frac{\log T}{\chosenCounter_t(x)}}} 
    & = \expect{\sum\nolimits_{x\in\discTyepSpace}
    \sum\nolimits_{t=\numDiscType + 1}^T 
    \expect{X_t(x)
    \sqrt{\frac{\log T}{\chosenCounter_t(x)}} \Big|~ \chosenCounter_t(x)}}\\
    & \le  \expect{\sum\nolimits_{x\in\discTyepSpace}
    \sum\nolimits_{t=\numDiscType + 1}^T X_t(x) \sqrt{\frac{2\log T}{\chosenCounter_{t+1}(x)}}}\\
    & = \expect{\sum\nolimits_{x\in\discTyepSpace} \sum\nolimits_{n=2}^{\chosenCounter_{T+1}(x)} \sqrt{\frac{2\log T}{n}}}\\
    & \le \expect{\sum\nolimits_{x\in\discTyepSpace}  \sqrt{ 8 \chosenCounter_{T+1}(x) \log(T)}}
    \le 2 \sqrt{2 \numDiscType T \log(T)}~,
\end{align*}
Substituting this bound in \eqref{eq:regret upper bound helper multiple states}, we obtain that with probability $1-O(1/T)$,
$$ \Reg{T} \le \numDiscType + 2\varepsilon T +
    10\pupper \sqrt{ 2\numDiscType T \log(T)}~.$$
Now, by construction, the set $\discTyepSpace$ has the 
cardinality of $O(\sfrac{\stateNum \pupper}{\varepsilon})$. 
Optimizing $\varepsilon = \Theta((\sfrac{\stateNum\log T}{T})^{\sfrac{1}{3}} )$ in the above regret bound, we have 
$\Reg{T} 
\le 
O\left(T^{\sfrac{2}{3}}(\stateNum\log T)^{\sfrac{1}{3}}\right)$.
\end{proof}

%% file: Paper/6-improvement.tex
In this section, we show that we can improve the regret bounds for \Cref{algo:dynamic pricing and advertising} in the case when valuation function is additive, i.e., $\buyerUtility(\type, \quality) = \type + \quality$. 

First we consider an additional assumption that the product quality domain $\qualitySpace$ is an `equally-spaced set', which includes many natural discrete ordered sets like $\qualitySpace =\{0,1\}$ or $\qualitySpace=[\stateNum]$ that are commonly used in the information design/Bayesian persuasion literature.


\begin{definition}[Equally-spaced sets]
\label{defn:uniformly-spaced}
\label{defn:equally-spaced}
We say that a discrete ordered set $\qualitySpace = \{\qualityVal_1, \ldots, \qualityVal_\stateNum\}$ 
is equally-spaced
if for all $i\in [\stateNum-1]$, $\qualityVal_{i+1} - \qualityVal_{i} = c$ for some constant $c$.
\end{definition}
With this definition, we prove the following improved regret bound.
\begin{restatable}{proposition}{additiveanduniformlyspaced}
\label{additive and uniformly-spaced}
Given an additive valuation function,
$\buyerUtility(\type, \quality) = \type + \quality$, 
and equally-spaced product quality domain, $\qualitySpace$, 
 \Cref{algo:dynamic pricing and advertising} with parameter $\varepsilon = \Theta((\sfrac{\log T}{T})^{\sfrac{1}{3}} \wedge \sfrac{1}{\stateNum})$ has an expected regret of 
$O(T^{\sfrac{2}{3}}(\log T)^{\sfrac{1}{3}} + \sqrt{\stateNum T\log T})$. 
\end{restatable}
Note that a corollary of the above theorem is that the regret is bounded by $O(T^{\sfrac{2}{3}}(\log T)^{\sfrac{1}{3}})$ when $\stateNum \le (\sfrac{ T }{\log T })^{\sfrac{1}{3}}$ and by $O(\sqrt{\stateNum T \log T})$ for larger $\stateNum$.
The high-level idea behind the above result is as follows. In the previous section (see \Cref{subsec:finalproof}) we show that the expected regret of \Cref{algo:dynamic pricing and advertising} is bounded by $O(T\varepsilon + \sqrt{\numDiscType T\log T})$. To prove \Cref{additive and uniformly-spaced} we show that in case of additive valuation
and the equally-spaced qualities, there exists
a discretization parameter $\varepsilon = \Theta((\sfrac{\log T}{T})^{\sfrac{1}{3}} \wedge \sfrac{1}{\stateNum})$ 
such that $\{\inverseVal(p, \quality)\}_{p\in\discPrice, \quality\in\qualitySpace} \subset \{0, \varepsilon, 2\varepsilon, \ldots, 1\}$.
Thus, the constructed set $\discTyepSpace$ satisfies $\numDiscType 
= O(\stateNum + \sfrac{1}{\varepsilon})$. Substituting the value of $\varepsilon$ then gives the result in \Cref{additive and uniformly-spaced}. A formal proof of \Cref{additive and uniformly-spaced} is provided in \Cref{proof of additive and uniformly-spaced}. 

Furthermore, for additive valuation functions, we can  also handle arbitrarily large or continuous quality spaces to obtain an $\tilde O(T^{3/4})$ regret independent of size of quality space $\stateNum$. 
\begin{proposition}
\label{additive and arbitrary}
Given an additive valuation function $\buyerUtility(\type, \quality) = \type + \quality$, and arbitrary (discrete or continuous) product quality space $\qualitySpace$, there exists an algorithm (\Cref{algo:large m} in \Cref{apx:algo large m})
that has expected regret of 
$O(\threeForthsRegret)$. 
\end{proposition}

The proposed \Cref{algo:large m} that achieves the above result is essentially a combination of a pre-processing 
step and \Cref{algo:dynamic pricing and advertising}.
In this pre-processing step, we pool the product qualities that are ``close enough''. This gives us a  
new problem instance with a smaller discrete product quality space 
so that we can apply \Cref{algo:dynamic pricing and advertising}. With additive valuation function, 
we show that this reduction does not incur too much loss in revenue.
A formal description of the algorithm and proof of \Cref{additive and arbitrary} is provided in \Cref{apx:algo large m}.

%% file: Paper/apx-proof-improvement.tex
In the remainder of this section, we present all proofs in \Cref{subsec:improvement}.
\subsection{Proof of \texorpdfstring{\Cref{additive and uniformly-spaced}}{}}
\label{proof of additive and uniformly-spaced}
\additiveanduniformlyspaced*
\begin{proof}[Proof of \Cref{additive and uniformly-spaced}]
For additive valuation, we know 
$\inverseVal(p, \posteriorMean) = ((p - \posteriorMean) \wedge 1) \vee 0$.
Since $\posteriorMean\in[0, 1]$, we know that 
$\highestVal = 2, \lowestVal = 0$.

We first prove the regret $O(T^{\sfrac{2}{3}}(\log T)^{\sfrac{1}{3}})$ 
when $\stateNum \le (\sfrac{ T }{\log T })^{\sfrac{1}{3}} + 1$.
Define the following discretization parameter 
that will be used to define the 
discretized price space $\discPrice$ and the discretized 
type space $\discTyepSpace$ in Eqn.~\eqref{discretized set defn}.
\begin{align}
    \label{eq:integer eps}
    \varepsilon = \max\left\{\varepsilon' \ge 0: \frac{\sfrac{1}{\stateNum}}{\varepsilon'} \in \N^+ \wedge \varepsilon' \le  \left(\frac{\log  T }{ T }\right)^{\sfrac{1}{3}}\right\}~.
\end{align}
We now argue that the above $\varepsilon = 
\Theta(\targetPrecision)$. 
To see this, let the integers $k_1, k_2\in\N^+$ satisfy
\begin{align*}
    \left \lfloor \frac{\sfrac{1}{(\stateNum-1)}}{(\frac{\log  T }{ T })^{\sfrac{1}{3}}}\right \rfloor
    = k_1, \quad
    \left \lfloor \frac{\sfrac{1}{(\stateNum-1)}}{\frac{1}{2}(\frac{\log  T }{ T })^{\sfrac{1}{3}}}\right \rfloor
    = k_2~. 
\end{align*}
By assumption, we have $\frac{1}{\stateNum-1} \ge \targetPrecision$, 
implying $k_1 \ge 1$,  and $k_2 \ge 2$.
Thus, there must exist an $\varepsilon'\in[
(\frac{1}{2}\targetPrecision, 
(\sfrac{\log  T }{ T })^{\sfrac{1}{3}}]$ such that 
$\frac{\sfrac{1}{(\stateNum-1)}}{\varepsilon'} \in [k_1:k_2]$, which
implies that the above defined $\varepsilon = 
\Theta(\targetPrecision)$.
Suppose $K_\varepsilon\in\N^+$ such that $K_\varepsilon\varepsilon = \sfrac{1}{(\stateNum-1)}$.
By definition of uniformly-spaced qualities, we know 
that $\qualityVal_i = \frac{i-1}{(\stateNum-1)}, \forall i \ge 2$. 
For a discretized price space 
$\discPrice = \{\varepsilon, 2\varepsilon, 
\ldots, 2-\varepsilon, 2\}$, 
we know that for any price $p = k_p\varepsilon \in \discPrice$ 
for some integer $k_p\in\N^+$, we have
$\inverseVal(k_p\varepsilon, \qualityVal_i) = k_p\varepsilon - \qualityVal_i 
= k_p\varepsilon - (i-1)K_\varepsilon\varepsilon
\in \{0, \varepsilon, \ldots, 1\}$.
Thus, for the set $\discTyepSpace$ defined in Eqn.~\eqref{discretized set defn}
we have $|\discTyepSpace| = O(\sfrac{1}{\varepsilon})$. 
With $\varepsilon$ defined in  Eqn.~\eqref{eq:integer eps},  
\Cref{algo:dynamic pricing and advertising} 
has the desired regret upper bound.

We now prove the regret 
$O(\sqrt{\stateNum T\log T})$
when number of qualities $\stateNum> (\sfrac{ T }{\log T })^{\sfrac{1}{3}} + 1$. 
For this case, we can simple feed \Cref{algo:dynamic pricing and advertising}
with discretization parameter $\varepsilon = \sfrac{1}{(\stateNum - 1)}$. 
Then, according to the proof of \Cref{thm:regret upper bound}, the regret of \Cref{algo:dynamic pricing and advertising} can be bounded as
$O(\sfrac{T}{\stateNum} + \sqrt{T\stateNum\log T})
= O(\sqrt{T\stateNum\log T})$ as desired.
\end{proof}

\subsection{Algorithm and Proof of \texorpdfstring{\Cref{additive and arbitrary}}{}}
\label{apx:algo large m}
The detailed algorithm description when the number of qualities is large
is provided in \Cref{algo:large m}.
\begin{algorithm2e}
\caption{Algorithm for arbitrary size $\stateNum$ 
of product quality space.}
\label{algo:large m}
\SetAlgoLined\DontPrintSemicolon
\textbf{Input:} Discretization parameter 
$\discrePrecision$ and pooling precision parameter $\poolPrecision$. \\
\textbf{Input:} Instance $\instance$ with quality space $\qualitySpace$ 
and prior $\prior$.\\ 
Construct instance $\newInstance$ as follows:
Let the quality space $\newQualitySpace = \{\pooledQuality_i\}_{i\in[\left \lceil \sfrac{1}{\poolPrecision} \right \rceil +1]}$ where $\pooledQuality_1 = 0, \newPrior_1 = \prior_1$; and 
$\pooledQuality_{i+1}  = \expect[\quality\sim\prior]{\quality\mid \quality \in ((i-1)\poolPrecision, i\poolPrecision]}$, 
and let the prior $\newPrior = (\newPrior_i)_{i\in[\left \lceil \sfrac{1}{\poolPrecision} \right \rceil + 1]}$ where $\newPrior_{i+1} =
\prob[\quality\sim\prior]{\quality\in ((i-1)\poolPrecision, i\poolPrecision]}$ for all $1 \le i \le \left \lceil \sfrac{1}{\poolPrecision} \right \rceil$.\label{line:construction new instance} \\
Run \Cref{algo:dynamic pricing and advertising} on instance $\newInstance$ with discretization parameter $\varepsilon$.
\end{algorithm2e}

Below, we provide a regret bound that is independent of the size of quality space
and it holds for a valuation function beyond the additive one as long as it satisfies 
the following assumption:
\begin{assumption}
\label{assump:kappa Lipschitz on q}
Function $\inverseVal(p, \cdot)$ satisfies that
for any price $p\in [0, \pupper]$, 
for any $\posteriorMean_1, \posteriorMean_2$ 
where $\posteriorMean_1 \le \posteriorMean_2$, 
$\inverseVal(p, \posteriorMean_1) - \inverseVal(p, \posteriorMean_2)
\le \posteriorMean_2 - \posteriorMean_1$. \footnote{We can also 
relax the assumption to be $\inverseVal(p, \posteriorMean_1) - \inverseVal(p, \posteriorMean_1)
\le L(\posteriorMean_2 - \posteriorMean_1)$ where an arbitrary
constant $L \in \R^+$ can be treated similarly.}
\end{assumption}
Notice that the additive valuation  $\buyerUtility(\type, \quality) = \type + \quality$, which has $\inverseVal(p, \posteriorMean) = p - \posteriorMean$,
satisfies the above assumption.

\begin{proposition}
\label{prop:large number qualities}
With \Cref{assump:valuation} and \Cref{assump:kappa Lipschitz on q},
\Cref{algo:large m} with $\poolPrecision = \varepsilon =\poolThreshold$ has an expected regret of $O(\threeForthsRegret)$
independent of the size $\stateNum$ of quality space.
\end{proposition}
Given the above \Cref{prop:large number qualities},
\Cref{additive and arbitrary} simply follows as the additive valuation function satisfies \Cref{assump:kappa Lipschitz on q}.
\begin{proof}[Proof of \Cref{prop:large number qualities}]
We fix a small $\poolPrecision \in(0, 1)$.
Let $\instance$ be an instance 
with quality space $\qualitySpace$ 
and prior $\prior\in\Delta^\qualitySpace$.
For expository simplicity, let us assume that 
for each $i \in [\left \lceil \sfrac{1}{\poolPrecision} \right \rceil]$,
there exists at least one quality $\quality\in \qualitySpace$
such that $\quality \in ((i-1)\poolPrecision, i\poolPrecision]$.
We now construct a new instance 
$\newInstance$ with quality space 
$\newQualitySpace = (\pooledQuality_i)_{i\in[\left \lceil \sfrac{1}{\poolPrecision} \right \rceil + 1]}$ and prior $\newPrior = (\newPrior_i)_{i\in[\left \lceil \sfrac{1}{\poolPrecision} \right \rceil  + 1]}$ as follows: 
\begin{itemize}[leftmargin=*]
    \item for $i = 1$: $\pooledQuality_i = 0, \newPrior_i = \prior_1$; 
    \item for $2 \le i \le 
    \left \lceil \sfrac{1}{\poolPrecision} \right \rceil + 1$: 
    $\pooledQuality_i = \expect[\quality\sim\prior]{\quality\mid \quality\in((i-2)\poolPrecision, (i-1)\poolPrecision]}, \newPrior_i = \prob[\quality\sim\prior]{\quality\in((i-2)\poolPrecision, (i-1)\poolPrecision]}$.
\end{itemize}
Essentially, the instance $\newInstance$ is constructed 
by pooling all product qualities that are ``close enough'' to each other (i.e., qualities in a grid $((i-1)\poolPrecision, i\poolPrecision]$). By construction, we know that $|\newQualitySpace| = O(\sfrac{1}{\poolPrecision})$.
Given a price $p$ and an advertising $\distOfMean$, 
let $\RevProb[\instance]{p, \distOfMean}$ be the seller's revenue
for problem instance $\instance$.
Below, we have the following revenue guarantee between 
these two problem instances $\instance, \newInstance$.
\begin{lemma}
\label{lem:pooling revenue loss}
Let  $\optPrice, \optAdver$ be the optimal price
and optimal advertising for instance $\instance$, 
with \Cref{assump:kappa Lipschitz on q},
there exists a price $\newPrice$ and advertising
$\newAdver$ for instance $\newInstance$ such that 
$\RevProb[\instance]{\optPrice, \optAdver} \le 
\RevProb[\newInstance]{\newPrice, \newAdver} + \poolPrecision$.
\end{lemma}
The proof of the above \Cref{lem:pooling revenue loss}
utilizes \Cref{assump:kappa Lipschitz on q}
and is provided subsequently.
With \Cref{lem:pooling revenue loss}, 
by feeding \Cref{algo:dynamic pricing and advertising} with new 
instance $\newInstance$, the total expected regret for instance $\instance$
can be bounded as follows
\begin{align*}
    \Reg[\instance]{ T }
    \le O\left(
     T  \poolPrecision +  T  \varepsilon + \sqrt{\numDiscType T \log  T }
    \right)
    = 
    O\left(
     T  \poolPrecision +  T  \varepsilon + \sqrt{\frac{1}{\varepsilon\poolPrecision} T \log  T }
    \right)
    \le O\left(\threeForthsRegret\right)
\end{align*}
where the term $ T  \poolPrecision$ is from \Cref{lem:pooling revenue loss} 
and due to reducing the instance $\instance$ to the new instance $\newInstance$, 
the term $ T  \varepsilon + \sqrt{\numDiscType T \log  T }$ is the incurred regret of \Cref{algo:dynamic pricing and advertising}
for the new instance $\newInstance$ where 
the number of discretized types
$\numDiscType$ for the new instance $\newInstance$ equals $\frac{1}{\poolPrecision\varepsilon}$, and in the last inequality, we 
choose $\poolPrecision = \varepsilon = \poolThreshold$.
\end{proof}

Below, we provide the proof for \Cref{lem:pooling revenue loss}.
\begin{proof}[Proof of \Cref{lem:pooling revenue loss}]
Let us fix the problem instance $\instance$
with quality space $\qualitySpace, |\qualitySpace| = m$
and prior distribution $\prior$. 
Let $\newInstance$ be the constructed instance 
(see \Cref{line:construction new instance}).
In the proof, we construct a price 
$\newPrice$ and an advertising strategy
$\newAdver$ for instance $\newInstance$ 
based on $\optPrice, \optAdver$.
Consider a price $\newPrice = \optPrice - \poolPrecision$. 
Below, we show that how to construct advertising strategy $\newAdver$
from the advertising strategy $\optAdver$. 
In particular, for each posterior mean $\posteriorMean\in\supp(\optAdver)$, 
we construct a corresponding posterior mean $\newPosteriorMean\in\supp(\newAdver)$, and furthermore, 
with \Cref{assump:valuation} and \Cref{assump:kappa Lipschitz on q}, 
we also show that we always have
$\inverseVal(\optPrice, \posteriorMean) \ge \inverseVal(\newPrice, \newPosteriorMean)$. 
Recall that from \Cref{lem:binary support}, 
the advertising strategy $\optAdver$ satisfies 
$\{i\in[\stateNum]: \optAdver_{i}(\posteriorMean) > 0\} \le 2$
for all $\posteriorMean\in\supp(\optAdver)$.
Our construction is based on threes cases of 
$\{i\in[\stateNum]: \optAdver_{i}(\posteriorMean) > 0\}$.
\begin{itemize}[leftmargin=*]
    \item \textbf{Case 1 --}
    if $\{i\in[\stateNum]: \optAdver_{i}(\posteriorMean) > 0\} = \{i'\}$, in this case, suppose 
    $\qualityVal_{i'} \in ((j-1)\poolPrecision, j\poolPrecision]$
    for some $j\in[\left \lceil \sfrac{1}{\poolPrecision} \right \rceil]$, 
    then consider 
    \begin{align*}
        \newAdver_{j+1}(\newPosteriorMean) = \frac{\prior_{i'}\optAdver_{i'}(\posteriorMean)}{\newPrior_{j+1}}; \quad
        \text{where }
        \newPosteriorMean = \pooledQuality_{j+1}.
    \end{align*}
    From the above construction, we know that 
    $\inverseVal(\optPrice, \posteriorMean) = \inverseVal(\optPrice, \qualityVal_{i'})$, and 
    \begin{align*}
        \inverseVal(\optPrice, \qualityVal_{i'})
        \overset{(a)}{\ge}
        \inverseVal(\newPrice, \qualityVal_{i'}) + \poolPrecision 
        \overset{(b)}{\ge}
        \inverseVal(\newPrice, \pooledQuality_{j+1})
        =\inverseVal(\newPrice, \newPosteriorMean)~,
    \end{align*}
    where inequality (a) holds since 
    $\poolPrecision = \optPrice - \newPrice \le \inverseVal(\optPrice, \qualityVal_{i'}) - \inverseVal(\newPrice, \qualityVal_{i'})$
    due to Assumption \ref{assump:1a},
    and inequality (b) holds since 
    $|\inverseVal(\newPrice, \pooledQuality_{j+1}) - \inverseVal(\newPrice, \qualityVal_{i'})|
    \le |\pooledQuality_{j+1} - \qualityVal_{i'}| \le \poolPrecision$
    due to \Cref{assump:kappa Lipschitz on q}.
    
    \item  \textbf{Case 2 --}
    if $\{i\in[\stateNum]: \optAdver_{i}(\posteriorMean) > 0\} = \{i', i''\}$ where $i' < i''$, in this case, suppose both
    $\qualityVal_{i'}, \qualityVal_{i''} \in ((j-1)\poolPrecision, j\poolPrecision]$
    for some $j \in[\left \lceil \sfrac{1}{\poolPrecision} \right \rceil]$,
    then consider 
    \begin{align*}
        \newAdver_{j+1}(\newPosteriorMean) = \frac{\prior_{i'}\optAdver_{i'}(\posteriorMean) + \prior_{i''}\optAdver_{i''}(\posteriorMean)}{\newPrior_{j+1}}; \quad
        \text{where }
        \newPosteriorMean = \pooledQuality_{j+1}.
    \end{align*}
    From the above construction, we know that 
    \begin{align*}
        \inverseVal(\optPrice, \posteriorMean)
        \overset{(a)}{\ge}
        \inverseVal(\newPrice, \posteriorMean) + \poolPrecision 
        \overset{(b)}{\ge}
        \inverseVal(\newPrice, \pooledQuality_{j+1})
        =\inverseVal(\newPrice, \newPosteriorMean)
    \end{align*}
    where inequality (a) holds since 
    $\poolPrecision = \optPrice - \newPrice \le \inverseVal(\optPrice, \posteriorMean) - \inverseVal(\newPrice, \posteriorMean)$
    due to Assumption \ref{assump:1a},
    and inequality (b) holds since 
    $|\inverseVal(\newPrice, \pooledQuality_{j+1}) - \inverseVal(\newPrice, \posteriorMean)|
    \le |\pooledQuality_{j+1} - \posteriorMean| \le \poolPrecision$
    due to \Cref{assump:kappa Lipschitz on q} and 
    the fact that 
    $\posteriorMean = \frac{\prior_{i'}\optAdver_{i'}(\posteriorMean)\qualityVal_{i'} + \prior_{i''}\optAdver_{i''}(\posteriorMean)\qualityVal_{i''}}{\prior_{i'}\optAdver_{i'}(\posteriorMean) + \prior_{i''}\optAdver_{i''}(\posteriorMean)} \in ((j-1)\poolPrecision, j\poolPrecision]$.
    
    \item  \textbf{Case 3 --}
    if $\{i\in[\stateNum]: \optAdver_{i}(\posteriorMean) > 0\} = \{i', i''\}$ where $i' < i''$, in this case, suppose 
    $\qualityVal_{i'} \in ((j'-1)\poolPrecision, j'\poolPrecision]$
    and 
    $\qualityVal_{i''} \in ((j''-1)\poolPrecision, j''\poolPrecision]$
    for some $j', j''\in[\left \lceil \sfrac{1}{\poolPrecision} \right \rceil]$ where $j' < j''$,
    then consider 
    \begin{align*}
        \newAdver_{j'+1}(\newPosteriorMean) = \frac{\prior_{i'}\optAdver_{i'}(\posteriorMean)}{\newPrior_{j'+1}}~, \quad
        \newAdver_{j''+1}(\newPosteriorMean) = \frac{\prior_{i''}\optAdver_{i''}(\posteriorMean)}{\newPrior_{j''+1}}~; 
        \quad 
        \text{where }
        \newPosteriorMean = \frac{\newPrior_{j'+1}\newAdver_{j'+1}(\newPosteriorMean)\pooledQuality_{j'+1}+\newPrior_{j''+1}\newAdver_{j''+1}(\newPosteriorMean)\pooledQuality_{j''+1}}{\newPrior_{j'+1}\newAdver_{j'+1}(\newPosteriorMean)+\newPrior_{j''+1}\newAdver_{j''+1}(\newPosteriorMean)}~.
    \end{align*}
    From the above construction, we know that 
    \begin{align*}
        \inverseVal(\optPrice, \posteriorMean)
        \overset{(a)}{\ge}
        \inverseVal(\newPrice, \posteriorMean) + \poolPrecision 
        \overset{(b)}{\ge}
        \inverseVal(\newPrice, \newPosteriorMean)~,
    \end{align*}
    where inequality (a) holds since 
    $\poolPrecision = \optPrice - \newPrice \le \inverseVal(\optPrice, \posteriorMean) - \inverseVal(\newPrice, \posteriorMean)$
    due to Assumption \ref{assump:1a},
    and inequality (b) holds due to Assumption \ref{assump:kappa Lipschitz on q}
    and the following fact:
    \begin{align*}
        |\posteriorMean - \newPosteriorMean|
        & = 
        \left|
        \frac{\prior_{i'}\optAdver_{i'}(\posteriorMean)\qualityVal_{i'} + 
        \prior_{i''}\optAdver_{i''}(\posteriorMean)\qualityVal_{i''}}{\prior_{i'}\optAdver_{i'}(\posteriorMean) + 
        \prior_{i''}\optAdver_{i''}(\posteriorMean)}
        -\frac{\newPrior_{j'+1}\newAdver_{j'+1}(\newPosteriorMean)\pooledQuality_{j'+1}+\newPrior_{j''+1}\newAdver_{j''+1}(\newPosteriorMean)\pooledQuality_{j''+1}}{\newPrior_{j'+1}\newAdver_{j'+1}(\newPosteriorMean)+\newPrior_{j''+1}\newAdver_{j''+1}(\newPosteriorMean)}\right| \\
        & = 
        \left|
        \frac{\prior_{i'}\optAdver_{i'}(\posteriorMean)\qualityVal_{i'} + 
        \prior_{i''}\optAdver_{i''}(\posteriorMean)\qualityVal_{i''}}{\prior_{i'}\optAdver_{i'}(\posteriorMean) + 
        \prior_{i''}\optAdver_{i''}(\posteriorMean)}
        -\frac{\prior_{i'}\optAdver_{i'}(\posteriorMean)\pooledQuality_{j'+1}+\prior_{i''}\optAdver_{i''}(\posteriorMean)\pooledQuality_{j''+1}}{\prior_{i'}\optAdver_{i'}(\posteriorMean) + 
        \prior_{i''}\optAdver_{i''}(\posteriorMean)}\right| \\
        & \le 
        \frac{\prior_{i'}\optAdver_{i'}(\posteriorMean) |\pooledQuality_{j'+1} - \qualityVal_{i'}|+ 
        \prior_{i''}\optAdver_{i''}(\posteriorMean)
        |\pooledQuality_{j''+1} - \qualityVal_{i''}|}{\prior_{i'}\optAdver_{i'}(\posteriorMean) + 
        \prior_{i''}\optAdver_{i''}(\posteriorMean)}  \\
        & \le 
        \frac{\prior_{i'}\optAdver_{i'}(\posteriorMean) \poolPrecision+ 
        \prior_{i''}\optAdver_{i''}(\posteriorMean)
        \poolPrecision}{\prior_{i'}\optAdver_{i'}(\posteriorMean) + 
        \prior_{i''}\optAdver_{i''}(\posteriorMean)} = \poolPrecision~.
    \end{align*}
\end{itemize}
We also note that by construction, for any 
posterior mean $\posteriorMean\in\supp(\optAdver)$, the corresponding constructed posterior mean $\newPosteriorMean\in\supp(\newAdver)$ satisfies: 
\begin{align}
    \label{eq:equal prob}
    \newAdver(\newPosteriorMean) 
    = \sum\nolimits_{i\in [\left \lceil \sfrac{1}{\poolPrecision} \right \rceil +1]} \newAdver_i(\newPosteriorMean) \newPrior_i
    = \optAdver(\posteriorMean)~.
\end{align}
Armed with the above observation $\inverseVal(\optPrice, \posteriorMean) \ge 
\inverseVal(\newPrice, \newPosteriorMean)$, 
we are now ready to show 
$\RevProb[\instance]{\optPrice, \optAdver} \le 
\RevProb[\newInstance]{\newPrice, \newAdver} + \poolPrecision$:
\begin{align*}
    \RevProb[\instance]{\optPrice, \optAdver} - 
    \RevProb[\newInstance]{\newPrice, \newAdver}
    & =  \optPrice 
    \int_\posteriorMean\optAdver(\posteriorMean)\demand(\inverseVal(\optPrice, \posteriorMean))\ \mathrm{d} \posteriorMean  -
    \newPrice \int_{\newPosteriorMean}\newAdver(\newPosteriorMean)\demand(\inverseVal(\newPrice, \newPosteriorMean))\ \mathrm{d} \newPosteriorMean\\
    & \overset{(a)}{\le} 
    \optPrice 
    \int_\posteriorMean\optAdver(\posteriorMean)\demand(\inverseVal(\optPrice, \posteriorMean))\ \mathrm{d} \posteriorMean  -
    \optPrice \int_{\newPosteriorMean}\newAdver(\newPosteriorMean)\demand(\inverseVal(\newPrice, \newPosteriorMean))\ \mathrm{d} \newPosteriorMean  + \poolPrecision \\
    & = \optPrice \left(\int_\posteriorMean\optAdver(\posteriorMean)\demand(\inverseVal(\optPrice, \posteriorMean))\ \mathrm{d} \posteriorMean 
    - \int_{\newPosteriorMean}\newAdver(\newPosteriorMean)\demand(\inverseVal(\newPrice, \newPosteriorMean))\ \mathrm{d} \newPosteriorMean \right)
    + \poolPrecision \\
    & \overset{(b)}{\le}  \poolPrecision
\end{align*}
where inequality (a) holds since we have $\newPrice = \optPrice - \poolPrecision$,
and inequality (b) holds by the observation  
$\inverseVal(\optPrice, \posteriorMean) \ge 
\inverseVal(\newPrice, \newPosteriorMean)$
and Eqn.~\eqref{eq:equal prob}.
\end{proof}

%% file: mybib.bib
@article{GKP-26,
  author  = {Gurkan, Huseyin and Keskin, N. Bora and Parker, Rodney P.},
  title   = {Dynamic Learning for Joint Pricing, Advertising, and Inventory Management},
  journal = {Manufacturing \& Service Operations Management},
  year    = {2026},
  note    = {Forthcoming}
}

@inproceedings{FT-25a,
  title={Persuasive calibration},
  author={Feng, Yiding and Tang, Wei},
  booktitle={Proceedings of the 2026 Annual ACM-SIAM Symposium on Discrete Algorithms (SODA)},
  pages={4660--4691},
  year={2026},
  organization={SIAM}
}

@MISC{AmazonCTR,
  author       = {Saggar, Suresh and Chen, Bole and Chen, Wei},
  title        = {{Advertising click-prediction modeling on Amazon EKS}},
  year         = 2021,
  url          = {https://aws.amazon.com/cn/blogs/containers/advertising-click-prediction-modeling-on-amazon-eks/},
  note         = {Accessed: 2025-11}
}

@inproceedings{YQWBN-22,
  title={Scale calibration of deep ranking models},
  author={Yan, Le and Qin, Zhen and Wang, Xuanhui and Bendersky, Michael and Najork, Marc},
  booktitle={Proceedings of the 28th ACM SIGKDD Conference on Knowledge Discovery and Data Mining},
  pages={4300--4309},
  year={2022}
}

@inproceedings{KLST-23,
  title={U-calibration: Forecasting for an unknown agent},
  author={Kleinberg, Bobby and Leme, Renato Paes and Schneider, Jon and Teng, Yifeng},
  booktitle={The Thirty Sixth Annual Conference on Learning Theory},
  pages={5143--5145},
  year={2023},
  organization={PMLR}
}

@inproceedings{JP-24,
  title={Calibrated Forecasting and Persuasion},
  author={Jain, Atulya and Perchet, Vianney},
  booktitle={The Twenty-Fifth ACM Conference on Economics and Computation (EC'24)},
  year={2024}
}

@article{FV-98,
  title={Asymptotic calibration},
  author={Foster, Dean P and Vohra, Rakesh V},
  journal={Biometrika},
  volume={85},
  number={2},
  pages={379--390},
  year={1998},
  publisher={Oxford University Press}
}

@article{ES-07,
  title={Optimal information disclosure in auctions and the handicap auction},
  author={Es{\H{o}}, P{\'e}ter and Szentes, Balazs},
  journal={The Review of Economic Studies},
  volume={74},
  number={3},
  pages={705--731},
  year={2007},
  publisher={Wiley-Blackwell}
}

@inproceedings{BSX-20,
  title={Online pricing with offline data: Phase transition and inverse square law},
  author={Bu, Jinzhi and Simchi-Levi, David and Xu, Yunzong},
  booktitle={International Conference on Machine Learning},
  pages={1202--1210},
  year={2020},
  organization={PMLR}
}

@article{CSW-22,
  title={Privacy-preserving dynamic personalized pricing with demand learning},
  author={Chen, Xi and Simchi-Levi, David and Wang, Yining},
  journal={Management Science},
  volume={68},
  number={7},
  pages={4878--4898},
  year={2022},
  publisher={INFORMS}
}

@article{HKZ-12,
  title={Bayesian dynamic pricing policies: Learning and earning under a binary prior distribution},
  author={Harrison, J Michael and Keskin, N Bora and Zeevi, Assaf},
  journal={Management Science},
  volume={58},
  number={3},
  pages={570--586},
  year={2012},
  publisher={INFORMS}
}

@article{CS-19,
  title={Joint pricing and inventory management with strategic customers},
  author={Chen, Yiwei and Shi, Cong},
  journal={Operations Research},
  volume={67},
  number={6},
  pages={1610--1627},
  year={2019},
  publisher={INFORMS}
}

@article{CF-18,
  title={Robust dynamic pricing with strategic customers},
  author={Chen, Yiwei and Farias, Vivek F},
  journal={Mathematics of Operations Research},
  volume={43},
  number={4},
  pages={1119--1142},
  year={2018},
  publisher={INFORMS}
}

@article{RS-17,
  title={Buyer-optimal learning and monopoly pricing},
  author={Roesler, Anne-Katrin and Szentes, Bal{\'a}zs},
  journal={American Economic Review},
  volume={107},
  number={7},
  pages={2072--2080},
  year={2017},
  publisher={American Economic Association 2014 Broadway, Suite 305, Nashville, TN 37203}
}

@article{AC-09,
  title={Dynamic pricing for nonperishable products with demand learning},
  author={Araman, Victor F and Caldentey, Ren{\'e}},
  journal={Operations research},
  volume={57},
  number={5},
  pages={1169--1188},
  year={2009},
  publisher={INFORMS}
}

@article{BR-12,
  title={Dynamic pricing under a general parametric choice model},
  author={Broder, Josef and Rusmevichientong, Paat},
  journal={Operations Research},
  volume={60},
  number={4},
  pages={965--980},
  year={2012},
  publisher={INFORMS}
}

@article{BZ-15,
  title={On the (surprising) sufficiency of linear models for dynamic pricing with demand learning},
  author={Besbes, Omar and Zeevi, Assaf},
  journal={Management Science},
  volume={61},
  number={4},
  pages={723--739},
  year={2015},
  publisher={INFORMS}
}

@article{CSWZ-22,
  title={Dynamic pricing and inventory control with fixed ordering cost and incomplete demand information},
  author={Chen, Boxiao and Simchi-Levi, David and Wang, Yining and Zhou, Yuan},
  journal={Management Science},
  volume={68},
  number={8},
  pages={5684--5703},
  year={2022},
  publisher={INFORMS}
}

@article{CWZ-24,
  title={Optimal policies for dynamic pricing and inventory control with nonparametric censored demands},
  author={Chen, Boxiao and Wang, Yining and Zhou, Yuan},
  journal={Management Science},
  volume={70},
  number={5},
  pages={3362--3380},
  year={2024},
  publisher={INFORMS}
}

@inproceedings{TH-21,
  title={On the bayesian rational assumption in information design},
  author={Tang, Wei and Ho, Chien-Ju},
  booktitle={Proceedings of the AAAI Conference on Human Computation and Crowdsourcing},
  volume={9},
  pages={120--130},
  year={2021}
}

@article{BCW-22,
  title={Third-degree price discrimination versus uniform pricing},
  author={Bergemann, Dirk and Castro, Francisco and Weintraub, Gabriel},
  journal={Games and Economic Behavior},
  volume={131},
  pages={275--291},
  year={2022},
  publisher={Elsevier}
}

@article{AFT-23,
  title={Dynamic pricing and learning with bayesian persuasion},
  author={Agrawal, Shipra and Feng, Yiding and Tang, Wei},
  journal={Advances in Neural Information Processing Systems},
  volume={36},
  pages={59273--59285},
  year={2023}
}

@article{EH-21,
  title={The power of opaque products in pricing},
  author={Elmachtoub, Adam N and Hamilton, Michael L},
  journal={Management Science},
  volume={67},
  number={8},
  pages={4686--4702},
  year={2021},
  publisher={INFORMS}
}

@article{AK-20,
  title={Competitive information disclosure by multiple senders},
  author={Au, Pak Hung and Kawai, Keiichi},
  journal={Games and Economic Behavior},
  volume={119},
  pages={56--78},
  year={2020},
  publisher={Elsevier}
}

@article{BHM-26,
  title={Screening with persuasion},
  author={Bergemann, Dirk and Heumann, Tibor and Morris, Stephen},
  journal={Journal of Political Economy},
  volume={134},
  number={2},
  pages={570--625},
  year={2026},
  publisher={The University of Chicago Press Chicago, IL}
}

@article{BDHN-22,
  title={Selling Information in Competitive Environments},
  author={Bonatti, Alessandro and Dahleh, Munther and Horel, Thibaut and Nouripour, Amir},
  journal={arXiv preprint arXiv:2202.08780},
  year={2022}
}

@article{CZ-20,
  title={Signalling by bayesian persuasion and pricing strategy},
  author={Chen, Yanlin and Zhang, Jun},
  journal={The Economic Journal},
  volume={130},
  number={628},
  pages={976--1007},
  year={2020},
  publisher={Oxford University Press}
}

@article{FTX-22,
  title={No-Regret Bayesian Recommendation to Homogeneous Users},
  author={Feng, Yiding and Tang, Wei and Xu, Haifeng},
  journal={Operations Research},
  year={2026},
  publisher={INFORMS}
}

@inproceedings{APS-12,
  title={Online-to-confidence-set conversions and application to sparse stochastic bandits},
  author={Abbasi-Yadkori, Yasin and Pal, David and Szepesvari, Csaba},
  booktitle={Artificial Intelligence and Statistics},
  pages={1--9},
  year={2012},
  organization={PMLR}
}

@inproceedings{KSU-08,
  title={Multi-armed bandits in metric spaces},
  author={Kleinberg, Robert and Slivkins, Aleksandrs and Upfal, Eli},
  booktitle={Proceedings of the fortieth annual ACM symposium on Theory of computing},
  pages={681--690},
  year={2008}
}

@inproceedings{S-11,
  title={Contextual bandits with similarity information},
  author={Slivkins, Aleksandrs},
  booktitle={Proceedings of the 24th annual Conference On Learning Theory},
  pages={679--702},
  year={2011},
  organization={JMLR Workshop and Conference Proceedings}
}

@inproceedings{L-22,
  title={Selling data to an agent with endogenous information},
  author={Li, Yingkai},
  booktitle={Proceedings of the 23rd ACM Conference on Economics and Computation},
  pages={664--665},
  year={2022}
}

@inproceedings{ZC-21,
  title={Optimal advertising for information products},
  author={Zheng, Shuran and Chen, Yiling},
  booktitle={Proceedings of the 22nd ACM Conference on Economics and Computation},
  pages={888--906},
  year={2021}
}

@inproceedings{CXZ-20,
  title={Selling information through consulting},
  author={Chen, Yiling and Xu, Haifeng and Zheng, Shuran},
  booktitle={Proceedings of the Fourteenth Annual ACM-SIAM Symposium on Discrete Algorithms},
  pages={2412--2431},
  year={2020},
  organization={SIAM}
}

@inproceedings{CV-21,
  title={How to Sell Information Optimally: An Algorithmic Study},
  author={Cai, Yang and Velegkas, Grigoris},
  booktitle={Proceedings of the12th Innovations in Theoretical Computer Science Conference},
  volume={185},
  year={2021}
}

@article{BB-15,
  title={Selling cookies},
  author={Bergemann, Dirk and Bonatti, Alessandro},
  journal={American Economic Journal: Microeconomics},
  volume={7},
  number={3},
  pages={259--94},
  year={2015}
}

@article{BBS-18,
  title={The design and price of information},
  author={Bergemann, Dirk and Bonatti, Alessandro and Smolin, Alex},
  journal={American economic review},
  volume={108},
  number={1},
  pages={1--48},
  year={2018}
}

@inproceedings{LSX-21,
  title={Optimal Pricing of Information},
  author={Liu, Shuze and Shen, Weiran and Xu, Haifeng},
  booktitle={Proceedings of the 22nd ACM Conference on Economics and Computation},
  pages={693--693},
  year={2021}
}

@article{BCVZ-22,
  title={Is Selling Complete Information (Approximately) Optimal?},
  author={Bergemann, Dirk and Cai, Yang and Velegkas, Grigoris and Zhao, Mingfei},
  journal={arXiv preprint arXiv:2202.09013},
  year={2022}
}

@inproceedings{BKP-12,
  title={Optimal mechanisms for selling information},
  author={Babaioff, Moshe and Kleinberg, Robert and Paes Leme, Renato},
  booktitle={Proceedings of the 13th ACM Conference on Electronic Commerce},
  pages={92--109},
  year={2012}
}

@article{KZ-14,
  title={Dynamic pricing with an unknown demand model: Asymptotically optimal semi-myopic policies},
  author={Keskin, N Bora and Zeevi, Assaf},
  journal={Operations research},
  volume={62},
  number={5},
  pages={1142--1167},
  year={2014},
  publisher={INFORMS}
}

@article{BZ-09,
  title={Dynamic pricing without knowing the demand function: Risk bounds and near-optimal algorithms},
  author={Besbes, Omar and Zeevi, Assaf},
  journal={Operations Research},
  volume={57},
  number={6},
  pages={1407--1420},
  year={2009},
  publisher={INFORMS}
}

@article{BDKS-15,
  author       = {Moshe Babaioff and
                  Shaddin Dughmi and
                  Robert D. Kleinberg and
                  Aleksandrs Slivkins},
  title        = {Dynamic Pricing with Limited Supply},
  journal      = {{ACM} Trans. Economics and Comput.},
  volume       = {3},
  number       = {1},
  pages        = {4:1--4:26},
  year         = {2015},
  url          = {https://doi.org/10.1145/2559152},
  doi          = {10.1145/2559152},
  bibsource    = {dblp computer science bibliography, https://dblp.org}
}

@inproceedings{KL-03,
  title={The value of knowing a demand curve: Bounds on regret for online posted-price auctions},
  author={Kleinberg, Robert and Leighton, Tom},
  booktitle={44th Annual IEEE Symposium on Foundations of Computer Science, 2003. Proceedings.},
  pages={594--605},
  year={2003},
  organization={IEEE}
}

@article{CIMS-17,
  title={Monopoly pricing in the presence of social learning},
  author={Crapis, Davide and Ifrach, Bar and Maglaras, Costis and Scarsini, Marco},
  journal={Management Science},
  volume={63},
  number={11},
  pages={3586--3608},
  year={2017},
  publisher={INFORMS}
}

@article{SVZ-22,
  title={Dynamic pricing with online reviews},
  author={Shin, Dongwook and Vaccari, Stefano and Zeevi, Assaf},
  journal={Management Science},
  year={2022},
  publisher={INFORMS}
}

@article{IMSZ-19,
  title={Bayesian social learning from consumer reviews},
  author={Ifrach, Bar and Maglaras, Costis and Scarsini, Marco and Zseleva, Anna},
  journal={Operations Research},
  volume={67},
  number={5},
  pages={1209--1221},
  year={2019},
  publisher={INFORMS}
}

@misc{stat-digital-ads,
  author       = {{Statista}},
  title        = {Digital Advertising---Worldwide},
  year         = {2026},
  howpublished = {\url{https://www.statista.com/outlook/dmo/digital-advertising/worldwide}},
  note         = {Statista Market Forecast, accessed July 2026}
}

@article{CEL-22,
  title={Price discrimination with fairness constraints},
  author={Cohen, Maxime C and Elmachtoub, Adam N and Lei, Xiao},
  journal={Management Science},
  volume={68},
  number={12},
  pages={8536--8552},
  year={2022},
  publisher={INFORMS}
}

@inproceedings{BMSW-24,
  title={Fair price discrimination},
  author={Banerjee, Siddhartha and Munagala, Kamesh and Shen, Yiheng and Wang, Kangning},
  booktitle={Proceedings of the 2024 Annual ACM-SIAM Symposium on Discrete Algorithms (SODA)},
  pages={2679--2703},
  year={2024},
  organization={SIAM}
}

@article{BBM-15,
  title={The limits of price discrimination},
  author={Bergemann, Dirk and Brooks, Benjamin and Morris, Stephen},
  journal={American Economic Review},
  volume={105},
  number={3},
  pages={921--957},
  year={2015},
  publisher={American Economic Association 2014 Broadway, Suite 305, Nashville, TN 37203}
}

@article{MDGC-06,
  title={The effect of banner advertising on internet purchasing},
  author={Manchanda, Puneet and Dub{\'e}, Jean-Pierre and Goh, Khim Yong and Chintagunta, Pradeep K},
  journal={Journal of Marketing Research},
  volume={43},
  number={1},
  pages={98--108},
  year={2006},
  publisher={SAGE Publications Sage CA: Los Angeles, CA}
}

@article{N-70,
  title={Information and consumer behavior},
  author={Nelson, Phillip},
  journal={Journal of political economy},
  volume={78},
  number={2},
  pages={311--329},
  year={1970},
  publisher={The University of Chicago Press}
}

@article{N-74,
  title={Advertising as information},
  author={Nelson, Phillip},
  journal={Journal of political economy},
  volume={82},
  number={4},
  pages={729--754},
  year={1974},
  publisher={The University of Chicago Press}
}

@article{HKB-19,
  title={Competitive Advertising and Pricing},
  author={Hwang, Ilwoo and Kim, Kyungmin and Boleslavsky, Raphael},
    journal={working paper},
  year={2019}
}

@inproceedings{BS-12,
  title={Send mixed signals: earn more, work less},
  author={Bro Miltersen, Peter and Sheffet, Or},
  booktitle={Proceedings of the 13th ACM Conference on Electronic Commerce},
  pages={234--247},
  year={2012}
}

@article{AB-19,
  title={Private bayesian persuasion},
  author={Arieli, Itai and Babichenko, Yakov},
  journal={Journal of Economic Theory},
  volume={182},
  pages={185--217},
  year={2019},
  publisher={Elsevier}
}

@article{EFGPT-14,
  title={Signaling schemes for revenue maximization},
  author={Emek, Yuval and Feldman, Michal and Gamzu, Iftah and PaesLeme, Renato and Tennenholtz, Moshe},
  journal={ACM Transactions on Economics and Computation (TEAC)},
  volume={2},
  number={2},
  pages={1--19},
  year={2014},
  publisher={ACM New York, NY, USA}
}

@article{JR-94,
  title={Price and quality in a new product monopoly},
  author={Judd, Kenneth L and Riordan, Michael H},
  journal={The Review of Economic Studies},
  volume={61},
  number={4},
  pages={773--789},
  year={1994},
  publisher={Wiley-Blackwell}
}

@article{KOU-22,
  title={Signaling in online credit markets},
  author={Kawai, Kei and Onishi, Ken and Uetake, Kosuke},
  journal={Journal of Political Economy},
  volume={130},
  number={6},
  pages={1585--1629},
  year={2022},
  publisher={The University of Chicago Press Chicago, IL}
}

@article{KR-84,
  title={Advertising as a Signal},
  author={Kihlstrom, Richard E and Riordan, Michael H},
  journal={Journal of Political Economy},
  volume={92},
  number={3},
  pages={427--450},
  year={1984},
  publisher={The University of Chicago Press}
}

@article{MR-86,
  title={Price and advertising signals of product quality},
  author={Milgrom, Paul and Roberts, John},
  journal={Journal of political economy},
  volume={94},
  number={4},
  pages={796--821},
  year={1986},
  publisher={The University of Chicago Press}
}

@article{SN-20,
  title={Does advertising serve as a signal? Evidence from a field experiment in mobile search},
  author={Sahni, Navdeep S and Nair, Harikesh S},
  journal={The Review of Economic Studies},
  volume={87},
  number={3},
  pages={1529--1564},
  year={2020},
  publisher={Oxford University Press}
}

@inproceedings{CMCG-21,
  title={Multi-receiver online bayesian persuasion},
  author={Castiglioni, Matteo and Marchesi, Alberto and Celli, Andrea and Gatti, Nicola},
  booktitle={International Conference on Machine Learning},
  pages={1314--1323},
  year={2021},
  organization={PMLR}
}

@article{CCMG-20,
  title={Online Bayesian Persuasion},
  author={Castiglioni, Matteo and Celli, Andrea and Marchesi, Alberto and Gatti, Nicola},
  journal={Advances in Neural Information Processing Systems},
  volume={33},
  year={2020}
}

@article{ZIX-21,
  title={Learning to Persuade on the Fly: Robustness Against Ignorance},
  author={Zu, You and Iyer, Krishnamurthy and Xu, Haifeng},
  journal={Operations Research},
  year={2024},
  publisher={INFORMS}
}

@article{D-17,
  title={Algorithmic information structure design: a survey},
  author={Dughmi, Shaddin},
  journal={ACM SIGecom Exchanges},
  volume={15},
  number={2},
  pages={2--24},
  year={2017},
  publisher={ACM New York, NY, USA}
}

@article{BM-19,
  title={Information design: A unified perspective},
  author={Bergemann, Dirk and Morris, Stephen},
  journal={Journal of Economic Literature},
  volume={57},
  number={1},
  pages={44--95},
  year={2019}
}

@article{Kamenica-19,
  title={Bayesian persuasion and information design},
  author={Kamenica, Emir},
  journal={Annual Review of Economics},
  volume={11},
  pages={249--272},
  year={2019},
  publisher={Annual Reviews}
}

@book{AMS-95,
  title={Repeated games with incomplete information},
  author={Aumann, Robert J and Maschler, Michael and Stearns, Richard E},
  year={1995},
  publisher={MIT press}
}

@book{BG-79,
  title={Theory of games and statistical decisions},
  author={Blackwell, David A and Girshick, Meyer A},
  year={1979},
  publisher={Courier Corporation}
}

@inproceedings{CS-21,
  title={Optimal Disclosure of Information to a Privately Informed Receiver},
  author={Candogan, Ozan and Strack, Philipp},
  booktitle={Proceedings of the 22nd ACM Conference on Economics and Computation},
  pages={263--263},
  year={2021}
}

@article{KMZL-17,
  title={Persuasion of a privately informed receiver},
  author={Kolotilin, Anton and Mylovanov, Tymofiy and Zapechelnyuk, Andriy and Li, Ming},
  journal={Econometrica},
  volume={85},
  number={6},
  pages={1949--1964},
  year={2017},
  publisher={Wiley Online Library}
}

@article{KG-11,
  title={Bayesian persuasion},
  author={Kamenica, Emir and Gentzkow, Matthew},
  journal={American Economic Review},
  volume={101},
  number={6},
  pages={2590--2615},
  year={2011}
}

@inproceedings{ABSY-19,
  title={Optimal Persuasion via Bi-Pooling},
  author={Arieli, Itai and Babichenko, Yakov and Smorodinsky, Rann and Yamashita, Takuro},
  booktitle={Proceedings of the 21st ACM Conference on Economics and Computation},
  pages={641--641},
  year={2020}
}

@inproceedings{C-19,
  title={Persuasion in networks: Public signals and k-cores},
  author={Candogan, Ozan},
  booktitle={Proceedings of the 2019 ACM Conference on Economics and Computation},
  pages={133--134},
  year={2019}
}
